\documentclass[
  aps,
  prl,
  reprint,
  superscriptaddress,
  amsmath,
  amssymb,
  showkeys
]{revtex4-2}

\usepackage{graphicx}
\usepackage{bm}
\usepackage{microtype}
\usepackage[colorlinks=true,citecolor=blue,linkcolor=blue,urlcolor=blue]{hyperref}
\usepackage{bbm}
\usepackage{xcolor}

\usepackage{pdfpages}

\makeatletter
\AtBeginDocument{\let\LS@rot\@undefined}
\makeatother

\begin{document}

%\title{Quantum-Geometric Amplification of an Exponentially Small Interaction Gap}
\title{Quantum-Geometric Amplification of Nonperturbative Interaction Scales}

\author{Stefano Bilamour}
\affiliation{Department of Physics and Astronomy, University of Bologna, 40127 Bologna, Italy}
\affiliation{Institut f\"{u}r Theoretische Physik und Astrophysik and W\"{u}rzburg-Dresden Cluster of Excellence ctd.qmat, Julius-Maximilians-Universit\"{a}t W\"{u}rzburg, 97074 W\"{u}rzburg, Germany.}

\author{Ren\'{e} Meyer}
\affiliation{Institut f\"{u}r Theoretische Physik und Astrophysik and W\"{u}rzburg-Dresden Cluster of Excellence ctd.qmat, Julius-Maximilians-Universit\"{a}t W\"{u}rzburg, 97074 W\"{u}rzburg, Germany.}
\affiliation{Shanghai Institute for Mathematics and Interdisciplinary Sciences (SIMIS),
Shanghai 200433,
China}

\author{Johanna Erdmenger}
\affiliation{Institut f\"{u}r Theoretische Physik und Astrophysik and W\"{u}rzburg-Dresden Cluster of Excellence ctd.qmat, Julius-Maximilians-Universit\"{a}t W\"{u}rzburg, 97074 W\"{u}rzburg, Germany.}

\author{Domenico Di Sante}\email{domenico.disante@unibo.it}
\affiliation{Department of Physics and Astronomy, University of Bologna, 40127 Bologna, Italy}

\date{\today}

\begin{abstract}
Weak interactions at a two-dimensional quadratic band touching (QBT) can generate an exponentially small symmetry-breaking gap, $m\sim e^{-A/V_1}$, that cannot be captured by every finite order of perturbation theory. We show that the singular quantum geometry of the gapped QBT converts this beyond-all-orders scale into an algebraically enhanced response. The quantum metric develops a momentum-space hot spot whose Brillouin-zone integral diverges logarithmically, and therefore grows algebraically when the gap is dynamically generated by interactions. Through the inverse-frequency Souza--Wilkens--Martin optical sum rule, the same amplification appears in the negative-first longitudinal optical moment, even though the ordinary optical conductivity remains $O(e^2/\hbar)$ and its absorption threshold is exponentially small. We establish these results analytically for a massive QBT and verify them microscopically in an interacting kagome lattice model known to develop a spontaneous quantum anomalous Hall mass at its QBT from loop currents order. Our results identify quantum geometry as an asymptotic amplifier of nonperturbative interaction scales.
\end{abstract}

\maketitle

\textit{Introduction --} Weak interactions can generate energy scales that are invisible to every finite order of perturbation theory. A paradigmatic example occurs at a two-dimensional quadratic band touching (QBT), where short-range interactions are marginal and can destabilize the semimetal at an exponentially small scale. Sun \textit{et al.}~\cite{Sun2009} showed that a symmetry-protected QBT is marginally unstable to arbitrarily weak repulsive interactions, with quantum anomalous Hall (QAH) and nematic states among the possible broken-symmetry phases. Interaction-driven topological phases at QBTs were subsequently studied in microscopic kagome and checkerboard models~\cite{Liu2010,Wen2010,Sur2018,Ren2018,baum2026exotic}. Although such instabilities can determine the ground-state topology, their characteristic scale $m\sim e^{-A/V_1}$ with $V_1$ the interaction strength and $A$ a dimensional constant, becomes exponentially difficult to resolve as the weak-coupling $V_1 \rightarrow 0$ limit is approached. This raises a broader question: can an observable retain a parametrically strong signature of an interaction-generated scale that is itself beyond all algebraic orders in the interaction?
In this letter, we find that quantum geometry~\cite{provost1980riemannian,Fubini1904,Study1905} provides a natural setting in which such an amplification may occur.

The quantum metric characterizes the variation of Bloch wave functions in momentum space and has emerged as an important ingredient in a broad range of physical responses, including superfluid transport, correlation effects and optical phenomena~\cite{Peotta2015,TormaPRLessay2023,Yu2025,verma2026quantum,Witt2026}. Its Brillouin-zone integral is closely connected to electronic localization~\cite{resta2011insulating,RestaSorella1999} and, through the Souza--Wilkens--Martin sum rule, to an inverse-frequency moment of the optical conductivity~\cite{Souza2000}. Closely related fluctuation--response relations connect the quantum metric to equilibrium current noise, providing an independent probe of
the geometry of Bloch states~\cite{Neupert2013}. More recently, Verma and Queiroz~\cite{Verma2025} emphasized this connection in the context of direct quantum-metric measurements through step response. At the same time, Oh \textit{et al.}~\cite{Oh2026} established a universal quantum-geometric optical conductivity for two-dimensional QBT systems and analyzed the effect of opening a small gap. These developments have largely addressed the two ingredients separately: quantum-geometric response is typically analyzed for a prescribed band gap, whereas studies of interaction-driven QBT instabilities emphasize the ordered phases and the nonperturbatively generated gap. Here we focus on their composition, how the singular quantum geometry acts on the exponentially small scale generated by the same marginal interaction that destabilizes the QBT.

\begin{figure*}[!t]
\centering
\includegraphics[width=\textwidth]{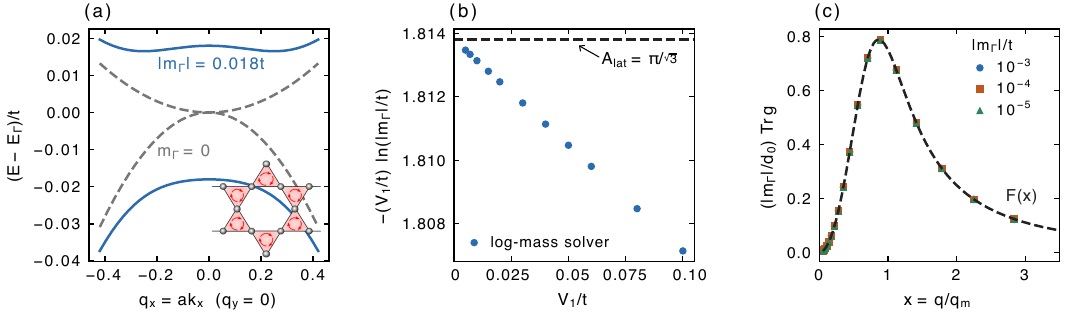}
\caption{
(a) Low-energy dispersion along \(q_y=0\) near the \(\Gamma\)-point QBT for the regularized kagome lattice at \(r/t=0.05\). Energies are measured relative to \(E_\Gamma=2t\). Dashed curves show the ungapped case \(m_\Gamma=0\); solid curves show a representative QAH mass \(|m_\Gamma|=0.018t\) from loop currents (see inset) corresponding to (a deliberately large for plotting purposes) \(V_1/t\simeq0.44\) from the weak-coupling lattice mass relation Eq.~\eqref{eq:lattice_mass}. 
(b) Weak-coupling scaling of the interaction-generated lattice mass. The quantity $-(V_1/t)\ln(|m_\Gamma|/t)$, obtained from the scalar log-mass self-consistency equation, approaches the parameter-free asymptotic coefficient $A_{\rm lat}=\pi/\sqrt3$ (dashed line) as $V_1/t\rightarrow0$; no exponential coefficient is fitted.
(c) Scaling collapse of the full-lattice quantum-metric trace near the QBT. With $q_m=\sqrt{|m_\Gamma|/d}$ and $x=q/q_m$, lattice results for $|m_\Gamma|/t=10^{-3},10^{-4},10^{-5}$ collapse onto the continuum scaling function $F(x)=x^2(2+x^4)/(1+x^4)^2$, demonstrating the $1/|m_\Gamma|$ growth and $\sqrt{|m_\Gamma|}$ narrowing of the geometric hot spot.
}
\label{fig:mass_geometry}
\end{figure*}

We show that this combination changes the asymptotic class (behavior) of the response. For a massive QBT, the quantum metric develops a momentum-space hot spot whose height diverges as $1/|m|$, while its characteristic width shrinks as $|m|^{1/2}$. Momentum integration yields a logarithmic infrared singularity of the form 
$G(m)\sim C\ln\frac{1}{|m|}$,
where the proprotionality constant $C$ depends just on the unit cell size, and therefore is finite. When the mass is generated by a marginal interaction $V_1$, $m(V_1)\sim M e^{-A/V_1}$, the logarithm converts the nonperturbative scale into an algebraic response
$G(V_1)\sim\frac{1}{V_1}$. 

Through the inverse-frequency optical sum rule~\cite{Souza2000,Verma2025}, the same amplification is directly transferred to the negative-first (inverse-frequency) longitudinal optical moment
\begin{equation}
W_{-1}^{\mathrm{tr}}
\equiv
\int_0^\infty\frac{d\omega}{\omega}
\,\mathrm{Re}\!\left[
\sigma_{xx}(\omega)+\sigma_{yy}(\omega)
\right]
\sim\frac{\mathrm{const}}{V_1},
\end{equation}
where the subscript $-1$ denotes the frequency moment $\omega^{-1}$, and $\mathrm{tr}$ denotes the Cartesian trace over the longitudinal components. Remarkably, the ordinary optical conductivity itself does not become exponentially large. Instead, it remains $O(e^2/\hbar)$, consistent with the quantum-geometric QBT conductivity~\cite{Oh2026}, while the absorption threshold is pushed to the exponentially small scale $\hbar\omega_g\sim e^{-A/V_1}$. The algebraic enhancement results from the logarithmically expanding frequency interval over which this finite optical continuum contributes.%\rene{What sets the upper cutoff of this integration region?}

We establish this mechanism analytically for a generic massive QBT and verify it microscopically in a regularized spinless kagome model with repulsive interactions. The full lattice calculation generates a spontaneous QAH mass and reproduces the continuum infrared geometry, while the integrated metric obeys $G_{\rm cell}=d_0/V_1+O(1)$, with the leading coefficient $d_0=t/8$ depending on the hopping $t$ of the microscopic kagome model. We further calculate the optical response of the model, finding $W_{-1}^{\mathrm{tr}} = \left(\frac{e^2}{\hbar}\right)\pi t/4\sqrt3 V_1 + O(1)$ and showing that the singular contribution originates from the low-energy QBT sector, whereas remote-band transitions remain finite.
Our results demonstrate that quantum geometry can act as an \emph{asymptotic amplifier}, converting an exponentially hidden interaction scale into an algebraically enhanced measurable response.

\textit{Microscopic model --} We realize this setting microscopically using spinless fermions on the kagome lattice with repulsive nearest-neighbor interactions,
\begin{equation}
H=
\sum_{\mathbf k}\Psi_{\mathbf k}^{\dagger}
H_{\rm kag}(\mathbf k;r)\Psi_{\mathbf k}
+
V_1\sum_{\langle ij\rangle}
(n_i-\bar n)(n_j-\bar n).
\label{eq:lattice_model}
\end{equation}
For spinless fermions with a single orbital per site, an on-site density interaction is trivial because \(n_i^2=n_i\); \(V_1\) is therefore the shortest-range nontrivial interaction.
Here $r$ denotes a weak symmetry-preserving regularization of the exactly-flat limit; its explicit implementation and our lattice conventions are given in the Supplemental Material Sec.~I~\cite{suppmat}.

\textit{Results --} At filling two for the model in Eq.~\eqref{eq:lattice_model}, Hartree--Fock self-consistent theory admits a cyclic complex bond order $\chi_b=\chi_R+i\nu_b\chi_I$ (Ref.~\cite{suppmat} Sec.~II and inset in Fig.~\ref{fig:mass_geometry}(a)). Its imaginary component breaks time reversal and projects onto the $\Gamma$-point QBT as a mass $m_\Gamma=-2\sqrt3V_1\chi_I$~\cite{Wen2010}, producing the two time-reversal-related QAH states with Chern number $C=\pm1$ (Ref.~\cite{suppmat} Sec.~IV). Figure~\ref{fig:mass_geometry}(a) shows the resulting gap.

Near the touching, the low-energy Hamiltonian is,  according to Ref.~\cite{suppmat} Sec.~I, 
\begin{equation}
H_{\rm QBT}(\mathbf q)
=
-cq^2\mathbbm{1}
-d\left[
(q_x^2-q_y^2)\sigma_z
+
2q_xq_y\sigma_x
\right]
+
m\sigma_y .
\label{eq:qbt_hamiltonian}
\end{equation}
Hereafter, we use \(m\) and \(m_\Gamma\) to distinguish the generic continuum QBT mass from the microscopic self-consistent kagome mass generated by the interaction, respectively.
The scalar term $cq^2\mathbbm{1}=c(q_x^2+q_y^2)\mathbbm{1}$ changes the dispersion but not the eigenstates, and $c,d > 0$. For a marginally relevant interaction in the QAH channel, the familiar zero-temperature gap equation is~\cite{Sun2009}
\begin{eqnarray}
\frac{1}{\lambda}
&=&
\int^\Lambda\frac{d^2q}{(2\pi)^2}
\frac1{\sqrt{d^2q^4+m^2}}
=
\frac{1}{4\pi d}
\operatorname{asinh}\frac{d\Lambda^2}{|m|} \, , \nonumber \\
&\Longrightarrow&
|m|\sim M e^{-4\pi d/\lambda},
\label{eq:gap_equation}
\end{eqnarray}
with $\Lambda$ an ultraviolet momentum cutoff. Eq.~\eqref{eq:gap_equation} exposes the intrinsically nonperturbative scale generated by the
interaction. In the weak-coupling limit, the QBT susceptibility diverges
logarithmically, so that the self-consistency condition does not produce a
gap that is analytic in $\lambda$, but instead yields the essential
singularity $|m|\sim M\exp\!\left(-\frac{4\pi d}{\lambda}\right)$.
This is a beyond-all-orders scale: as $\lambda\to0^+$, $m$ vanishes faster
than any power of $\lambda$ and is therefore invisible to perturbation theory
at any finite order.  The interaction-driven QAH gap is thus controlled by
physics that cannot be inferred from a power-series expansion about the
noninteracting limit.

We next connect this continuum result to the microscopic kagome lattice.
For a prescribed QAH mass $m_\Gamma$, the full three-band lattice
Hamiltonian of Eq.~\eqref{eq:lattice_model} is diagonalized throughout the Brillouin zone and the induced
QAH bond response is used to define a lattice kernel
$K_{\rm lat}^{\rm cell}(m_\Gamma)$ (Ref.~\cite{suppmat} Sec.~III).  The Hartree--Fock self-consistency
condition can then be written as
$1/V_1=K_{\rm lat}^{\rm cell}(m_\Gamma)$.
We solve this equation numerically using the full Brillouin-zone lattice
kernel and a logarithmic mass variable
$u=\ln(t/|m_\Gamma|)$, which allows exponentially small gaps to be
resolved without directly sampling their exponentially small momentum
scale. The resulting self-consistent lattice solutions are shown as the
blue points in Fig.~\ref{fig:mass_geometry}(b).

In the weak-mass limit, the logarithmically divergent part of the kernel $K_{\rm lat}^{\rm cell}(m_\Gamma)$
originates entirely from the two bands forming the QBT near $\Gamma$,
whereas states at generic momenta and the remote third kagome band contribute
finite lattice corrections.  As derived in Ref.~\cite{suppmat} Sec.~III,
\begin{equation}
K_{\rm lat}^{\rm cell}(m_\Gamma)
=
\frac{A_c}{4\pi d_0}
\ln\frac{t}{|m_\Gamma|}
+
B_{\rm lat}
+
o(1),
\end{equation}
where $A_c=\sqrt{3}/2$ is the dimensionless kagome unit cell area and $B_{\rm lat}$ contains the finite full-Brillouin-zone contribution.
Inverting this relation gives
\begin{equation}
|m_\Gamma(V_1)|
\sim
M_{\rm lat}e^{-A_{\rm lat}/V_1},
\qquad
A_{\rm lat}
=
\frac{4\pi d_0}{A_c},
\label{eq:lattice_mass}
\end{equation}
with the nonuniversal prefactor $M_{\rm lat}$ determined by the finite
lattice contribution $B_{\rm lat}$.  Thus the coefficient controlling the
essential singularity is fixed solely by the infrared QBT, while the
prefactor retains information about the complete microscopic band
structure. Expanding the microscopic kagome Hamiltonian about \(\Gamma\) gives \(d_0=d=t/8\) (Ref.~\cite{suppmat} Sec.~I). Therefore, for a unit hopping parameter \(t=1\), we find \(A_{\rm lat}=\pi/\sqrt3\), in excellent agreement with the asymptotic behavior extracted from the full-lattice numerical Hartree–Fock solutions shown in Fig.~\ref{fig:mass_geometry}(b).

We now turn to the geometric consequence of the nonperturbative scale
generated by Eqs.~\eqref{eq:gap_equation} and
\eqref{eq:lattice_mass}.  For the continuum QBT Hamiltonian of
Eq.~\eqref{eq:qbt_hamiltonian}, the trace of the quantum metric takes
the scaling form (Ref.~\cite{suppmat} Sec. V)
\begin{equation}
\operatorname{Tr}g(\mathbf q,m)
=
\frac{d^2q^2(2m^2+d^2q^4)}
     {(m^2+d^2q^4)^2}
=
\frac{d}{|m|}
F\!\left(\frac{q}{q_m}\right),
\label{eq:metric_scaling}
\end{equation}
where
$F(x)=x^2(2+x^4)/(1+x^4)^2$ and $q_m=\sqrt{\frac{|m|}{d}}$.
The interaction-generated mass therefore introduces a charac\-teristic
momentum scale $q_m\propto |m|^{1/2}$.  As $m\to0$, the geometric hot
spot around the QBT becomes increasingly high,
while simultaneously collapsing into an
increasingly narrow region of momentum space.  The full three-band
kagome metric exhibits precisely this scaling collapse in the shrinking
neighborhood of $\Gamma$, as shown in
Fig.~\ref{fig:mass_geometry}(c) for several values of $m_{\Gamma}$.

\begin{figure}[!b]
\centering
\includegraphics[width=\columnwidth]{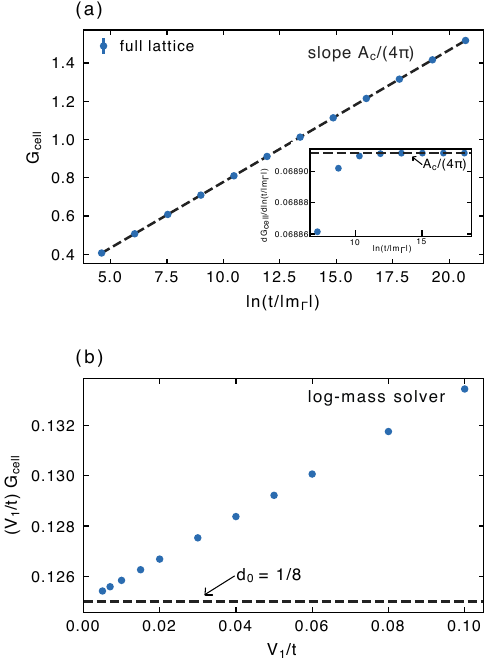}
\caption{
(a) Full-lattice integrated quantum metric $G_{\rm cell}$ as a function
of $\ln(t/|m_\Gamma|)$.  The weak-mass data approach the QBT prediction
with slope $A_c/(4\pi)$.  Inset: logarithmic derivative
$dG_{\rm cell}/d\ln(t/|m_\Gamma|)$, showing convergence to the analytic
coefficient $A_c/(4\pi) = 0.068916$.
(b) Interaction dependence of the integrated metric.  Combining the
logarithmic geometric singularity with the nonperturbative
self-consistent mass gives
$G_{\rm cell}\sim d_0/V_1$; the full-lattice numerical data approach
$(V_1/t)G_{\rm cell}=1/8$ for the kagome model.
}
\label{fig:integrated_geometry}
\end{figure}

The momentum integrated response has a subtler asymptotic behavior.  The central
region $q\lesssim q_m$ does not itself produce a divergence: its height
scales as $1/|m|$, but its phase-space area scales as
$q_m^2\sim |m|/d$, leaving an $O(1)$ contribution.  The divergence
instead originates from the parametrically broad intermediate regime
$q_m\ll q\ll \Lambda$, where the mass can be neglected and
$ \operatorname{Tr}g(\mathbf q,m) \simeq {1}/{q^2}$.
The two-dimensional phase-space integral then contains
$\int q\,dq/q^2=\int dq/q$, producing a logarithm.  The conti\-nuum theory
therefore gives
\begin{equation}
\begin{aligned}
G_{\rm QBT}(m)
&=
\int^{\Lambda}\frac{d^2q}{(2\pi)^2}\,
\operatorname{Tr}g
=
\frac{1}{4\pi}
\ln\frac{d\Lambda^2}{|m|}
+O(1).
\end{aligned}
\label{eq:integrated_metric}
\end{equation}
The lattice result is not a separate
singularity from the continuum one: it shares the same infrared logarithm,
multiplied by the primitive-cell normalization $A_c$ and supplemented by
finite contributions from the remainder of the Brillouin zone (Ref.~\cite{suppmat} Sec.~VI),
\begin{equation}
G_{\rm cell}(m_\Gamma)
=A_c\int_{\rm BZ}\frac{d^2k}{(2\pi)^2}\,\operatorname{Tr}g=\frac{A_c}{4\pi}\ln\frac{t}{|m_\Gamma|}
+B_g+o(1) \,.
\end{equation}
Here $B_g$ is a finite lattice contribution that depends on the
ultraviolet completion, whereas the coefficient of the logarithm is
fixed entirely by the QBT.  Indeed, the
full-lattice numerical results approach the analytically fixed slope
$A_c/(4\pi)$, as shown in Fig.~\ref{fig:integrated_geometry}(a), with
the corresponding logarithmic derivative displayed in the inset.

The integrated metric is the gauge-invariant contribution to the quadratic localization spread of the occupied subspace, namely the mean-square spatial extent of a Wannier state around its center~\cite{Marzari1997}. Its growth therefore signals increasing intrinsic delocalization as the interaction-generated QAH gap closes, providing a real-space interpretation of the geometric amplification. Related geometric constraints on Wannier localization have recently been shown to strongly influence correlated states and local-moment formation in flat-band systems~\cite{Witt2026}. In the present QAH phase this quantity remains well defined even though the nonzero Chern number obstructs the globally smooth Bloch frame required for exponentially localized Wannier functions~\cite{Brouder2007}. The logarithmic divergence found here is distinct from that topological obstruction: it reflects the singular growth of the gauge-invariant localization tensor as the QBT gap closes.

The crucial step is now to evaluate this logarithm at the
self-consistently generated mass.  From
Eq.~\eqref{eq:lattice_mass}, $ \ln\frac{t}{|m_\Gamma|}
=
\frac{A_{\rm lat}}{V_1}
-\ln\frac{M_{\rm lat}}{t}
+o(1)$,
such that we obtain
\begin{equation}
G_{\rm cell}(V_1)
=
\frac{d_0}{V_1}
+O(1).
\label{eq:metric_interaction_law}
\end{equation}
The finite lattice constant $B_g$, the mass prefactor $M_{\rm lat}$,
and the $O(V_1)$ Hartree--Fock renormalization of the quadratic
coefficient all modify only the $O(1)$ term (Ref.~\cite{suppmat} Sec.~VI); the coefficient of the
divergence is fixed by the weak-coupling value $d_0$.  For the kagome
model, the microscopic expansion gives $d_0=t/8$, and hence
\begin{equation}
\frac{V_1}{t}G_{\rm cell}
\longrightarrow
\frac18 \quad \text{for}
\quad
V_1/t\to0^+,
\end{equation}
where the weak-coupling limit removes the finite lattice corrections in
$G_{\rm cell}=d_0/V_1+O(1)$ and isolates the universal coefficient
$d_0/t=1/8$ fixed by the QBT,
in agreement with the full-lattice numerical Hartree--Fock asymptotics
shown in Fig.~\ref{fig:integrated_geometry}(b).

Momentum integration has therefore changed the asymptotic class of the
response.  The spectral gap is controlled by the beyond-all-orders scale
$m_\Gamma\sim e^{-A_{\rm lat}/V_1}$, but the singular QBT geometry
depends logarithmically on that scale.  Consequently,
\begin{equation}
 \ln\frac{1}{|m_\Gamma|}
\sim\frac{1}{V_1}   \, ,
\end{equation}
 so that an exponentially small interaction-generated spectral scale is
converted into an algebraically large integrated geometric response. Notably, the $1/V_1$ amplification is characteristic of a two-dimensional
QBT. In contrast to a two-dimensional Dirac cone, for which short-range
interactions are irrelevant at weak coupling because of the vanishing
low-energy density of states, $\rho_{\rm Dirac}(E)\sim |E|$~\cite{Gonzalez2001}, the finite
density of states of a QBT, $\rho_{\rm QBT}(E)\sim\mathrm{const}$, makes
short-range interactions marginal and generates the essential
nonperturbative scale $m\sim e^{-A/V_1}$.

\begin{figure}[!t]
\centering
\includegraphics[width=\columnwidth]{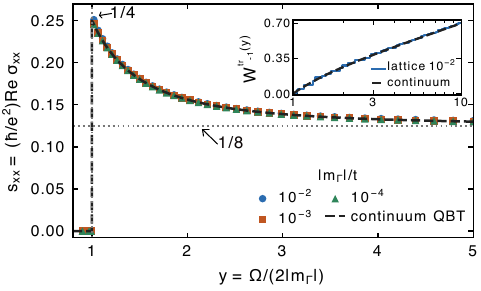}
\caption{
Longitudinal optical conductivity as a function of the scaled frequency
$y=\Omega/(2|m_\Gamma|)$. Full-lattice results for decreasing QAH masses
collapse onto the continuum QBT prediction above the absorption threshold
$y=1$, with threshold value $1/4$ and asymptotic plateau $1/8$.
Inset: cumulative inverse-frequency optical weight, showing the
logarithmic growth generated by the QBT optical plateau.
}
\label{fig:optical_threshold}
\end{figure}

This geometric amplification has a direct manifestation in the optical conductivity.  Writing
$\Omega=\hbar\omega$ and
$s_{ij}=(\hbar/e^2)\operatorname{Re}\sigma_{ij}$, the zero-temperature
interband Kubo formula for the massive QBT gives (Ref.~\cite{suppmat} Secs.~VII,VIII)
\begin{equation}
s_{xx}(\Omega)=s_{yy}(\Omega)
=
\frac18
\left(
1+\frac{4m^2}{\Omega^2}
\right)
\Theta(\Omega-2|m|)
.
\label{eq:optical_conductivity}
\end{equation}
The interaction-generated mass therefore sets the absorption edge,
$\Omega_g=2|m_\Gamma|\sim
2M_{\rm lat}e^{-A_{\rm lat}/V_1}$, but not the overall magnitude of
the optical response.  Introducing the scaled frequency
$y=\Omega/(2|m_\Gamma|)$, Eq.~\eqref{eq:optical_conductivity} becomes
$s_{xx}=(1+y^{-2})/8$ for $y>1$: the conductivity jumps to the threshold value $1/4$ and approaches the QBT plateau $1/8$
at larger $y$.  As shown in Fig.~\ref{fig:optical_threshold}, the
full three-band lattice results for progressively smaller masses collapse
onto this continuum scaling form.  Thus the exponentially small
nonperturbative scale appears primarily as an exponentially low frequency
threshold below an optical response that remains of order $e^2/\hbar$.

This finite low-energy conductivity makes the inverse-frequency optical
moment singular.  Defining $W_{-1}^{\mathrm{tr}}
=
\int_0^\infty
\frac{d\omega}{\omega}\,
\operatorname{Re}
\left[
\sigma_{xx}(\omega)+\sigma_{yy}(\omega)
\right]$,
the Souza--Wilkens--Martin relation gives~\cite{Souza2000,Verma2025}
\begin{equation}
W_{-1}^{\mathrm{tr}}(V_1)
=
\frac{\pi e^2}{\hbar A_c}
G_{\rm cell}(V_1)
.
\label{eq:swm}
\end{equation}
$W_{-1}^{\rm tr}$ is therefore the natural optical probe of the
amplification: it accumulates the finite QBT conductivity down to the
exponentially small absorption threshold and, through the
Souza--Wilkens--Martin relation, directly measures the integrated quantum
metric.  This enhancement is already visible within the QBT window:
because the conductivity remains finite for
$2|m|\ll\Omega\ll\Omega_{\Lambda}$, we have
\begin{equation}
\int_{2|m|}^{\Omega_{\Lambda}}
\frac{d\Omega}{\Omega}\,
s_{xx}(\Omega)
\propto
\ln\frac{\Omega_{\Lambda}}{2|m|}.
\label{SWM}
\end{equation}
The cumulative inverse-frequency weight shown in the inset of
Fig.~\ref{fig:optical_threshold} exhibits precisely this logarithmic
growth, and the numerical full-lattice curve agrees perfectly with the continuum QBT result.  Finite contributions from remote-band
transitions provide only the ultraviolet completion of this infrared
logarithm (Ref.~\cite{suppmat} Sec.~IX).

Combining the exact sum rule, Eq.~\eqref{eq:swm}, with the interaction
dependence of the integrated quantum metric,
Eq.~\eqref{eq:metric_interaction_law}, yields
\begin{equation}
W_{-1}^{\mathrm{tr}}(V_1)
=
\frac{\pi e^2 d_0}{\hbar A_c}
\frac{1}{V_1}
+
O(1)
.
\label{eq:optical_interaction_law}
\end{equation}
Hence the same singular quantum geometry that converts the
beyond-all-orders gap into $G_{\rm cell}\sim1/V_1$ produces an
algebraically enhanced optical moment.  To expose this asymptotic law,
we multiply $W_{-1}^{\rm tr}$ by $(V_1/t)(\hbar/e^2)$, which removes
both its dimensions and the predicted $1/V_1$ divergence.  For the
kagome model, $d_0=t/8$ and $A_c=\sqrt3/2$, giving
\begin{equation}
\frac{V_1}{t}\frac{\hbar}{e^2}
W_{-1}^{\mathrm{tr}}
\longrightarrow
\frac{\pi}{4\sqrt3} \quad \text{for}
\quad V_1/t\to0^+ .
\end{equation}
Thus the rescaled optical moment approaches a finite, parameter-free
constant, providing a direct test not only of the $1/V_1$ enhancement
but also of its QBT-determined prefactor.

\textit{Conclusions --}
We have shown that an interaction-generated scale that is exponentially
small in the weak-coupling limit can nevertheless leave an algebraically
strong imprint on measurable response.  
Importantly, an experimental test need not rely on direct control or
microscopic determination of the interaction strenght $V_1$. In fact, the optical threshold itself provides
the interaction-generated scale from $\Omega_g=2|m_\Gamma|$, while optical spectroscopy determines the corresponding inverse-frequency
moment.  Eliminating in Eq.~\eqref{SWM} the interaction in favor of the measured gap gives
$W_{-1}^{\mathrm{tr}}\propto\ln(1/\Omega_g)$, with a coefficient fixed by
the QBT geometry.  The predicted scaling collapse of the conductivity as
a function of $\Omega/\Omega_g$ (Fig.~\ref{fig:optical_threshold}), together with the logarithmic growth of
the integrated optical weight (inset of Fig.~\ref{fig:optical_threshold}) as the threshold is reduced, therefore
provides a direct route to observing quantum-geometric amplification even
when the microscopic interaction strength itself is not continuously
tunable.  Moreover, in engineered platforms, the effective ratio $V_1/t$ may
additionally be varied indirectly through bandwidth, screening, or
superlattice control; electrically tunable artificial kagome lattices and
kagome materials hosting $\Gamma$-centered quadratic touchings and flat bands provide
promising settings for such tests~\cite{wang2026correlated,di2026kagome,wilson2024v3sb5,wang2023quantum,yin2022topological,neupert2022charge}. It is however important to state that, in realistic scenarios, finite temperature, doping, or scattering may ultimately cut off this
infrared regime (Ref.~\cite{suppmat} Sec.~X).

More broadly, our results identify quantum geometry as an
\emph{asymptotic amplifier}: near a singular band degeneracy, momentum integration can transform a beyond-all-orders interaction scale
into an algebraic observable.  From this perspective, the scale
$e^{-A/V_1}$ is naturally viewed as a nonperturbative sector of a
transseries~\cite{Aniceto2019,dorigoni2019introduction,Aschenbrenner2017}, invisible to every finite order in the ordinary weak-coupling
expansion, while quantum geometry promotes information carried by this
exponentially small sector into an algebraically leading response.  This
suggests a broader non-Archimedean viewpoint on interacting band
structures, in which spectral, geometric, and response scales are
organized according to asymptotic dominance rather than ordinary power counting~\cite{Aschenbrenner2017}. Developing such a transseries
description systematically, and determining whether analogous changes of
asymptotic class occur near other interacting band degeneracies, provides
a natural direction for future work.

\textit{Acknowledgments --} The authors are
grateful for financial support by the Deutsche
Forschungsgemeinschaft (DFG, German Research
Foundation) through the W\"{u}rzburg-Dresden Cluster of Excellence ctd.qmat – Complexity, Topology
and Dynamics in Quantum Matter (EXC 2147,
project-id 390858490) as well as through the Collaborative Research Center SFB 1170 ToCoTronics
(Project ID 258499086).
D.D.S acknowledges MUR funding within the FIS2 (n. 1236, 01-08-2023) Project no. FIS-2023-00144 (CUP J53C25001880001). R.M.~furthermore acknowledges hospitality from the Shanghai Institute for Mathematics and Interdisciplinary Sciences (SIMIS)  and associated travel support under STCSM Grant 25HB2701900. The authors acknowledge the use of OpenAI's ChatGPT (GPT-5.6 Sol) for assistance with manuscript drafting, language refinement, and the organization of analytical derivations; all scientific content, calculations, interpretations, and conclusions were independently verified by the authors.  

\bibliography{references}

\onecolumngrid

\clearpage
\newpage



\section*{Supplementary material}

# Supplemental Material for "Quantum-Geometric Amplification of Nonperturbative Interaction Scales"

Stefano Bilamour,[1,2] René Meyer,[2,3] Johanna Erdmenger,[2] and Domenico Di Sante[1,*]

[1] *Department of Physics and Astronomy, University of Bologna, 40127 Bologna, Italy*
[2] *Institut für Theoretische Physik und Astrophysik and Würzburg-Dresden Cluster of Excellence ctd.qmat, Julius-Maximilians-Universität Würzburg, 97074 Würzburg, Germany.*
[3] *Shanghai Institute for Mathematics and Interdisciplinary Sciences (SIMIS), Shanghai 200433, China*

## CONTENTS



* domenico.disante@unibo.it

## I. KAGOME-LATTICE CONVENTIONS AND NONINTERACTING QUADRATIC BAND TOUCHING

In this section we fix all real-space, reciprocal-space, Fourier-transform, and band-basis conventions used throughout the main text and the Supplemental Material. We then derive explicitly the quadratic band touching (QBT) at the $\Gamma$ point and its two-band low-energy Hamiltonian.

Throughout, $t > 0$ denotes the nearest-neighbor hopping amplitude and $a$ the triangular Bravais-lattice constant. The fermions are spinless and there is one orbital on each kagome site. We distinguish the dimensionful crystal momentum $\mathbf{k}$ from the dimensionless momentum

$$\mathbf{q} = a\mathbf{k}. \tag{1}$$

Consequently, coefficients multiplying $q^2$ have dimensions of energy, whereas coefficients multiplying $k^2$ have dimensions of energy times length squared.

### A. Bravais lattice, basis, and reciprocal lattice

We choose primitive Bravais-lattice vectors

$$\mathbf{a}_1 = a(1,0), \tag{2}$$

$$\mathbf{a}_2 = a\left(\frac{1}{2}, \frac{\sqrt{3}}{2}\right). \tag{3}$$

A Bravais-lattice point is therefore

$$\mathbf{R} = n_1\mathbf{a}_1 + n_2\mathbf{a}_2, \qquad n_1, n_2 \in \mathbb{Z}. \tag{4}$$

The area of one primitive cell is

$$\begin{aligned} A_c &= |\mathbf{a}_1 \times \mathbf{a}_2| \\ &= \left| a(1,0) \times a\left(\frac{1}{2}, \frac{\sqrt{3}}{2}\right)\right| \\ &= \frac{\sqrt{3}}{2}a^2. \end{aligned} \tag{5}$$

We take the three kagome sites in each unit cell to be

$$\mathbf{r}_A = (0,0), \tag{6}$$

$$\mathbf{r}_B = \frac{\mathbf{a}_1}{2} = a\left(\frac{1}{2}, 0\right), \tag{7}$$

$$\mathbf{r}_C = \frac{\mathbf{a}_2}{2} = a\left(\frac{1}{4}, \frac{\sqrt{3}}{4}\right). \tag{8}$$

Thus the physical position of orbital $\alpha = A, B, C$ in cell $\mathbf{R}$ is

$$\mathbf{x}_{\mathbf{R}\alpha} = \mathbf{R} + \mathbf{r}_\alpha. \tag{9}$$

The reciprocal vectors $\mathbf{b}_1, \mathbf{b}_2$ are defined by

$$\mathbf{a}_i \cdot \mathbf{b}_j = 2\pi\delta_{ij}. \tag{10}$$

To obtain $\mathbf{b}_1 = (b_{1x}, b_{1y})$, the two conditions are

$$ab_{1x} = 2\pi, \tag{11}$$

$$a\left(\frac{1}{2}b_{1x} + \frac{\sqrt{3}}{2}b_{1y}\right) = 0. \tag{12}$$

The first gives

$$b_{1x} = \frac{2\pi}{a}, \tag{13}$$

and inserting this into the second gives

$$b_{1y} = -\frac{2\pi}{\sqrt{3}a}. \tag{14}$$

Hence

$$\boxed{\mathbf{b}_1 = \frac{2\pi}{a}\left(1, -\frac{1}{\sqrt{3}}\right)}. \tag{15}$$

Similarly, from

$$ab_{2x} = 0, \tag{16}$$

$$a\frac{\sqrt{3}}{2}b_{2y} = 2\pi, \tag{17}$$

we obtain

$$\boxed{\mathbf{b}_2 = \frac{2\pi}{a}\left(0, \frac{2}{\sqrt{3}}\right)}. \tag{18}$$

The Brillouin-zone area is therefore

$$A_{\mathrm{BZ}} = |\mathbf{b}_1 \times \mathbf{b}_2| = \frac{(2\pi)^2}{A_c} = \frac{8\pi^2}{\sqrt{3}a^2}. \tag{19}$$

With these conventions one convenient choice of high-symmetry points is

$$\Gamma = (0, 0), \tag{20}$$

$$K = \frac{2\mathbf{b}_1 + \mathbf{b}_2}{3} = \left(\frac{4\pi}{3a}, 0\right), \tag{21}$$

$$M = \frac{\mathbf{b}_1 + \mathbf{b}_2}{2} = \left(\frac{\pi}{a}, \frac{\pi}{\sqrt{3}a}\right). \tag{22}$$

### B. Nearest-neighbor bonds

For the basis in Eq. (8), each pair of sublattices is connected by two opposite nearest-neighbor displacement vectors. For the $A$–$B$ pair,

$$\boldsymbol{\delta}^{\pm}_{AB} = \pm\frac{\mathbf{a}_1}{2}. \tag{23}$$

For the $A$–$C$ pair,

$$\boldsymbol{\delta}^{\pm}_{AC} = \pm\frac{\mathbf{a}_2}{2}. \tag{24}$$

Finally, for the $B$–$C$ pair,

$$\boldsymbol{\delta}^{\pm}_{BC} = \pm\frac{\mathbf{a}_2 - \mathbf{a}_1}{2}. \tag{25}$$

A real-space nearest-neighbor hopping Hamiltonian in which every undirected bond is counted once can therefore be written as

$$\begin{aligned} H_{\mathrm{NN}} = -t\sum_{\mathbf{R}} \Big[ & c^{\dagger}_{\mathbf{R}A}\left(c_{\mathbf{R}B} + c_{\mathbf{R}-\mathbf{a}_1,B}\right) \\ & + c^{\dagger}_{\mathbf{R}A}\left(c_{\mathbf{R}C} + c_{\mathbf{R}-\mathbf{a}_2,C}\right) \\ & + c^{\dagger}_{\mathbf{R}B}\left(c_{\mathbf{R}C} + c_{\mathbf{R}+\mathbf{a}_1-\mathbf{a}_2,C}\right) + \mathrm{H.c.}\Big]. \end{aligned} \tag{26}$$

For example, the two $A$–$B$ bonds emanating from $A_{\mathbf{R}}$ terminate at

$$\mathbf{R} + \mathbf{r}_B = \mathbf{R} + \frac{\mathbf{a}_1}{2}, \tag{27}$$

$$\mathbf{R} - \mathbf{a}_1 + \mathbf{r}_B = \mathbf{R} - \frac{\mathbf{a}_1}{2}, \tag{28}$$

which confirms explicitly the two displacement vectors $\pm\mathbf{a}_1/2$.

### C. Embedded Fourier convention and Bloch Hamiltonian

We use the physical, or embedded, Fourier convention

$$c_{\mathbf{k}\alpha} = \frac{1}{\sqrt{N_c}} \sum_{\mathbf{R}} e^{-i\mathbf{k}\cdot(\mathbf{R}+\mathbf{r}_\alpha)} c_{\mathbf{R}\alpha}, \tag{29}$$

where $N_c$ is the number of unit cells. The inverse transform is

$$c_{\mathbf{R}\alpha} = \frac{1}{\sqrt{N_c}} \sum_{\mathbf{k}} e^{+i\mathbf{k}\cdot(\mathbf{R}+\mathbf{r}_\alpha)} c_{\mathbf{k}\alpha}. \tag{30}$$

Consider first the intracell $A$–$B$ hopping $c^\dagger_{\mathbf{R}A} c_{\mathbf{R}B}$. Inserting Eq. (30) gives

$$\begin{aligned}
&\sum_{\mathbf{R}} c^\dagger_{\mathbf{R}A} c_{\mathbf{R}B} \\
&= \frac{1}{N_c} \sum_{\mathbf{R}} \sum_{\mathbf{k},\mathbf{k}'} e^{-i\mathbf{k}\cdot(\mathbf{R}+\mathbf{r}_A)} e^{+i\mathbf{k}'\cdot(\mathbf{R}+\mathbf{r}_B)} c^\dagger_{\mathbf{k}A} c_{\mathbf{k}'B}.
\end{aligned} \tag{31}$$

Using

$$\sum_{\mathbf{R}} e^{i(\mathbf{k}'-\mathbf{k})\cdot\mathbf{R}} = N_c \delta_{\mathbf{k},\mathbf{k}'}, \tag{32}$$

and $\mathbf{r}_A = 0$, we obtain

$$\sum_{\mathbf{k}} e^{+i\mathbf{k}\cdot\mathbf{a}_1/2} c^\dagger_{\mathbf{k}A} c_{\mathbf{k}B}. \tag{33}$$

The second $A$–$B$ bond gives

$$\begin{aligned}
&\sum_{\mathbf{R}} c^\dagger_{\mathbf{R}A} c_{\mathbf{R}-\mathbf{a}_1,B} \\
&= \sum_{\mathbf{k}} e^{i\mathbf{k}\cdot(-\mathbf{a}_1+\mathbf{r}_B)} c^\dagger_{\mathbf{k}A} c_{\mathbf{k}B} \\
&= \sum_{\mathbf{k}} e^{-i\mathbf{k}\cdot\mathbf{a}_1/2} c^\dagger_{\mathbf{k}A} c_{\mathbf{k}B}.
\end{aligned} \tag{34}$$

Adding both contributions gives

$$e^{+i\mathbf{k}\cdot\mathbf{a}_1/2} + e^{-i\mathbf{k}\cdot\mathbf{a}_1/2} = 2\cos\frac{\mathbf{k}\cdot\mathbf{a}_1}{2}. \tag{35}$$

The same calculation for the other two sublattice pairs gives the three real structure factors

$$c_1(\mathbf{k}) = \cos\frac{\mathbf{k}\cdot\mathbf{a}_1}{2}, \tag{36}$$

$$c_2(\mathbf{k}) = \cos\frac{\mathbf{k}\cdot\mathbf{a}_2}{2}, \tag{37}$$

$$c_3(\mathbf{k}) = \cos\frac{\mathbf{k}\cdot(\mathbf{a}_2-\mathbf{a}_1)}{2}. \tag{38}$$

Defining

$$\Psi_{\mathbf{k}} = \begin{pmatrix} c_{\mathbf{k}A} \\ c_{\mathbf{k}B} \\ c_{\mathbf{k}C} \end{pmatrix}, \tag{39}$$

the nearest-neighbor Hamiltonian is

$$H_{\mathrm{NN}} = \sum_{\mathbf{k}} \Psi^\dagger_{\mathbf{k}} H_0(\mathbf{k}) \Psi_{\mathbf{k}}, \tag{40}$$

with

$$\boxed{H_0(\mathbf{k}) = -2t \begin{pmatrix} 0 & c_1 & c_2 \\ c_1 & 0 & c_3 \\ c_2 & c_3 & 0 \end{pmatrix}.} \tag{41}$$

Because the orbital positions have been included explicitly in the Fourier transform, Eq. (41) is in the embedded gauge. It is not strictly periodic under $\mathbf{k} \to \mathbf{k} + \mathbf{G}$ for an arbitrary reciprocal vector $\mathbf{G}$. Instead,

$$\Psi_{\mathbf{k}+\mathbf{G}} = S_{\mathbf{G}} \Psi_{\mathbf{k}}, \tag{42}$$

where

$$S_{\mathbf{G}} = \operatorname{diag}\left(e^{-i\mathbf{G}\cdot\mathbf{r}_A}, e^{-i\mathbf{G}\cdot\mathbf{r}_B}, e^{-i\mathbf{G}\cdot\mathbf{r}_C}\right). \tag{43}$$

Consequently,

$$H_0(\mathbf{k}+\mathbf{G}) = S_{\mathbf{G}} H_0(\mathbf{k}) S_{\mathbf{G}}^\dagger. \tag{44}$$

This sewing relation will be important when discussing topology and quantum geometry.

For reference, the periodic-gauge operators are

$$\widetilde{c}_{\mathbf{k}\alpha} = e^{+i\mathbf{k}\cdot\mathbf{r}_\alpha} c_{\mathbf{k}\alpha}. \tag{45}$$

If

$$D(\mathbf{k}) = \operatorname{diag}\left(e^{-i\mathbf{k}\cdot\mathbf{r}_A}, e^{-i\mathbf{k}\cdot\mathbf{r}_B}, e^{-i\mathbf{k}\cdot\mathbf{r}_C}\right), \tag{46}$$

then

$$\Psi_{\mathbf{k}}^{\text{emb}} = D(\mathbf{k}) \Psi_{\mathbf{k}}^{\text{per}}, \tag{47}$$

and hence

$$H_{\text{per}}(\mathbf{k}) = D^\dagger(\mathbf{k}) H_{\text{emb}}(\mathbf{k}) D(\mathbf{k}). \tag{48}$$

The physical consequences of this distinction for momentum derivatives, the quantum metric, and optical matrix elements are discussed separately in Sec. VII.

### D. Exact nearest-neighbor band structure

It is useful to derive the three eigenvalues analytically. Define

$$M(\mathbf{k}) = \begin{pmatrix} 0 & c_1 & c_2 \\ c_1 & 0 & c_3 \\ c_2 & c_3 & 0 \end{pmatrix}, \tag{49}$$

so that

$$H_0(\mathbf{k}) = -2t M(\mathbf{k}). \tag{50}$$

Let $\mu$ be an eigenvalue of $M$. Then

$$\begin{aligned} 0 &= \det[M - \mu \mathbb{1}] \\ &= \det \begin{pmatrix} -\mu & c_1 & c_2 \\ c_1 & -\mu & c_3 \\ c_2 & c_3 & -\mu \end{pmatrix}. \end{aligned} \tag{51}$$

Expanding along the first row,

$$\begin{aligned}\det[M-\mu\mathbb{1}] &= (-\mu)(\mu^2-c_3^2) \\ &\quad - c_1(-\mu c_1 - c_2c_3) \\ &\quad + c_2(c_1c_3+\mu c_2) \\ &= -\mu^3+\mu(c_1^2+c_2^2+c_3^2)+2c_1c_2c_3. \end{aligned} \tag{52}$$

Thus

$$\mu^3 - S\mu - 2P = 0, \tag{53}$$

where

$$S = c_1^2+c_2^2+c_3^2, \qquad P = c_1c_2c_3. \tag{54}$$

Now introduce

$$\theta_1 = \frac{\mathbf{k}\cdot\mathbf{a}_1}{2}, \qquad \theta_2 = \frac{\mathbf{k}\cdot\mathbf{a}_2}{2}. \tag{55}$$

Then

$$c_1 = \cos\theta_1, \qquad c_2 = \cos\theta_2, \tag{56}$$

and

$$c_3 = \cos(\theta_2-\theta_1). \tag{57}$$

The following trigonometric identity is central:

$$c_1^2+c_2^2+c_3^2-2c_1c_2c_3 = 1. \tag{58}$$

For completeness, we verify it explicitly. Write

$$s_i = \sin\theta_i. \tag{59}$$

Since

$$c_3 = c_1c_2 + s_1s_2, \tag{60}$$

we have

$$\begin{aligned} &c_1^2+c_2^2+c_3^2-2c_1c_2c_3 \\ &= c_1^2+c_2^2+(c_1c_2+s_1s_2)^2-2c_1c_2(c_1c_2+s_1s_2) \\ &= c_1^2+c_2^2-c_1^2c_2^2+s_1^2s_2^2 \\ &= c_1^2+c_2^2-c_1^2c_2^2+(1-c_1^2)(1-c_2^2) \\ &= 1. \end{aligned} \tag{61}$$

Therefore

$$S - 2P = 1. \tag{62}$$

Equation (53) therefore has the exact root

$$\mu_f = -1, \tag{63}$$

because

$$(-1)^3 - S(-1) - 2P = -1 + S - 2P = 0. \tag{64}$$

This is the origin of the exactly flat kagome band.

Using $1 - S = -2P$, the characteristic polynomial factors as

$$\mu^3 - S\mu - 2P = (\mu + 1)\left[\mu^2 - \mu + (1 - S)\right]. \tag{65}$$

The remaining roots are therefore

$$\mu_\pm = \frac{1 \pm \sqrt{4S - 3}}{2}. \tag{66}$$

Since $H_0 = -2tM$, the three energies are

$$E_f(\mathbf{k}) = 2t, \tag{67}$$

$$E_\pm(\mathbf{k}) = -t\left[1 \pm \sqrt{4S(\mathbf{k}) - 3}\right]. \tag{68}$$

For $t > 0$, $E_+$ is the lowest band. The other dispersive band, $E_-$, touches the flat band at $\Gamma$.

An explicit eigenvector of the flat band is also useful. With

$$\theta_3 = \theta_2 - \theta_1, \tag{69}$$

consider

$$v_f(\mathbf{k}) = \begin{pmatrix} \sin\theta_3 \\ -\sin\theta_2 \\ \sin\theta_1 \end{pmatrix}. \tag{70}$$

To verify that this has $\mu_f = -1$, we must show

$$[M + \mathbb{1}]v_f = 0. \tag{71}$$

The first component is

$$\sin(\theta_2 - \theta_1) - \cos\theta_1 \sin\theta_2 + \cos\theta_2 \sin\theta_1 = 0, \tag{72}$$

because

$$\sin(\theta_2 - \theta_1) = \sin\theta_2 \cos\theta_1 - \cos\theta_2 \sin\theta_1. \tag{73}$$

The second component is

$$\cos\theta_1 \sin(\theta_2 - \theta_1) - \sin\theta_2 + \cos(\theta_2 - \theta_1)\sin\theta_1 = 0, \tag{74}$$

because

$$\cos\theta_1 \sin(\theta_2 - \theta_1) + \sin\theta_1 \cos(\theta_2 - \theta_1) = \sin\theta_2. \tag{75}$$

Finally, the third component is

$$\cos\theta_2 \sin(\theta_2 - \theta_1) - \cos(\theta_2 - \theta_1)\sin\theta_2 + \sin\theta_1 = 0, \tag{76}$$

since the first two terms are

$$\sin[(\theta_2 - \theta_1) - \theta_2] = -\sin\theta_1. \tag{77}$$

Away from points at which $v_f$ vanishes, the normalized flat-band eigenvector is therefore

$$|u_f(\mathbf{k})\rangle = \frac{1}{\mathcal{N}_f(\mathbf{k})}\begin{pmatrix} \sin\theta_3 \\ -\sin\theta_2 \\ \sin\theta_1 \end{pmatrix}, \tag{78}$$

with

$$\mathcal{N}_f^2 = \sin^2\theta_1 + \sin^2\theta_2 + \sin^2\theta_3. \tag{79}$$

At $\Gamma$, $\theta_1 = \theta_2 = \theta_3 = 0$ and this numerator vanishes. The inability to define the isolated flat-band eigenvector there reflects the fact that the flat band touches a dispersive band at $\Gamma$.

### E. Exact Γ-point eigenspaces

At Γ, all three structure factors are unity,

$$c_1 = c_2 = c_3 = 1. \tag{80}$$

Hence

$$H_0(\Gamma) = -2t \begin{pmatrix} 0 & 1 & 1 \\ 1 & 0 & 1 \\ 1 & 1 & 0 \end{pmatrix}. \tag{81}$$

The fully symmetric normalized state

$$|s\rangle = \frac{1}{\sqrt{3}} \begin{pmatrix} 1 \\ 1 \\ 1 \end{pmatrix} \tag{82}$$

satisfies

$$\begin{pmatrix} 0 & 1 & 1 \\ 1 & 0 & 1 \\ 1 & 1 & 0 \end{pmatrix} |s\rangle = 2|s\rangle. \tag{83}$$

Therefore

$$E_s = -4t. \tag{84}$$

Every vector orthogonal to $|s\rangle$ satisfies

$$v_A + v_B + v_C = 0. \tag{85}$$

For such a vector,

$$(Mv)_A = v_B + v_C = -v_A, \tag{86}$$

and similarly for the other components. Thus the entire two-dimensional subspace orthogonal to $|s\rangle$ has eigenvalue $\mu = -1$, or energy

$$E_\Gamma = 2t. \tag{87}$$

We choose the following real orthonormal basis for this QBT doublet:

$$|u_1\rangle = \frac{1}{\sqrt{2}} \begin{pmatrix} 1 \\ -1 \\ 0 \end{pmatrix}, \tag{88}$$

$$|u_2\rangle = \frac{1}{\sqrt{6}} \begin{pmatrix} 1 \\ 1 \\ -2 \end{pmatrix}. \tag{89}$$

One verifies directly that

$$\langle u_1|u_1\rangle = 1, \tag{90}$$
$$\langle u_2|u_2\rangle = 1, \tag{91}$$
$$\langle u_1|u_2\rangle = 0, \tag{92}$$

and

$$\langle s|u_1\rangle = \langle s|u_2\rangle = 0. \tag{93}$$

Define the $3 \times 2$ isometry

$$U = \begin{pmatrix} |u_1\rangle & |u_2\rangle \end{pmatrix}, \tag{94}$$

so that

$$U^\dagger U = \mathbb{1}_2 \tag{95}$$

and the projector onto the QBT doublet is

$$P = UU^\dagger. \tag{96}$$

### F. Expansion around Γ

We now expand the exact three-band Hamiltonian around Γ. Introduce the dimensionless momentum

$$\mathbf{q} = (q_x, q_y) = a(k_x, k_y). \tag{97}$$

The relevant scalar products are

$$\mathbf{k} \cdot \mathbf{a}_1 = q_x, \tag{98}$$

$$\mathbf{k} \cdot \mathbf{a}_2 = \frac{q_x + \sqrt{3}q_y}{2}, \tag{99}$$

$$\mathbf{k} \cdot (\mathbf{a}_2 - \mathbf{a}_1) = \frac{-q_x + \sqrt{3}q_y}{2}. \tag{100}$$

Therefore

$$c_1 = \cos \frac{q_x}{2}, \tag{101}$$

$$c_2 = \cos \frac{q_x + \sqrt{3}q_y}{4}, \tag{102}$$

$$c_3 = \cos \frac{-q_x + \sqrt{3}q_y}{4}. \tag{103}$$

Using

$$\cos x = 1 - \frac{x^2}{2} + O(x^4), \tag{104}$$

we obtain

$$c_1 = 1 - \frac{q_x^2}{8} + O(q^4), \tag{105}$$

$$c_2 = 1 - \frac{(q_x + \sqrt{3}q_y)^2}{32} + O(q^4), \tag{106}$$

$$c_3 = 1 - \frac{(-q_x + \sqrt{3}q_y)^2}{32} + O(q^4). \tag{107}$$

There are no terms linear in $\mathbf{q}$ because all three structure factors are cosine functions and are therefore even around Γ.

Write

$$H_0(\mathbf{q}) = H_0(\Gamma) + \delta H^{(2)}(\mathbf{q}) + O(q^4). \tag{108}$$

Since the diagonal elements vanish, the nonzero matrix elements of $\delta H^{(2)}$ are

$$\delta H^{(2)}_{AB} = -2t\left(-\frac{q_x^2}{8}\right) = \frac{tq_x^2}{4}, \tag{109}$$

$$\delta H^{(2)}_{AC} = -2t\left[-\frac{(q_x + \sqrt{3}q_y)^2}{32}\right] = \frac{t}{16}(q_x + \sqrt{3}q_y)^2, \tag{110}$$

$$\delta H^{(2)}_{BC} = \frac{t}{16}(-q_x + \sqrt{3}q_y)^2. \tag{111}$$

For compactness, define

$$A = \frac{tq_x^2}{4}, \tag{112}$$

$$B = \frac{t}{16}(q_x + \sqrt{3}q_y)^2, \tag{113}$$

$$C = \frac{t}{16}(-q_x + \sqrt{3}q_y)^2. \tag{114}$$

Then

$$\delta H^{(2)} = \begin{pmatrix} 0 & A & B \\ A & 0 & C \\ B & C & 0 \end{pmatrix}. \tag{115}$$

We now project this perturbation onto the doublet $\{|u_1\rangle, |u_2\rangle\}$.
The $(1,1)$ matrix element is

$$\begin{aligned} \langle u_1|\delta H^{(2)}|u_1\rangle &= 2\left(\frac{1}{\sqrt{2}}\right)\left(-\frac{1}{\sqrt{2}}\right)A \\ &= -A \\ &= -\frac{t}{4}q_x^2. \end{aligned} \tag{116}$$

For the off-diagonal element, the $A$ contribution cancels:

$$\begin{aligned} &A\left[(u_1)_A(u_2)_B + (u_1)_B(u_2)_A\right] \\ &= A\left[\frac{1}{\sqrt{2}}\frac{1}{\sqrt{6}} - \frac{1}{\sqrt{2}}\frac{1}{\sqrt{6}}\right] = 0. \end{aligned} \tag{117}$$

The $B$ contribution is

$$B\left[(u_1)_A(u_2)_C + (u_1)_C(u_2)_A\right] = -\frac{B}{\sqrt{3}}, \tag{118}$$

while the $C$ contribution is

$$C\left[(u_1)_B(u_2)_C + (u_1)_C(u_2)_B\right] = \frac{C}{\sqrt{3}}. \tag{119}$$

Therefore

$$\langle u_1|\delta H^{(2)}|u_2\rangle = \frac{C-B}{\sqrt{3}}. \tag{120}$$

Now

$$\begin{aligned} C - B &= \frac{t}{16}\left[(-q_x + \sqrt{3}q_y)^2 - (q_x + \sqrt{3}q_y)^2\right] \\ &= \frac{t}{16}(-4\sqrt{3}q_xq_y) \\ &= -\frac{\sqrt{3}t}{4}q_xq_y. \end{aligned} \tag{121}$$

Thus

$$\langle u_1|\delta H^{(2)}|u_2\rangle = -\frac{t}{4}q_xq_y. \tag{122}$$

Finally,

$$\begin{aligned} \langle u_2|\delta H^{(2)}|u_2\rangle &= 2\left[\frac{A}{6} - \frac{2B}{6} - \frac{2C}{6}\right] \\ &= \frac{A}{3} - \frac{2(B+C)}{3}. \end{aligned} \tag{123}$$

We have

$$\begin{aligned} B + C &= \frac{t}{16}\left[(q_x + \sqrt{3}q_y)^2 + (-q_x + \sqrt{3}q_y)^2\right] \\ &= \frac{t}{16}(2q_x^2 + 6q_y^2) \\ &= \frac{t}{8}(q_x^2 + 3q_y^2). \end{aligned} \tag{124}$$

Hence

$$\langle u_2|\delta H^{(2)}|u_2\rangle = \frac{tq_x^2}{12} - \frac{t(q_x^2+3q_y^2)}{12} \\ = -\frac{t}{4}q_y^2. \tag{125}$$

Collecting Eqs. (116), (122), and (125),

$$\boxed{U^\dagger H_0(\mathbf{q})U = 2t\,\mathbb{1} - \frac{t}{4}\begin{pmatrix} q_x^2 & q_xq_y \\ q_xq_y & q_y^2 \end{pmatrix} + O(q^4).} \tag{126}$$

No Löwdin correction is required at quadratic order. Indeed, $H_0(\mathbf{q}) - H_0(0) = O(q^2)$, so the matrix element coupling the doublet to the symmetric state is itself $O(q^2)$. The second-order Schrieffer–Wolff/Löwdin correction consequently scales as

$$P\delta HQ\frac{1}{E_\Gamma - QH_0Q}Q\delta HP \sim \frac{O(q^2)O(q^2)}{2t-(-4t)} = O(q^4). \tag{127}$$

The direct projection in Eq. (126) is therefore exact through order $q^2$.

The matrix in Eq. (126) may be decomposed as

$$\begin{pmatrix} q_x^2 & q_xq_y \\ q_xq_y & q_y^2 \end{pmatrix} = \frac{q^2}{2}\mathbb{1} \\ + \frac{1}{2}\left[(q_x^2-q_y^2)\sigma_z + 2q_xq_y\sigma_x\right]. \tag{128}$$

Thus

$$\boxed{H_{\mathrm{QBT}}^{\mathrm{NN}} = 2t\,\mathbb{1} - \frac{tq^2}{8}\mathbb{1} - \frac{t}{8}\left[(q_x^2-q_y^2)\sigma_z + 2q_xq_y\sigma_x\right] + O(q^4).} \tag{129}$$

It is convenient to define the positive curvature magnitude

$$d_0 \equiv \frac{t}{8}. \tag{130}$$

In dimensionful momentum variables, Eq. (129) becomes

$$H_{\mathrm{QBT}}^{\mathrm{NN}} = 2t\,\mathbb{1} - \frac{ta^2k^2}{8}\mathbb{1} \\ - \frac{ta^2}{8}\left[(k_x^2-k_y^2)\sigma_z + 2k_xk_y\sigma_x\right] + O(k^4a^4). \tag{131}$$

To display the winding of the QBT explicitly, write

$$q_x = q\cos\varphi, \qquad q_y = q\sin\varphi. \tag{132}$$

Then

$$q_x^2 - q_y^2 = q^2\cos 2\varphi, \tag{133}$$

$$2q_xq_y = q^2\sin 2\varphi. \tag{134}$$

The pseudospin part of the Hamiltonian is therefore proportional to

$$q^2\left[\cos 2\varphi\,\sigma_z + \sin 2\varphi\,\sigma_x\right]. \tag{135}$$

As $\mathbf{q}$ winds once around $\Gamma$, the pseudospin vector winds twice. This winding-two texture is the characteristic low-energy structure of the quadratic band touching.

The matrix

$$\begin{pmatrix} q_x^2 & q_x q_y \\ q_x q_y & q_y^2 \end{pmatrix} = \mathbf{q}\,\mathbf{q}^T \tag{136}$$

has eigenvalues $q^2$ and 0. Its normalized eigenvectors are, respectively,

$$v_\parallel = \begin{pmatrix} \cos\varphi \\ \sin\varphi \end{pmatrix}, \tag{137}$$

$$v_\perp = \begin{pmatrix} -\sin\varphi \\ \cos\varphi \end{pmatrix}. \tag{138}$$

Hence the two bands emerging from the $\Gamma$ doublet have energies

$$E_\parallel(\mathbf{q}) = 2t - \frac{t}{4}q^2 + O(q^4), \tag{139}$$

$$E_\perp(\mathbf{q}) = 2t + O(q^4). \tag{140}$$

For the pure nearest-neighbor model the latter is, in fact, exactly flat to all orders and equals $2t$ throughout the Brillouin zone.

### G. Cross-check from the exact band energies

The same quadratic dispersion can be obtained directly from the exact eigenvalues in Eq. (68). For small $\mathbf{q}$,

$$\cos^2 x = 1 - x^2 + O(x^4). \tag{141}$$

Using

$$\theta_1 = \frac{q_x}{2}, \tag{142}$$

$$\theta_2 = \frac{q_x + \sqrt{3}q_y}{4}, \tag{143}$$

$$\theta_3 = \frac{-q_x + \sqrt{3}q_y}{4}, \tag{144}$$

we find

$$\begin{aligned} \theta_1^2 + \theta_2^2 + \theta_3^2 &= \frac{q_x^2}{4} + \frac{(q_x + \sqrt{3}q_y)^2 + (-q_x + \sqrt{3}q_y)^2}{16} \\ &= \frac{q_x^2}{4} + \frac{2q_x^2 + 6q_y^2}{16} \\ &= \frac{3}{8}(q_x^2 + q_y^2) \\ &= \frac{3q^2}{8}. \end{aligned} \tag{145}$$

Therefore

$$S = c_1^2 + c_2^2 + c_3^2 = 3 - \frac{3q^2}{8} + O(q^4). \tag{146}$$

It follows that

$$\begin{aligned} 4S - 3 &= 9 - \frac{3q^2}{2} + O(q^4) \\ &= 9\left(1 - \frac{q^2}{6}\right) + O(q^4). \end{aligned} \tag{147}$$

Using

$$\sqrt{1-x} = 1 - \frac{x}{2} + O(x^2), \tag{148}$$

we obtain

$$\begin{aligned}\sqrt{4S-3} &= 3\left(1 - \frac{q^2}{12}\right) + O(q^4) \\ &= 3 - \frac{q^2}{4} + O(q^4).\end{aligned} \tag{149}$$

The lower dispersive band is consequently

$$\begin{aligned}E_+(\mathbf{q}) &= -t\left[1 + 3 - \frac{q^2}{4}\right] + O(q^4) \\ &= -4t + \frac{t}{4}q^2 + O(q^4),\end{aligned} \tag{150}$$

whereas the dispersive member of the QBT doublet is

$$\begin{aligned}E_-(\mathbf{q}) &= -t\left[1 - 3 + \frac{q^2}{4}\right] + O(q^4) \\ &= 2t - \frac{t}{4}q^2 + O(q^4).\end{aligned} \tag{151}$$

Together with the exactly flat band $E_f = 2t$, this agrees precisely with the projected Hamiltonian Eq. (126).

### H. Exactly-flat endpoint and regularized QBT

The microscopic regulator implemented in the lattice Hamiltonian is

$$\boxed{H_{\mathrm{reg}}(\mathbf{k}; r) = 2r\left[3 - \cos(\mathbf{k}\cdot\mathbf{a}_1) - \cos(\mathbf{k}\cdot\mathbf{a}_2) - \cos(\mathbf{k}\cdot(\mathbf{a}_2 - \mathbf{a}_1))\right] I_3.}$$

No factor of the nearest-neighbour hopping $t$ multiplies $r$; both $t$ and $r$ have dimensions of energy. The regulator is proportional to the identity in sublattice space and therefore has the same matrix in the embedded and periodic orbital gauges.

Its real-space form is

$$\begin{aligned}\widehat{H}_{\mathrm{reg}} &= 6r\sum_{\mathbf{R},\alpha} c^\dagger_{\mathbf{R}\alpha} c_{\mathbf{R}\alpha} \\ &\quad - r\sum_{\mathbf{R},\alpha}\sum_{\boldsymbol{\delta}\in\{\mathbf{a}_1,\mathbf{a}_2,\mathbf{a}_2-\mathbf{a}_1\}}\left(c^\dagger_{\mathbf{R}\alpha} c_{\mathbf{R}+\boldsymbol{\delta},\alpha} + \text{h.c.}\right).\end{aligned}$$

Thus it consists of identical same-sublattice hoppings $-r$ along

$$\boldsymbol{\delta} = \pm\mathbf{a}_1, \quad \pm\mathbf{a}_2, \quad \pm(\mathbf{a}_2 - \mathbf{a}_1),$$

together with the onsite term $6r$. Indeed, each positive displacement and its Hermitian conjugate give

$$-r\left(e^{i\mathbf{k}\cdot\boldsymbol{\delta}} + e^{-i\mathbf{k}\cdot\boldsymbol{\delta}}\right) = -2r\cos(\mathbf{k}\cdot\boldsymbol{\delta}),$$

which reproduces the Bloch expression above.

For $\mathbf{q} = a\mathbf{k}$, define

$$x_1 = q_x, \qquad x_2 = \frac{q_x + \sqrt{3}q_y}{2}, \qquad x_3 = \frac{-q_x + \sqrt{3}q_y}{2}.$$

Then

$$H_{\mathrm{reg}}(\mathbf{q}; r) = 2r[3 - \cos x_1 - \cos x_2 - \cos x_3] I_3.$$

Using $\cos x = 1 - x^2/2 + O(x^4)$ gives

$$H_{\rm reg}(\mathbf{q}; r) = r(x_1^2 + x_2^2 + x_3^2)I_3 + O(q^4).$$

The quadratic form is

$$\begin{aligned} x_1^2 + x_2^2 + x_3^2 &= q_x^2 + \frac{(q_x + \sqrt{3}q_y)^2 + (-q_x + \sqrt{3}q_y)^2}{4} \\ &= \frac{3}{2}(q_x^2 + q_y^2), \end{aligned}$$

and hence

$$\boxed{H_{\rm reg}(\mathbf{q}; r) = \frac{3r}{2}(q_x^2 + q_y^2)I_3 + O(q^4).}$$

At $\Gamma$ the doublet basis is

$$u_1 = \frac{(1, -1, 0)^T}{\sqrt{2}}, \qquad u_2 = \frac{(1, 1, -2)^T}{\sqrt{6}}, \qquad U = (u_1, u_2).$$

Since

$$u_1^\dagger u_1 = u_2^\dagger u_2 = 1, \qquad u_1^\dagger u_2 = 0,$$

one obtains

$$\begin{aligned} U^\dagger H_{\rm reg} U &= 2r[3 - \cos x_1 - \cos x_2 - \cos x_3] \begin{pmatrix} u_1^\dagger u_1 & u_1^\dagger u_2 \\ u_2^\dagger u_1 & u_2^\dagger u_2 \end{pmatrix} \\ &= 2r[3 - \cos x_1 - \cos x_2 - \cos x_3] I_2 \\ &= \frac{3r}{2} q^2 I_2 + O(q^4). \end{aligned}$$

Therefore the regulator produces only an identity coefficient. Its coefficients multiplying $\sigma_x$, $\sigma_y$, and $\sigma_z$ all vanish. In particular,

$$H_{\rm reg}(0; r) = 0,$$

so the regulator neither shifts nor splits the $\Gamma$ doublet.

The nearest-neighbour projection in the same fixed basis is

$$U^\dagger H_0 U = 2tI - \frac{tq^2}{8} I - \frac{t}{8}\left[(q_x^2 - q_y^2)\sigma_z + 2q_x q_y \sigma_x\right] + O(q^4).$$

Adding the regulator gives

$$\begin{aligned} U^\dagger (H_0 + H_{\rm reg}) U = 2tI &+ \left(\frac{3r}{2} - \frac{t}{8}\right) q^2 I \\ &- \frac{t}{8}\left[(q_x^2 - q_y^2)\sigma_z + 2q_x q_y \sigma_x\right] + O(q^4). \end{aligned}$$

Equivalently,

$$U^\dagger (H_0 + H_{\rm reg}) U = \begin{pmatrix} 2t + \frac{3r}{2}q^2 - \frac{t}{4}q_x^2 & -\frac{t}{4}q_x q_y \\ -\frac{t}{4}q_x q_y & 2t + \frac{3r}{2}q^2 - \frac{t}{4}q_y^2 \end{pmatrix} + O(q^4).$$

The regulator is exactly proportional to the full-space identity, so it has no matrix element between the doublet and the normalized symmetric singlet $s = (1, 1, 1)^T/\sqrt{3}$. The nearest-neighbour quadratic term has

$$U^\dagger H_2 s = \begin{pmatrix} \sqrt{2}\, t q_x q_y / 8 \\ \sqrt{2}\, t (q_x^2 - q_y^2)/16 \end{pmatrix} = O(q^2).$$

Because the $\Gamma$ doublet–singlet separation is $6t$, the associated Löwdin correction is $O(q^4)$ for $t \neq 0$. Thus the fixed-subspace projection and the downfolded effective Hamiltonian agree through quadratic order.

Writing

$$H_{\mathrm{QBT}}(\mathbf{q}; r) = E_\Gamma(r) I - c_0(r) q^2 I - d_0(r) \left[(q_x^2 - q_y^2)\sigma_z + 2q_x q_y \sigma_x\right] + O(q^4),$$

with $d_0 > 0$ for $t > 0$, gives

$$\boxed{E_\Gamma(r) = 2t, \qquad c_0(r) = \frac{t}{8} - \frac{3r}{2}, \qquad d_0(r) = \frac{t}{8}.}$$

Since $H_{\mathrm{reg}}(\mathbf{k}; r) = f_r(\mathbf{k}) I_3$, the regulator shifts all three Bloch energies at a given momentum by the same amount while leaving their eigenvectors and direct interband splittings unchanged. It therefore modifies the scalar QBT curvature $c_0$, and hence removes the exactly-flat filling-two endpoint, without changing the pseudospin coefficient $d_0$. In particular, the infrared coefficients controlling the quantum metric and the logarithmic QAH kernel are independent of $r$; the regulator affects only the occupation structure and nonuniversal finite contributions.

For the benchmark $t = 1$ and $r = 0.05$,

$$\boxed{E_\Gamma = 2, \qquad c_0 = 0.05, \qquad d_0 = 0.125.}$$

The two quadratic branches may be written

$$E_\pm(\mathbf{q}) = E_\Gamma - [c_0(r) \pm d_0(r)] q^2 + O(q^4).$$

At $r = 0$,

$$c_0 = d_0 = \frac{t}{8},$$

so

$$E_+ = 2t - \frac{t}{4} q^2 + O(q^4), \qquad E_- = 2t + O(q^4).$$

The second branch is in fact exactly flat to all orders in the pure nearest-neighbour model, with energy $2t$ throughout the Brillouin zone.

The quadratic flatness condition is

$$|c_0| = |d_0|.$$

At $t = 1, r = 0.05$,

$$c_0 + d_0 = 0.175, \qquad c_0 - d_0 = -0.075,$$

and therefore

$$E_+ = 2 - 0.175 q^2 + O(q^4), \qquad E_- = 2 + 0.075 q^2 + O(q^4).$$

One branch bends below and the other above the touching energy, with $|c_0| < |d_0|$. The former flat band is exactly

$$E_{\text{former flat}}(\mathbf{k}) = 2t + 2r\left[3 - \cos(\mathbf{k}\cdot\mathbf{a}_1) - \cos(\mathbf{k}\cdot\mathbf{a}_2) - \cos(\mathbf{k}\cdot(\mathbf{a}_2 - \mathbf{a}_1))\right],$$

which is nonconstant for $r \neq 0$. The regulator therefore converts the singular exactly-flat endpoint into a genuine quadratic band-touching semimetal without breaking translations, time reversal, inversion, or the $C_3/C_6$ symmetry protecting the $\Gamma$ doublet.

## II. HARTREE–FOCK DECOUPLING AND QAH ORDER PARAMETER

In this section we derive the Hartree–Fock (HF) Hamiltonian used for the interacting kagome calculations and establish explicitly the relation between the imaginary nearest-neighbor bond expectation value and the QAH mass at the $\Gamma$-point quadratic band touching (QBT).

We consider spinless fermions with one orbital per kagome site. At filling two there are two particles per three-site unit cell, and hence the average density per site is

$$\bar{n} = \frac{2}{3}. \tag{152}$$

The nearest-neighbor interaction is written in density-shifted form as

$$H_{\text{int}} = V_1 \sum_{\langle ij \rangle} (n_i - \bar{n})(n_j - \bar{n}), \qquad V_1 > 0. \tag{153}$$

Here $\langle ij \rangle$ denotes an undirected nearest-neighbor bond, counted once.

For spinless fermions with a single orbital per site, an on-site density-density interaction would be trivial because

$$n_i^2 = c_i^\dagger c_i c_i^\dagger c_i = c_i^\dagger (1 - c_i^\dagger c_i) c_i = c_i^\dagger c_i = n_i. \tag{154}$$

Thus an apparent term $U \sum_i n_i^2$ only renormalizes the one-body chemical potential. The nearest-neighbor coupling $V_1$ in Eq. (153) is therefore the shortest-range nontrivial density interaction in the present spinless single-orbital model.

### A. General Hartree–Fock decomposition

Define the density deviation

$$\delta n_i = n_i - \bar{n} \tag{155}$$

and its mean-field expectation value

$$\delta \rho_i = \langle \delta n_i \rangle = \langle n_i \rangle - \bar{n}. \tag{156}$$

For a bond $i \neq j$, define also the complex Fock expectation value

$$\chi_{ij} = \langle c_i^\dagger c_j \rangle. \tag{157}$$

Hermiticity implies

$$\chi_{ji} = \chi_{ij}^*. \tag{158}$$

We first derive the HF factorization for the unshifted product $n_i n_j$. Explicitly,

$$n_i n_j = c_i^\dagger c_i c_j^\dagger c_j. \tag{159}$$

Retaining the direct and exchange contractions gives

$$\begin{aligned} n_i n_j \simeq\ & \langle n_i \rangle n_j + n_i \langle n_j \rangle - \langle n_i \rangle \langle n_j \rangle \\ & - \langle c_i^\dagger c_j \rangle c_j^\dagger c_i - c_i^\dagger c_j \langle c_j^\dagger c_i \rangle \\ & + \langle c_i^\dagger c_j \rangle \langle c_j^\dagger c_i \rangle. \end{aligned} \tag{160}$$

Using Eq. (157),

$$\begin{aligned} n_i n_j \simeq\ & \rho_i n_j + \rho_j n_i - \rho_i \rho_j \\ & - \chi_{ij} c_j^\dagger c_i - \chi_{ji} c_i^\dagger c_j + |\chi_{ij}|^2, \end{aligned} \tag{161}$$

where

$$\rho_i = \langle n_i \rangle. \tag{162}$$

We now apply this decomposition to

$$\delta n_i \delta n_j = n_i n_j - \bar{n}(n_i + n_j) + \bar{n}^2. \tag{163}$$

Writing

$$\rho_i = \bar{n} + \delta\rho_i, \tag{164}$$

the Hartree terms combine into

$$\rho_j n_i - \bar{n} n_i = \delta\rho_j n_i, \tag{165}$$
$$\rho_i n_j - \bar{n} n_j = \delta\rho_i n_j. \tag{166}$$

It is convenient to rewrite these in terms of $\delta n_i = n_i - \bar{n}$. After collecting all constants, one obtains

$$\begin{aligned}\delta n_i \delta n_j \simeq\ & \delta\rho_j\, \delta n_i + \delta\rho_i\, \delta n_j - \delta\rho_i \delta\rho_j \\ & - \chi_{ji} c_i^\dagger c_j - \chi_{ij} c_j^\dagger c_i + |\chi_{ij}|^2.\end{aligned} \tag{167}$$

Equation (167) contains both the Hartree density channel and the Fock bond channel.

For the translationally invariant $C_3$-symmetric phases considered below,

$$\langle n_A \rangle = \langle n_B \rangle = \langle n_C \rangle = \frac{2}{3}, \tag{168}$$

and hence

$$\delta\rho_A = \delta\rho_B = \delta\rho_C = 0. \tag{169}$$

The density-shifted Hartree contribution therefore vanishes identically in this symmetry sector.

The interaction reduces to the exchange decoupling

$$H_{\rm int}^{\rm HF} = -V_1 \sum_{\langle ij \rangle} \left( \chi_{ji} c_i^\dagger c_j + \chi_{ij} c_j^\dagger c_i \right) + V_1 \sum_{\langle ij \rangle} |\chi_{ij}|^2. \tag{170}$$

The final term is the usual HF double-counting correction.

### B. Translation-invariant cyclic bond ansatz

The kagome lattice has six nearest-neighbor bonds per primitive cell. For the translation-invariant and $C_3$-symmetric QAH state, all bonds have the same real part and the same magnitude of the imaginary part.

We choose the cyclic ordered-sublattice convention

$$A \to B \to C \to A. \tag{171}$$

Define

$$\nu_{AB} = \nu_{BC} = \nu_{CA} = +1, \tag{172}$$

and, by antisymmetry,

$$\nu_{BA} = \nu_{CB} = \nu_{AC} = -1. \tag{173}$$

The same sublattice sign is assigned to the two translationally equivalent nearest-neighbor bonds connecting a given ordered pair.

The cyclic bond expectation value is parametrized as

$$\boxed{\chi_{ij} = \chi_R + i\nu_{ij}\chi_I.} \tag{174}$$

Hermiticity follows immediately:

$$\begin{aligned}\chi_{ji} &= \chi_R + i\nu_{ji}\chi_I \\ &= \chi_R - i\nu_{ij}\chi_I \\ &= \chi_{ij}^*.\end{aligned} \tag{175}$$

The real part $\chi_R$ preserves time reversal and renormalizes the ordinary hopping amplitude. The imaginary part $\chi_I$ changes sign under complex conjugation,

$$\mathcal{T}: \qquad \chi_I \to -\chi_I, \tag{176}$$

and therefore breaks time reversal.

For a spinless system, time reversal acts as complex conjugation in the real-space orbital basis. Thus the two states

$$+\chi_I \qquad \text{and} \qquad -\chi_I \tag{177}$$

are exactly related by time reversal.

### C. Effective complex hopping

Consider one oriented bond $i \to j$. The bare kinetic term is

$$-t\left(c_i^\dagger c_j + c_j^\dagger c_i\right). \tag{178}$$

From Eq. (170), the coefficient multiplying $c_i^\dagger c_j$ after Fock decoupling is

$$\begin{aligned} -t - V_1\chi_{ji} &= -t - V_1\left(\chi_R - i\nu_{ij}\chi_I\right) \\ &= -\left(t + V_1\chi_R\right) + i\nu_{ij}V_1\chi_I. \end{aligned} \tag{179}$$

We therefore define

$$\boxed{\tau \equiv t + V_1\chi_R,} \tag{180}$$

$$\boxed{\eta \equiv -V_1\chi_I.} \tag{181}$$

Then

$$-t - V_1\chi_{ji} = -\left(\tau + i\nu_{ij}\eta\right). \tag{182}$$

Thus the self-consistent NN hopping on an oriented bond is

$$T_{ij} = \tau + i\nu_{ij}\eta. \tag{183}$$

It is sometimes convenient to introduce an amplitude and phase,

$$t_{\text{eff}} = \sqrt{\tau^2 + \eta^2}, \tag{184}$$

$$\phi = \operatorname{atan2}(\eta, \tau), \tag{185}$$

so that

$$\tau + i\nu_{ij}\eta = t_{\text{eff}}e^{i\nu_{ij}\phi}. \tag{186}$$

Equivalently,

$$t_{\text{eff}}\cos\phi = \tau, \tag{187}$$

$$t_{\text{eff}}\sin\phi = \eta. \tag{188}$$

Combining Eqs. (181) and (188),

$$\boxed{t_{\text{eff}}\sin\phi = \eta = -V_1\chi_I.} \tag{189}$$

### D. Bloch Hamiltonian of the cyclic QAH state

Using the structure factors defined in Sec. I,

$$c_1 = \cos\frac{\mathbf{k}\cdot\mathbf{a}_1}{2}, \tag{190}$$

$$c_2 = \cos\frac{\mathbf{k}\cdot\mathbf{a}_2}{2}, \tag{191}$$

$$c_3 = \cos\frac{\mathbf{k}\cdot(\mathbf{a}_2-\mathbf{a}_1)}{2}, \tag{192}$$

the NN HF hopping Hamiltonian in the embedded gauge is

$$\boxed{H^{\mathrm{HF}}_{\mathrm{hop}}(\mathbf{k}) = -2\begin{pmatrix} 0 & (\tau+i\eta)c_1 & (\tau-i\eta)c_2 \\ (\tau-i\eta)c_1 & 0 & (\tau+i\eta)c_3 \\ (\tau+i\eta)c_2 & (\tau-i\eta)c_3 & 0 \end{pmatrix}.} \tag{193}$$

Equivalently, using Eq. (186),

$$H^{\mathrm{HF}}_{\mathrm{hop}}(\mathbf{k}) = -2t_{\mathrm{eff}}\begin{pmatrix} 0 & e^{i\phi}c_1 & e^{-i\phi}c_2 \\ e^{-i\phi}c_1 & 0 & e^{i\phi}c_3 \\ e^{i\phi}c_2 & e^{-i\phi}c_3 & 0 \end{pmatrix}. \tag{194}$$

The full one-body HF Hamiltonian also contains the scalar regulator derived in the previous section,

$$H_{\mathrm{HF}}(\mathbf{k}) = H^{\mathrm{HF}}_{\mathrm{hop}}(\mathbf{k}) + H_{\mathrm{reg}}(\mathbf{k};r). \tag{195}$$

Since

$$H_{\mathrm{reg}}(\mathbf{k};r) = f_r(\mathbf{k})\mathbb{1}_3, \tag{196}$$

the regulator does not modify the chiral sublattice structure.

Separating the real and imaginary pieces of Eq. (193) gives

$$H^{\mathrm{HF}}_{\mathrm{hop}} = H_R + H_I, \tag{197}$$

where

$$H_R(\mathbf{k}) = -2\tau\begin{pmatrix} 0 & c_1 & c_2 \\ c_1 & 0 & c_3 \\ c_2 & c_3 & 0 \end{pmatrix} \tag{198}$$

and

$$H_I(\mathbf{k}) = -2i\eta\begin{pmatrix} 0 & c_1 & -c_2 \\ -c_1 & 0 & c_3 \\ c_2 & -c_3 & 0 \end{pmatrix}. \tag{199}$$

Equation (199) is odd under time reversal because complex conjugation sends

$$H_I \to -H_I. \tag{200}$$

### E. Projection of the QAH order onto the Γ doublet

At Γ,

$$c_1 = c_2 = c_3 = 1 \tag{201}$$

and the regulator vanishes,

$$H_{\mathrm{reg}}(\Gamma;r) = 0. \tag{202}$$

Define

$$A = \begin{pmatrix} 0 & 1 & 1 \\ 1 & 0 & 1 \\ 1 & 1 & 0 \end{pmatrix} \tag{203}$$

and

$$C = \begin{pmatrix} 0 & 1 & -1 \\ -1 & 0 & 1 \\ 1 & -1 & 0 \end{pmatrix}. \tag{204}$$

Then

$$H_{\rm hop}^{\rm HF}(\Gamma) = -2\tau A - 2i\eta C. \tag{205}$$

The QBT basis is

$$|u_1\rangle = \frac{1}{\sqrt{2}} \begin{pmatrix} 1 \\ -1 \\ 0 \end{pmatrix}, \tag{206}$$

$$|u_2\rangle = \frac{1}{\sqrt{6}} \begin{pmatrix} 1 \\ 1 \\ -2 \end{pmatrix}, \tag{207}$$

with

$$U = (|u_1\rangle, |u_2\rangle). \tag{208}$$

Both basis vectors obey

$$(u_\ell)_A + (u_\ell)_B + (u_\ell)_C = 0, \tag{209}$$

so

$$A|u_\ell\rangle = -|u_\ell\rangle. \tag{210}$$

Therefore

$$U^\dagger A U = -\mathbb{1}_2. \tag{211}$$

We next project the chiral matrix $C$. Acting on $u_1$,

$$\begin{aligned} C|u_1\rangle &= \frac{1}{\sqrt{2}} \begin{pmatrix} -1 \\ -1 \\ 2 \end{pmatrix} \\ &= -\sqrt{3}\,|u_2\rangle. \end{aligned} \tag{212}$$

Similarly,

$$\begin{aligned} C|u_2\rangle &= \frac{1}{\sqrt{6}} \begin{pmatrix} 3 \\ -3 \\ 0 \end{pmatrix} \\ &= \sqrt{3}\,|u_1\rangle. \end{aligned} \tag{213}$$

Thus, in the $(u_1, u_2)$ basis,

$$U^\dagger C U = \begin{pmatrix} 0 & \sqrt{3} \\ -\sqrt{3} & 0 \end{pmatrix}. \tag{214}$$

Since

$$i\sigma_y = \begin{pmatrix} 0 & 1 \\ -1 & 0 \end{pmatrix}, \tag{215}$$

we obtain

$$\boxed{U^\dagger C U = i\sqrt{3}\,\sigma_y.} \tag{216}$$

Using Eqs. (211) and (216),

$$\begin{aligned} U^\dagger H^{\rm HF}_{\rm hop}(\Gamma) U &= -2\tau(-\mathbb{1}_2) - 2i\eta(i\sqrt{3}\sigma_y) \\ &= 2\tau\mathbb{1}_2 + 2\sqrt{3}\,\eta\sigma_y. \end{aligned} \tag{217}$$

Therefore the QAH mass generated by the imaginary bond order is

$$\boxed{m_\Gamma = 2\sqrt{3}\,\eta.} \tag{218}$$

Combining with Eq. (181),

$$\boxed{m_\Gamma = -2\sqrt{3}\,V_1\chi_I.} \tag{219}$$

This relation fixes both the normalization and the sign convention for the microscopic QAH order parameter used throughout the remainder of the Supplemental Material.

Equivalently, in the phase parametrization,

$$\boxed{m_\Gamma = 2\sqrt{3}\,t_{\rm eff}\sin\phi.} \tag{220}$$

Under time reversal,

$$\chi_I \to -\chi_I, \tag{221}$$
$$\eta \to -\eta, \tag{222}$$
$$m_\Gamma \to -m_\Gamma. \tag{223}$$

Thus the two self-consistent QAH solutions have opposite masses and opposite chirality.

### F. Low-energy HF Hamiltonian around Γ

The real hopping part of the HF Hamiltonian has exactly the same momentum dependence as the noninteracting nearest-neighbor kagome Hamiltonian, with

$$t \to \tau. \tag{224}$$

Consequently, repeating the projection derived in Sec. I gives

$$\begin{aligned} U^\dagger H_R(\mathbf{q}) U &= 2\tau\mathbb{1} - \frac{\tau}{8}q^2\mathbb{1} \\ &\quad - \frac{\tau}{8}\left[(q_x^2 - q_y^2)\sigma_z + 2q_xq_y\sigma_x\right] + O(q^4). \end{aligned} \tag{225}$$

The regulator contributes

$$U^\dagger H_{\rm reg} U = \frac{3r}{2}q^2\mathbb{1} + O(q^4). \tag{226}$$

Finally, the imaginary hopping term satisfies

$$H_I(\mathbf{q}) = H_I(0) + O(\eta q^2), \tag{227}$$

because each structure factor is

$$c_i(\mathbf{q}) = 1 + O(q^2). \tag{228}$$

Its leading projection is therefore

$$U^\dagger H_I(\mathbf{q}) U = m_\Gamma \sigma_y + O(\eta q^2). \tag{229}$$

Combining Eqs. (225), (226), and (229), the HF QBT Hamiltonian is

$$\begin{aligned} H_{\mathrm{QBT}}^{\mathrm{HF}}(\mathbf{q}) = 2\tau\,\mathbb{1} &- \left(\frac{\tau}{8} - \frac{3r}{2}\right) q^2 \mathbb{1} \\ &- \frac{\tau}{8}\left[(q_x^2 - q_y^2)\sigma_z + 2q_x q_y \sigma_x\right] \\ &+ m_\Gamma \sigma_y + O(q^4, \eta q^2). \end{aligned} \tag{230}$$

Thus, in the notation of the main text,

$$\boxed{E_\Gamma^{\mathrm{HF}} = 2\tau,} \tag{231}$$

$$\boxed{c_{\mathrm{eff}} = \frac{\tau}{8} - \frac{3r}{2},} \tag{232}$$

$$\boxed{d_{\mathrm{eff}} = \frac{\tau}{8},} \tag{233}$$

and

$$\boxed{m_\Gamma = 2\sqrt{3}\,\eta = -2\sqrt{3}\,V_1 \chi_I.} \tag{234}$$

For the bare even background,

$$\tau = t, \tag{235}$$

so

$$c_0 = \frac{t}{8} - \frac{3r}{2}, \tag{236}$$

$$d_0 = \frac{t}{8}. \tag{237}$$

For the benchmark

$$t = 1, \qquad r = 0.05, \tag{238}$$

this gives

$$c_0 = 0.05, \qquad d_0 = 0.125. \tag{239}$$

If the time-reversal-even real Fock channel is allowed to relax, $\chi_R$ renormalizes the hopping through

$$\tau = t + V_1 \chi_R, \tag{240}$$

and therefore

$$d_{\mathrm{eff}} = d_0 + \frac{V_1 \chi_R}{8}, \tag{241}$$

while the regulator continues to contribute only to the scalar curvature.

### G. HF energy and self-consistency

For the uniform cyclic state all six NN bonds per cell have

$$|\chi_{ij}|^2 = \chi_R^2 + \chi_I^2. \tag{242}$$

Hence the exchange double-counting contribution is

$$E_{\rm dc} = 6N_c V_1(\chi_R^2 + \chi_I^2). \tag{243}$$

Using

$$\chi_R = \frac{\tau - t}{V_1}, \tag{244}$$

$$\chi_I = -\frac{\eta}{V_1}, \tag{245}$$

this may equivalently be written

$$\frac{E_{\rm dc}}{N_c} = \frac{6}{V_1}\left[(\tau - t)^2 + \eta^2\right]. \tag{246}$$

At zero temperature and filling two, the HF energy per cell is therefore

$$\frac{E_{\rm HF}}{N_c} = \frac{1}{N_c}\sum_{\mathbf{k}}\sum_{n\in{\rm occ}} \varepsilon_{n\mathbf{k}} + 6V_1(\chi_R^2 + \chi_I^2), \tag{247}$$

where the sum runs over the two occupied HF bands.

The self-consistency condition for an individual real-space bond $i = (\mathbf{R}, \alpha)$ and $j = (\mathbf{R} + \mathbf{\Delta}, \beta)$ is

$$\chi_{\alpha\beta}(\mathbf{\Delta}) = \left\langle c^\dagger_{\mathbf{R}\alpha} c_{\mathbf{R}+\mathbf{\Delta},\beta} \right\rangle. \tag{248}$$

Using the embedded Fourier convention,

$$c_{\mathbf{R}\alpha} = \frac{1}{\sqrt{N_c}}\sum_{\mathbf{k}} e^{i\mathbf{k}\cdot(\mathbf{R}+\mathbf{r}_\alpha)} c_{\mathbf{k}\alpha}, \tag{249}$$

we obtain

$$\begin{aligned}\chi_{\alpha\beta}(\mathbf{\Delta}) &= \frac{1}{N_c}\sum_{\mathbf{k}} e^{i\mathbf{k}\cdot(\mathbf{\Delta}+\mathbf{r}_\beta-\mathbf{r}_\alpha)} \left\langle c^\dagger_{\mathbf{k}\alpha} c_{\mathbf{k}\beta}\right\rangle \\ &= \frac{1}{N_c}\sum_{\mathbf{k}}\sum_{n\in{\rm occ}} e^{i\mathbf{k}\cdot(\mathbf{\Delta}+\mathbf{r}_\beta-\mathbf{r}_\alpha)} u^*_{n\alpha}(\mathbf{k}) u_{n\beta}(\mathbf{k}).\end{aligned} \tag{250}$$

Equations (180), (181), and (250) close the HF equations.

In particular, the QAH equation may be expressed as

$$\eta = -V_1 \chi_I(\eta), \tag{251}$$

or, using $m_\Gamma = 2\sqrt{3}\eta$,

$$m_\Gamma = -2\sqrt{3}V_1\chi_I(m_\Gamma). \tag{252}$$

The weak-coupling analysis of this equation is developed in the next section, where it is recast in terms of the symmetry-resolved QAH kernel and solved directly in logarithmic mass variables.

## III. WEAK-COUPLING LATTICE KERNEL AND ESSENTIAL SINGULARITY

We now reformulate the QAH Hartree–Fock self-consistency condition as a scalar response equation. An important point is that the resulting lattice kernel is a full-Brillouin-zone quantity: its value at finite mass depends on the Bloch eigenstates throughout the Brillouin zone and on all occupied-to-unoccupied interband processes. Only its singular dependence on the QAH mass is controlled by the quadratic band touching (QBT) near Γ.

We first consider the time-reversal-even bare background, for which $\tau = t$, and include the scalar regularization introduced in Sec. I. The effect of relaxing the time-reversal-even Fock channel is discussed below.

### A. Prescribed QAH source and lattice mass operator

Instead of imposing the QAH Hartree–Fock self-consistency condition immediately, it is convenient to regard the imaginary bond field $\eta$ as a prescribed source. With the cyclic convention

$$\nu_{AB} = \nu_{BC} = \nu_{CA} = +1, \qquad \nu_{ji} = -\nu_{ij}, \tag{253}$$

the corresponding real-space perturbation is

$$\delta H_\eta = -i\eta \sum_{\langle ij \rangle} \nu_{ij} c_i^\dagger c_j + \text{H.c.} \tag{254}$$

where one orientation is chosen for each nearest-neighbor bond.

In the embedded Bloch basis $(A, B, C)$, Eq. (254) becomes

$$\delta H_\eta(\mathbf{k}) = -2i\eta \begin{pmatrix} 0 & c_1 & -c_2 \\ -c_1 & 0 & c_3 \\ c_2 & -c_3 & 0 \end{pmatrix}, \tag{255}$$

with

$$c_1 = \cos \frac{\mathbf{k} \cdot \mathbf{a}_1}{2}, \tag{256}$$

$$c_2 = \cos \frac{\mathbf{k} \cdot \mathbf{a}_2}{2}, \tag{257}$$

$$c_3 = \cos \frac{\mathbf{k} \cdot (\mathbf{a}_2 - \mathbf{a}_1)}{2}. \tag{258}$$

At $\Gamma$ one has $c_1 = c_2 = c_3 = 1$, and therefore

$$\delta H_\eta(\Gamma) = -2i\eta \begin{pmatrix} 0 & 1 & -1 \\ -1 & 0 & 1 \\ 1 & -1 & 0 \end{pmatrix}. \tag{259}$$

Using the QBT basis

$$u_1 = \frac{1}{\sqrt{2}}(1, -1, 0)^T, \tag{260}$$

$$u_2 = \frac{1}{\sqrt{6}}(1, 1, -2)^T, \tag{261}$$

and

$$U^\dagger \begin{pmatrix} 0 & 1 & -1 \\ -1 & 0 & 1 \\ 1 & -1 & 0 \end{pmatrix} U = i\sqrt{3}\, \sigma_y, \tag{262}$$

we obtain

$$U^\dagger \delta H_\eta(\Gamma) U = 2\sqrt{3}\, \eta\, \sigma_y. \tag{263}$$

Thus the microscopic cyclic bond field produces the QBT mass

$$\boxed{m_\Gamma = 2\sqrt{3}\, \eta.} \tag{264}$$

It is useful to normalize the lattice source directly by $m_\Gamma$. We therefore define

$$\widehat{O}_m(\mathbf{k}) \equiv \frac{\partial H(\mathbf{k}; m_\Gamma)}{\partial m_\Gamma}. \tag{265}$$

Since $m_\Gamma = 2\sqrt{3}\,\eta$, Eqs. (255) and (264) give

$$\widehat{O}_m(\mathbf{k}) = -\frac{i}{\sqrt{3}}\begin{pmatrix} 0 & c_1 & -c_2 \\ -c_1 & 0 & c_3 \\ c_2 & -c_3 & 0 \end{pmatrix}. \tag{266}$$

By construction,

$$\boxed{U^\dagger \widehat{O}_m(\Gamma) U = \sigma_y.} \tag{267}$$

The full one-body Hamiltonian at prescribed mass can therefore be written as

$$H_{\text{lat}}(\mathbf{k}; m_\Gamma) = H_{\text{even}}(\mathbf{k}) + m_\Gamma \widehat{O}_m(\mathbf{k}). \tag{268}$$

Its eigenvectors are defined by

$$H_{\text{lat}}(\mathbf{k}; m_\Gamma)|u_{n\mathbf{k}}(m_\Gamma)\rangle = \varepsilon_{n\mathbf{k}}(m_\Gamma)|u_{n\mathbf{k}}(m_\Gamma)\rangle. \tag{269}$$

For $m_\Gamma \neq 0$ these eigenvectors already contain the cyclic complex QAH deformation. Importantly, Eq. (269) is solved over the entire Brillouin zone, not only near $\Gamma$.

### B. Exact full-Brillouin-zone lattice kernel

Let

$$A_c^{\text{phys}} = \frac{\sqrt{3}}{2} a^2 \tag{270}$$

be the physical primitive-cell area, and introduce the momentum measure per primitive cell

$$d\mu_k \equiv A_c^{\text{phys}} \frac{d^2 k}{(2\pi)^2}. \tag{271}$$

It satisfies

$$\int_{\text{BZ}} d\mu_k = 1. \tag{272}$$

The QAH order parameter conjugate to $m_\Gamma$ is

$$\Phi_{\text{cell}}^{\text{obs}}(m_\Gamma) = \int_{\text{BZ}} d\mu_k \sum_{n\in\text{occ}} \langle u_{n\mathbf{k}}(m_\Gamma)|\widehat{O}_m(\mathbf{k})|u_{n\mathbf{k}}(m_\Gamma)\rangle. \tag{273}$$

Equation (273) makes explicit that the order parameter receives contributions from occupied Bloch states at every momentum in the Brillouin zone.

With the bond normalization of Sec. II, this quantity is equivalently

$$\Phi_{\text{cell}}^{\text{obs}} = 2\sqrt{3}\,\chi_I^{\text{obs}}. \tag{274}$$

Indeed, the expectation value of $\partial H/\partial\eta$ per cell is $12\chi_I$, while $\partial\eta/\partial m_\Gamma = 1/(2\sqrt{3})$, giving

$$\frac{\partial E_{\text{band}}^{\text{cell}}}{\partial m_\Gamma} = 2\sqrt{3}\,\chi_I. \tag{275}$$

This is also an immediate consequence of the Hellmann–Feynman theorem applied to Eq. (268).

The finite-source lattice kernel is defined by

$$\boxed{K_{\text{lat}}^{\text{cell}}(m_\Gamma) \equiv -\frac{\Phi_{\text{cell}}^{\text{obs}}(m_\Gamma)}{m_\Gamma}.} \tag{276}$$

Combining Eqs. (274) and (264),

$$K_{\text{lat}}^{\text{cell}}(m_\Gamma) = -\frac{\chi_I^{\text{obs}}(\eta)}{\eta}. \tag{277}$$

Substitution of Eq. (273) gives the explicit full-lattice expression

$$K_{\text{lat}}^{\text{cell}}(m_\Gamma) = -\frac{1}{m_\Gamma}\int_{\text{BZ}} d\mu_k \sum_{n\in\text{occ}} \langle u_{n\mathbf{k}}(m_\Gamma)|\widehat{O}_m(\mathbf{k})|u_{n\mathbf{k}}(m_\Gamma)\rangle. \tag{278}$$

Equation (278) is the central microscopic definition of the lattice kernel. It contains the eigenvectors $|u_{n\mathbf{k}}\rangle$ throughout the Brillouin zone. The QBT analysis introduced below is not a replacement for this full-lattice formula; it is used only to extract analytically its singular $m_\Gamma \to 0$ contribution.

The Hartree–Fock relation

$$\eta = -V_1\chi_I \tag{279}$$

is equivalent to

$$m_\Gamma = -V_1\Phi_{\text{cell}}. \tag{280}$$

For a nonzero QAH solution, division by $m_\Gamma$ gives the exact scalar self-consistency condition

$$\boxed{\frac{1}{V_1} = K_{\text{lat}}^{\text{cell}}(m_\Gamma).} \tag{281}$$

At finite $m_\Gamma$, Eq. (276) is a nonlinear response ratio rather than an infinitesimal susceptibility. Time reversal implies

$$\Phi_{\text{cell}}^{\text{obs}}(-m_\Gamma) = -\Phi_{\text{cell}}^{\text{obs}}(m_\Gamma), \tag{282}$$

and consequently

$$K_{\text{lat}}^{\text{cell}}(-m_\Gamma) = K_{\text{lat}}^{\text{cell}}(m_\Gamma). \tag{283}$$

### C. Spectral representation away from the QBT

The role of Bloch states away from $\Gamma$ is particularly transparent in the $m_\Gamma \to 0$ limit. At a momentum for which the occupied and unoccupied bands are nondegenerate, first-order perturbation theory in $m_\Gamma$ gives

$$|u_{n\mathbf{k}}(m_\Gamma)\rangle = |u_{n\mathbf{k}}^{(0)}\rangle + m_\Gamma \sum_{\ell\neq n} |u_{\ell\mathbf{k}}^{(0)}\rangle \frac{\langle u_{\ell\mathbf{k}}^{(0)}|\widehat{O}_m(\mathbf{k})|u_{n\mathbf{k}}^{(0)}\rangle}{\varepsilon_{n\mathbf{k}}^{(0)} - \varepsilon_{\ell\mathbf{k}}^{(0)}} + O(m_\Gamma^2). \tag{284}$$

For the time-reversal-symmetric reference Hamiltonian,

$$\langle u_{n\mathbf{k}}^{(0)}|\widehat{O}_m(\mathbf{k})|u_{n\mathbf{k}}^{(0)}\rangle = 0. \tag{285}$$

Using Eq. (284), the contribution linear in $m_\Gamma$ to the total occupied expectation value is

$$\sum_{n\in\text{occ}} \langle \widehat{O}_m\rangle_{n\mathbf{k}} = -2m_\Gamma \sum_{\substack{n\in\text{occ}\\ \ell\in\text{unocc}}} \frac{\left|\langle u_{\ell\mathbf{k}}^{(0)}|\widehat{O}_m(\mathbf{k})|u_{n\mathbf{k}}^{(0)}\rangle\right|^2}{\varepsilon_{\ell\mathbf{k}}^{(0)} - \varepsilon_{n\mathbf{k}}^{(0)}} + O(m_\Gamma^3). \tag{286}$$

Terms connecting two occupied bands cancel in the sum over the occupied subspace, leaving only occupied-to-unoccupied transitions.

It is therefore natural to define the zero-mass lattice kernel density away from the QBT as

$$\boxed{\kappa_{\text{lat}}(\mathbf{k},0) = 2 \sum_{\substack{n\in\text{occ}\\ \ell\in\text{unocc}}} \frac{\left|\langle u_{\ell\mathbf{k}}^{(0)}|\widehat{O}_m(\mathbf{k})|u_{n\mathbf{k}}^{(0)}\rangle\right|^2}{\varepsilon_{\ell\mathbf{k}}^{(0)} - \varepsilon_{n\mathbf{k}}^{(0)}}.} \tag{287}$$

This expression makes two features explicit. First, the kernel depends not only on the band energies but also on the full lattice eigenvectors through the QAH matrix elements. Second, at filling two the sum contains both the low-energy transition between the two bands forming the QBT and finite-energy transitions involving the remote kagome band.

At a generic momentum away from $\Gamma$, all denominators in Eq. (287) remain finite. Hence these momenta produce a finite contribution to $K_{\rm lat}^{\rm cell}$ as $m_\Gamma \to 0$. The divergence can arise only from the vanishing interband splitting at the QBT.

### D. Infrared contribution from the quadratic band touching

We next isolate the singular contribution near $\Gamma$. Introduce the dimensionless momentum

$$\mathbf{q} = a\mathbf{k} \tag{288}$$

measured from $\Gamma$, and the dimensionless cell area

$$\mathcal{A}_c \equiv \frac{A_c^{\rm phys}}{a^2} = \frac{\sqrt{3}}{2}. \tag{289}$$

The corresponding measure per primitive cell is

$$d\mu_q = \mathcal{A}_c \frac{d^2q}{(2\pi)^2}. \tag{290}$$

The massive QBT Hamiltonian is

$$\begin{aligned} H_{\rm QBT}(\mathbf{q}) = &\left(E_\Gamma - c_0 q^2\right) \mathbb{1} \\ &- d_0 \left[(q_x^2 - q_y^2)\sigma_z + 2q_x q_y \sigma_x\right] + m_\Gamma \sigma_y, \end{aligned} \tag{291}$$

with

$$d_0 = \frac{t}{8} > 0. \tag{292}$$

The scalar term proportional to $\mathbb{1}$ changes the band energies but not the QBT eigenvectors or their direct splitting. The pseudospin vector is

$$\mathbf{h}(\mathbf{q}) = \left(-2d_0 q_x q_y,\, m_\Gamma,\, -d_0(q_x^2 - q_y^2)\right), \tag{293}$$

with magnitude

$$R(q) = \sqrt{d_0^2 q^4 + m_\Gamma^2}. \tag{294}$$

The occupied member of the QBT doublet has

$$\langle \sigma_y \rangle_- = -\frac{m_\Gamma}{\sqrt{d_0^2 q^4 + m_\Gamma^2}}. \tag{295}$$

Since the lattice mass operator reduces to $\sigma_y$ at $\Gamma$, the singular QBT contribution to the kernel density is

$$\boxed{\kappa_{\rm QBT}(\mathbf{q}, m_\Gamma) = \frac{1}{\sqrt{d_0^2 q^4 + m_\Gamma^2}}.} \tag{296}$$

Choose a fixed disk

$$D_\Gamma(q_c) = \{\mathbf{q} : q < q_c\} \tag{297}$$

within the regime where the QBT expansion is valid. The continuum contribution from this disk is

$$K_{\rm QBT}^{\rm cell} = \mathcal{A}_c \int_{q<q_c} \frac{d^2q}{(2\pi)^2} \frac{1}{\sqrt{d_0^2 q^4 + m_\Gamma^2}}. \tag{298}$$

Using polar coordinates,

$$K_{\rm QBT}^{\rm cell} = \frac{\mathcal{A}_c}{2\pi}\int_0^{q_c}\frac{q\,dq}{\sqrt{d_0^2q^4+m_\Gamma^2}}. \tag{299}$$

With

$$x = d_0q^2, \qquad q\,dq = \frac{dx}{2d_0}, \tag{300}$$

this becomes

$$K_{\rm QBT}^{\rm cell} = \frac{\mathcal{A}_c}{4\pi d_0}\int_0^{d_0q_c^2}\frac{dx}{\sqrt{x^2+m_\Gamma^2}}, \tag{301}$$

and therefore

$$\boxed{K_{\rm QBT}^{\rm cell}(m_\Gamma) = \frac{\mathcal{A}_c}{4\pi d_0}\,\mathrm{arsinh}\left(\frac{d_0q_c^2}{|m_\Gamma|}\right).} \tag{302}$$

For $|m_\Gamma| \ll d_0q_c^2$,

$$\mathrm{arsinh}\,X = \ln(2X) + O(X^{-2}), \tag{303}$$

so

$$K_{\rm QBT}^{\rm cell} = \rho_{\rm lat}^{\rm cell}\ln\left(\frac{2d_0q_c^2}{|m_\Gamma|}\right) + o(1), \tag{304}$$

with

$$\boxed{\rho_{\rm lat}^{\rm cell} = \frac{\mathcal{A}_c}{4\pi d_0}.} \tag{305}$$

For the kagome QBT,

$$d_0 = \frac{t}{8}, \qquad \mathcal{A}_c = \frac{\sqrt{3}}{2}, \tag{306}$$

and hence

$$\boxed{\rho_{\rm lat}^{\rm cell} = \frac{\sqrt{3}}{\pi t}.} \tag{307}$$

The logarithmic coefficient is thus determined entirely by the long-wavelength QBT. The remainder of the lattice kernel, however, cannot be obtained from the continuum Hamiltonian alone.

### E. Full-lattice ultraviolet completion and the constant $B_{\rm lat}^{\rm cell}$

We now make explicit how states away from $\Gamma$ enter the weak-mass asymptotics.

For finite $m_\Gamma$, define the full-lattice kernel density by

$$\kappa_{\rm lat}(\mathbf{k}, m_\Gamma) \equiv -\frac{1}{m_\Gamma}\sum_{n\in{\rm occ}}\langle u_{n\mathbf{k}}(m_\Gamma)|\widehat{O}_m(\mathbf{k})|u_{n\mathbf{k}}(m_\Gamma)\rangle, \tag{308}$$

so that

$$K_{\rm lat}^{\rm cell}(m_\Gamma) = \int_{\rm BZ} d\mu_k\,\kappa_{\rm lat}(\mathbf{k}, m_\Gamma). \tag{309}$$

Split the Brillouin zone into the QBT disk and its complement:

$$K_{\rm lat}^{\rm cell} = \int_{D_\Gamma} d\mu_q\, \kappa_{\rm lat}(\mathbf{q}, m_\Gamma) + \int_{\rm BZ\setminus D_\Gamma} d\mu_k\, \kappa_{\rm lat}(\mathbf{k}, m_\Gamma). \tag{310}$$

The second term contains explicitly all momenta away from the QBT.

Inside $D_\Gamma$ we add and subtract the singular continuum kernel, obtaining the exact decomposition

$$K_{\rm lat}^{\rm cell} = \int_{D_\Gamma} d\mu_q \left[\kappa_{\rm lat}(\mathbf{q}, m_\Gamma) - \kappa_{\rm QBT}(\mathbf{q}, m_\Gamma)\right] + \int_{\rm BZ\setminus D_\Gamma} d\mu_k\, \kappa_{\rm lat}(\mathbf{k}, m_\Gamma) + \int_{D_\Gamma} d\mu_q\, \kappa_{\rm QBT}(\mathbf{q}, m_\Gamma). \tag{311}$$

Only the last term is infrared singular. The first term remains integrable as $m_\Gamma \to 0$, because the leading QBT singularity has been explicitly removed. The second term also has a regular zero-mass limit because the occupied-to-unoccupied energy separations remain finite outside the QBT disk.

Consequently,

$$\boxed{K_{\rm lat}^{\rm cell}(m_\Gamma) = \rho_{\rm lat}^{\rm cell} \ln \frac{t}{|m_\Gamma|} + B_{\rm lat}^{\rm cell} + o(1).} \tag{312}$$

We use $t$ as the reference energy appearing inside the logarithm.

The finite lattice constant is therefore

$$\boxed{B_{\rm lat}^{\rm cell} = \lim_{m_\Gamma \to 0} \left[K_{\rm lat}^{\rm cell}(m_\Gamma) - \rho_{\rm lat}^{\rm cell} \ln \frac{t}{|m_\Gamma|}\right].} \tag{313}$$

More importantly, Eq. (311) gives an explicit decomposition of this constant:

$$B_{\rm lat}^{\rm cell} = B_{\rm near}^{\rm reg}(q_c) + B_{\rm far}(q_c) + B_{\rm match}(q_c), \tag{314}$$

where

$$B_{\rm near}^{\rm reg}(q_c) \equiv \lim_{m_\Gamma \to 0} \int_{D_\Gamma} d\mu_q \left[\kappa_{\rm lat}(\mathbf{q}, m_\Gamma) - \kappa_{\rm QBT}(\mathbf{q}, m_\Gamma)\right], \tag{315}$$

$$B_{\rm far}(q_c) \equiv \int_{\rm BZ\setminus D_\Gamma} d\mu_k\, \kappa_{\rm lat}(\mathbf{k}, 0), \tag{316}$$

and

$$B_{\rm match}(q_c) \equiv \rho_{\rm lat}^{\rm cell} \ln\left(\frac{2d_0 q_c^2}{t}\right). \tag{317}$$

Using the spectral representation Eq. (287), the contribution from momenta away from $\Gamma$ is explicitly

$$B_{\rm far}(q_c) = 2\int_{\rm BZ\setminus D_\Gamma} d\mu_k \sum_{\substack{n\in{\rm occ}\\ \ell\in{\rm unocc}}} \frac{\left|\langle u_{\ell\mathbf{k}}^{(0)} | \widehat{O}_m(\mathbf{k}) | u_{n\mathbf{k}}^{(0)}\rangle\right|^2}{\varepsilon_{\ell\mathbf{k}}^{(0)} - \varepsilon_{n\mathbf{k}}^{(0)}}. \tag{318}$$

Equation (318) displays explicitly how Bloch eigenvectors at momenta far from $\Gamma$ enter the weak-coupling mass scale. They do not modify the coefficient of the infrared logarithm, but they contribute to the finite constant $B_{\rm lat}^{\rm cell}$ through their QAH matrix elements and their interband energy denominators.

The three terms in Eq. (314) have a simple interpretation. $B_{\rm near}^{\rm reg}$ contains finite corrections to the leading QBT description within the neighborhood of $\Gamma$; $B_{\rm far}$ contains the rest of the Brillouin zone, including finite-energy processes involving the remote kagome band; and $B_{\rm match}$ converts the arbitrary continuum momentum cutoff $q_c$ into the reference energy $t$ used in Eq. (312).

The three contributions separately depend on $q_c$, but their sum does not. Thus $q_c$ is only an intermediate matching scale and has no physical significance.

This decomposition also makes clear the distinct infrared and lattice roles:

$$\rho_{\rm lat}^{\rm cell} \quad \text{is fixed by the QBT near } \Gamma, \tag{319}$$

whereas

$$B_{\rm lat}^{\rm cell} \quad \text{contains information from the full lattice problem.} \tag{320}$$

For the bare-even benchmark

$$t = 1, \qquad r = 0.05, \tag{321}$$

we obtain

$$\boxed{B_{\rm lat}^{\rm cell} = 0.036667506(6)\, t^{-1}.} \tag{322}$$

### F. Essential interaction scale

Combining the exact self-consistency condition Eq. (281) with the lattice asymptotics Eq. (312) gives

$$\frac{1}{V_1} = \rho_{\rm lat}^{\rm cell} \ln \frac{t}{|m_\Gamma|} + B_{\rm lat}^{\rm cell} + o(1). \tag{323}$$

Solving for the logarithm,

$$\ln \frac{t}{|m_\Gamma|} = \frac{1}{\rho_{\rm lat}^{\rm cell} V_1} - \frac{B_{\rm lat}^{\rm cell}}{\rho_{\rm lat}^{\rm cell}} + o(1). \tag{324}$$

Exponentiation yields

$$\boxed{|m_\Gamma(V_1)| \sim M_{\rm lat} \exp\left(-\frac{A_{\rm lat}}{V_1}\right),} \tag{325}$$

where

$$\boxed{A_{\rm lat} = \frac{1}{\rho_{\rm lat}^{\rm cell}} = \frac{4\pi d_0}{\mathcal{A}_c}} \tag{326}$$

and

$$\boxed{M_{\rm lat} = t \exp\left(\frac{B_{\rm lat}^{\rm cell}}{\rho_{\rm lat}^{\rm cell}}\right).} \tag{327}$$

For

$$d_0 = \frac{t}{8}, \qquad \mathcal{A}_c = \frac{\sqrt{3}}{2}, \tag{328}$$

one obtains

$$\boxed{A_{\rm lat} = \frac{\pi t}{\sqrt{3}}.} \tag{329}$$

Equivalently, in terms of the dimensionless coupling $V_1/t$,

$$\frac{|m_\Gamma|}{t} \sim \frac{M_{\rm lat}}{t} \exp\left[-\frac{\pi/\sqrt{3}}{V_1/t}\right]. \tag{330}$$

Using Eq. (322),

$$\boxed{M_{\rm lat} = 1.0687690(1)\, t.} \tag{331}$$

Equations (326) and (327) summarize the infrared–ultraviolet separation. The coefficient $A_{\rm lat}$ is fixed by the QBT coefficient $d_0$, whereas $M_{\rm lat}$ contains the finite contribution of the entire kagome band structure through $B^{\rm cell}_{\rm lat}$. In particular, the Bloch eigenvectors away from $\Gamma$ affect the prefactor $M_{\rm lat}$ even though they do not change the essential-singularity coefficient $A_{\rm lat}$.

### G. Effect of the time-reversal-even Fock channel

The derivation above used the bare-even background $\tau = t$. When the real Fock channel is allowed to relax, the quadratic QBT coefficient becomes

$$d_{\rm eff}(V_1) = \frac{\tau}{8} = \frac{t + V_1\chi_R}{8}. \tag{332}$$

Since $\chi_R$ approaches a finite value as $V_1 \to 0$,

$$d_{\rm eff}(V_1) = d_0 + O(V_1), \qquad d_0 = \frac{t}{8}. \tag{333}$$

For the relaxed-even solution,

$$\chi_R \longrightarrow \frac{1}{6} \qquad (V_1 \to 0), \tag{334}$$

so

$$d_{\rm eff} = d_0 + \frac{V_1}{48} + O(V_1^2). \tag{335}$$

If the logarithmic coefficient is expressed in terms of $d_{\rm eff}$, the corresponding inverse coefficient is

$$A_{\rm eff}(V_1) = \frac{4\pi d_{\rm eff}(V_1)}{\mathcal{A}_c} = A_{\rm lat} + O(V_1). \tag{336}$$

Therefore

$$\frac{A_{\rm eff}(V_1)}{V_1} = \frac{A_{\rm lat}}{V_1} + O(1). \tag{337}$$

The $O(1)$ correction in the exponent modifies the finite, nonuniversal prefactor but not the coefficient multiplying $1/V_1$. The leading essential singularity is therefore controlled by the weak-coupling limiting value

$$d_0 = \lim_{V_1\to 0} d_{\rm eff}(V_1) = \frac{t}{8}. \tag{338}$$

### H. Logarithmic mass variable

At weak coupling the self-consistent mass becomes exponentially small, making direct solution in terms of $m_\Gamma$ numerically inconvenient. We therefore introduce

$$\boxed{u \equiv \ln\frac{t}{|m_\Gamma|},} \tag{339}$$

so that

$$|m_\Gamma| = te^{-u}. \tag{340}$$

The exact scalar equation Eq. (281) becomes

$$F(u;V_1) \equiv \frac{1}{V_1} - K^{\rm cell}_{\rm lat}(te^{-u}) = 0. \tag{341}$$

Writing

$$K^{\rm cell}_{\rm lat}(m) = \rho^{\rm cell}_{\rm lat} \ln\frac{t}{|m|} + B^{\rm cell}_{\rm lat} + \Delta K(m), \tag{342}$$

with

$$\Delta K(m) \to 0 \qquad (m \to 0), \tag{343}$$

gives

$$F(u;V_1) = \frac{1}{V_1} - \rho^{\rm cell}_{\rm lat} u - B^{\rm cell}_{\rm lat} - \Delta K(te^{-u}). \tag{344}$$

The asymptotic solution is

$$u = \frac{A_{\rm lat}}{V_1} - \ln\frac{M_{\rm lat}}{t} + o(1). \tag{345}$$

Thus the numerically represented variable grows only linearly with $1/V_1$, even though the physical mass decreases exponentially.

The momentum scale associated with the massive QBT is

$$q_m = \sqrt{\frac{|m_\Gamma|}{d_0}}, \tag{346}$$

and consequently

$$q_m \sim \exp\left(-\frac{A_{\rm lat}}{2V_1}\right). \tag{347}$$

A uniform momentum mesh capable of directly resolving this scale would require

$$\Delta q \lesssim q_m, \tag{348}$$

so that its required linear resolution grows exponentially as $V_1 \to 0$. Working with the logarithmic variable $u$ together with the analytic subtraction of the QBT singularity avoids this resolution problem. Numerical quadrature and convergence tests are discussed in Sec. XI.

## IV. TOPOLOGY OF THE INTERACTION-GENERATED STATE

We now establish the topological character of the interaction-generated QAH state. The imaginary Hartree–Fock bond order derived above produces a mass term $m_\Gamma\sigma_y$ in the low-energy QBT Hamiltonian. This term breaks time-reversal symmetry and gaps the quadratic touching between the occupied and unoccupied states at $\Gamma$. The two signs of $m_\Gamma$ describe the two degenerate, time-reversal-related chiral states and carry opposite Chern numbers.

### A. Time-reversal breaking by the QAH mass

In the real QBT basis

$$u_1 = \frac{1}{\sqrt{2}}(1,-1,0)^T, \qquad u_2 = \frac{1}{\sqrt{6}}(1,1,-2)^T, \tag{349}$$

the spinless time-reversal operation is represented by complex conjugation,

$$\mathcal{T} = \mathcal{K}, \qquad \mathcal{T}^2 = +1. \tag{350}$$

The matrices $\sigma_x$ and $\sigma_z$ are real, whereas

$$\sigma_y^* = -\sigma_y. \tag{351}$$

Consequently, the massless QBT Hamiltonian is time-reversal invariant, while

$$\mathcal{T}\,(m_\Gamma\sigma_y)\,\mathcal{T}^{-1} = -m_\Gamma\sigma_y. \tag{352}$$

Thus a nonzero $m_\Gamma$ spontaneously breaks time reversal.

Microscopically,

$$m_\Gamma = 2\sqrt{3}\,\eta = -2\sqrt{3}\,V_1\chi_I, \tag{353}$$

so time reversal acts equivalently as

$$\chi_I \to -\chi_I, \qquad \eta \to -\eta, \qquad m_\Gamma \to -m_\Gamma. \tag{354}$$

The two self-consistent Hartree–Fock solutions $\pm\chi_I$ are therefore degenerate and correspond to opposite orientations of the circulating bond currents.

### B. Berry curvature of the massive quadratic band touching

The part of the continuum Hamiltonian relevant for the eigenvectors is

$$h_{\mathrm{QBT}}(\mathbf{q}) = -d\left[(q_x^2-q_y^2)\sigma_z + 2q_xq_y\sigma_x\right] + m\sigma_y, \tag{355}$$

where $d>0$. The scalar contribution $(E_\Gamma - cq^2)\mathbb{1}$ has been omitted because it changes neither the eigenvectors nor the Berry curvature.

Writing

$$h_{\mathrm{QBT}}(\mathbf{q}) = \mathbf{h}(\mathbf{q})\cdot\boldsymbol{\sigma}, \tag{356}$$

the pseudospin vector is

$$\mathbf{h}(\mathbf{q}) = \left(-2dq_xq_y,\, m,\, -d(q_x^2-q_y^2)\right), \tag{357}$$

with

$$R(q) = |\mathbf{h}| = \sqrt{m^2+d^2q^4}. \tag{358}$$

The two eigenvalues are

$$E_\pm(\mathbf{q}) = E_\Gamma - cq^2 \pm R(q). \tag{359}$$

For the Berry-curvature convention

$$\Omega_n(\mathbf{q}) = -2\,\mathrm{Im}\left\langle \partial_{q_x}u_{n\mathbf{q}} \middle| \partial_{q_y}u_{n\mathbf{q}}\right\rangle, \tag{360}$$

the two-band expression can be written as

$$\Omega_\pm = \mp\frac{\mathbf{h}\cdot\left(\partial_{q_x}\mathbf{h}\times\partial_{q_y}\mathbf{h}\right)}{2|\mathbf{h}|^3}. \tag{361}$$

Here the $-$ label denotes the lower member of the QBT doublet.

The derivatives of Eq. (357) are

$$\partial_{q_x}\mathbf{h} = \left(-2dq_y,\, 0,\, -2dq_x\right), \tag{362}$$

$$\partial_{q_y}\mathbf{h} = \left(-2dq_x,\, 0,\, 2dq_y\right), \tag{363}$$

and therefore

$$\partial_{q_x}\mathbf{h} \times \partial_{q_y}\mathbf{h} = \big(0,\, 4d^2q^2,\, 0\big)\,. \tag{364}$$

Taking the scalar product with $\mathbf{h}$ gives

$$\mathbf{h}\cdot\big(\partial_{q_x}\mathbf{h} \times \partial_{q_y}\mathbf{h}\big) = 4d^2mq^2. \tag{365}$$

Hence

$$\boxed{\Omega_-(\mathbf{q}) = \frac{2d^2mq^2}{\left(m^2+d^2q^4\right)^{3/2}},} \tag{366}$$

while

$$\boxed{\Omega_+(\mathbf{q}) = -\frac{2d^2mq^2}{\left(m^2+d^2q^4\right)^{3/2}}.} \tag{367}$$

Several properties follow immediately. First,

$$\Omega_\pm(\mathbf{q};-m) = -\Omega_\pm(\mathbf{q};m), \tag{368}$$

as required by time reversal. Second, the curvature is independent of the sign of $d$, since $d$ appears quadratically. The chirality of the state is therefore controlled by the sign of the QAH mass $m$, not by the sign convention chosen for the quadratic pseudospin texture.

### C. Berry flux carried by the gapped QBT

The Berry curvature of the lower QBT band is concentrated on the momentum scale

$$q_m = \sqrt{\frac{|m|}{d}}. \tag{369}$$

Nevertheless, its integrated Berry flux remains finite as $m \to 0$.

Using polar coordinates, the Berry flux of the lower band is

$$\frac{1}{2\pi}\int_{\mathbb{R}^2} d^2q\,\Omega_-(\mathbf{q}) = \int_0^\infty q\,dq\,\frac{2d^2mq^2}{(m^2+d^2q^4)^{3/2}}. \tag{370}$$

Introduce

$$x = \frac{dq^2}{|m|}, \qquad q\,dq = \frac{|m|}{2d}\,dx. \tag{371}$$

Equation (370) then becomes

$$\frac{1}{2\pi}\int d^2q\,\Omega_- = \mathrm{sgn}(m)\int_0^\infty dx\,\frac{x}{(1+x^2)^{3/2}}. \tag{372}$$

Since

$$\int_0^\infty dx\,\frac{x}{(1+x^2)^{3/2}} = 1, \tag{373}$$

we obtain

$$\boxed{\frac{1}{2\pi}\int d^2q\,\Omega_-(\mathbf{q}) = \mathrm{sgn}(m).} \tag{374}$$

Thus opening the QBT by the QAH mass produces a Berry flux

$$\Phi_B = 2\pi\,\mathrm{sgn}(m) \tag{375}$$

in the lower member of the touching doublet, with the opposite flux in the upper member.

The integer Berry flux is a direct consequence of the winding-two structure of the quadratic touching. In polar coordinates, $q_x = q\cos\theta$ and $q_y = q\sin\theta$, the massless pseudospin texture contains

$$q_x^2 - q_y^2 = q^2\cos 2\theta, \tag{376}$$

$$2q_xq_y = q^2\sin 2\theta. \tag{377}$$

The in-plane pseudospin therefore winds twice as $\mathbf{q}$ encircles $\Gamma$. Gapping this winding-two touching produces an integer Berry flux, in contrast to the half-integer contribution associated with a single continuum Dirac cone.

The continuum calculation by itself describes only the neighborhood of the QBT and should not be regarded as a complete lattice topological classification. Its role is to identify the Berry flux transferred between the two bands when the touching is gapped. The global Chern number is determined below from the full lattice occupied subspace.

### D. Chern number of the full three-band lattice model

At filling two, the physical ground state contains two occupied bands per primitive cell. The appropriate gauge-invariant object is therefore the rank-two occupied projector

$$P_{\mathrm{occ}}(\mathbf{k}) = \sum_{n\in\mathrm{occ}} |u_{n\mathbf{k}}\rangle\langle u_{n\mathbf{k}}|. \tag{378}$$

The Berry curvature of the occupied subspace is

$$\Omega_{\mathrm{occ}}(\mathbf{k}) = i\,\mathrm{Tr}\left\{P_{\mathrm{occ}}\left[\partial_{k_x}P_{\mathrm{occ}}, \partial_{k_y}P_{\mathrm{occ}}\right]\right\}, \tag{379}$$

and the corresponding Chern number is

$$\boxed{C_{\mathrm{occ}} = \frac{1}{2\pi}\int_{\mathrm{BZ}} d^2k\,\Omega_{\mathrm{occ}}(\mathbf{k}).} \tag{380}$$

The projector formulation is particularly convenient at filling two. Possible band crossings or small gaps between the two occupied bands do not affect $C_{\mathrm{occ}}$ as long as the occupied two-dimensional subspace remains separated from the unoccupied band. Only a closing of the gap between the occupied and unoccupied sectors can change Eq. (380).

At $m_\Gamma = 0$, precisely such a closing occurs at the QBT at $\Gamma$. For any nonzero QAH mass, the touching is removed and the occupied subspace becomes separated from the upper band. The Berry flux generated near $\Gamma$ is then continuously embedded into the full lattice Brillouin zone.

For the lattice conventions used here, the full-Brillouin-zone calculation gives

$$\boxed{C_{\mathrm{occ}} = \mathrm{sgn}(m_\Gamma).} \tag{381}$$

Thus

$$m_\Gamma > 0 \quad\Longrightarrow\quad C_{\mathrm{occ}} = +1, \tag{382}$$

whereas

$$m_\Gamma < 0 \quad\Longrightarrow\quad C_{\mathrm{occ}} = -1. \tag{383}$$

The two self-consistent Hartree–Fock solutions are therefore QAH states with opposite topology,

$$\boxed{+m_\Gamma \;\longleftrightarrow\; C_{\mathrm{occ}} = +1, \qquad -m_\Gamma \;\longleftrightarrow\; C_{\mathrm{occ}} = -1.} \tag{384}$$

Because the complete three-orbital Hilbert space is topologically trivial,

$$P_{\rm occ} + P_{\rm unocc} = \mathbb{1}, \tag{385}$$

the Chern number of the single unoccupied band satisfies

$$C_{\rm unocc} = -C_{\rm occ}. \tag{386}$$

This complementary-projector relation will also be useful below for the quantum metric.

### E. Robustness to the scalar regulator and the even Fock channel

The scalar regularization used to remove the exactly flat-band endpoint takes the form

$$H_{\rm reg}(\mathbf{k}; r) = f_r(\mathbf{k})\mathbb{1}_3. \tag{387}$$

Since it is proportional to the identity at every momentum, it leaves all Bloch eigenvectors unchanged. It therefore changes neither the Berry curvature nor the Chern number at fixed QAH mass:

$$\Omega_{\rm occ}(\mathbf{k}; r) = \Omega_{\rm occ}(\mathbf{k}; 0), \tag{388}$$

and

$$C_{\rm occ}(r) = C_{\rm occ}(0), \tag{389}$$

provided the filling and the occupied–unoccupied gap remain well defined.

The time-reversal-even Fock channel similarly changes the quadratic coefficient according to

$$d_0 \to d_{\rm eff} = \frac{\tau}{8}, \qquad \tau = t + V_1\chi_R, \tag{390}$$

but does not alter the winding of the QBT. Since the Berry curvature depends on $d_{\rm eff}^2$, its sign remains controlled entirely by $m_\Gamma$. Continuous renormalization of $d_{\rm eff}$ therefore cannot change the Chern number without a closing of the occupied–unoccupied gap.

### F. Quantized anomalous Hall response

The interaction-generated state is consequently a Chern insulator. At zero temperature and with the chemical potential in the QAH gap, the dc Hall conductivity is

$$\boxed{\sigma_{xy} = C_{\rm occ}\frac{e^2}{h} = \mathrm{sgn}(m_\Gamma)\frac{e^2}{h},} \tag{391}$$

with the sign reversed between the two time-reversal-related Hartree–Fock solutions.

The exponentially small weak-coupling scale

$$|m_\Gamma| \sim M_{\rm lat} e^{-A_{\rm lat}/V_1} \tag{392}$$

therefore controls the spectral gap protecting a state whose zero-temperature topological invariant remains integer-valued for arbitrarily small nonzero $V_1$. In the limit $V_1 \to 0^+$ the gap collapses beyond all orders in the interaction, while the Chern number changes only at the singular point $m_\Gamma = 0$ where the QBT is restored.

## V. CONTINUUM QUANTUM METRIC

We now derive the quantum metric of the massive continuum QBT and its singular behavior as the interaction-generated mass tends to zero. The same mass that gaps the quadratic touching and produces the QAH state also regularizes a singular quantum-geometric structure concentrated around Γ. Locally, the metric develops an amplitude proportional to $1/|m|$ over a momentum scale $q_m \propto |m|^{1/2}$; after momentum integration, this singular structure produces the logarithm used in the main text.

Throughout this section we use the dimensionless momentum $\mathbf{q} = a\mathbf{k}$ introduced above and take $d > 0$. If derivatives are instead taken with respect to the physical momentum $\mathbf{k}$, the metric acquires the corresponding factors of $a^2$, while the combination $d^2k\ \mathrm{Tr}\, g$ entering the integrated metric is unchanged.

### A. Quantum metric of a two-band Hamiltonian

For a normalized Bloch eigenstate $|u_{n\mathbf{q}}\rangle$, the quantum geometric tensor is

$$Q_{ij}^{(n)} = \langle \partial_{q_i} u_{n\mathbf{q}} \big| (1 - |u_{n\mathbf{q}}\rangle\langle u_{n\mathbf{q}}|) \big| \partial_{q_j} u_{n\mathbf{q}} \rangle . \tag{393}$$

Its real part defines the quantum metric,

$$g_{ij}^{(n)} = \mathrm{Re}\, Q_{ij}^{(n)}, \tag{394}$$

whereas its antisymmetric imaginary part gives the Berry curvature.

For a two-band Hamiltonian

$$h(\mathbf{q}) = \mathbf{h}(\mathbf{q}) \cdot \boldsymbol{\sigma}, \tag{395}$$

define the unit pseudospin

$$\widehat{\mathbf{h}} = \frac{\mathbf{h}}{|\mathbf{h}|}. \tag{396}$$

The projectors onto the two bands are

$$P_{\pm} = \frac{1}{2}\left(\mathbb{1} \pm \widehat{\mathbf{h}} \cdot \boldsymbol{\sigma}\right). \tag{397}$$

Using the projector representation of the quantum metric,

$$g_{ij}^{(\pm)} = \frac{1}{2}\mathrm{Tr}\left[\partial_{q_i} P_{\pm} \partial_{q_j} P_{\pm}\right], \tag{398}$$

one finds

$$\boxed{g_{ij}^{(\pm)} = \frac{1}{4}\,\partial_{q_i}\widehat{\mathbf{h}} \cdot \partial_{q_j}\widehat{\mathbf{h}}.} \tag{399}$$

The two bands therefore have the same quantum metric,

$$g_{ij}^{(+)} = g_{ij}^{(-)}. \tag{400}$$

This is in contrast to the Berry curvature, whose sign is opposite for the two bands. The quantum metric is consequently insensitive to the sign of the QAH chirality.

### B. Metric tensor of the massive QBT

The part of the continuum QBT Hamiltonian that determines its eigenvectors is

$$h_{\mathrm{QBT}}(\mathbf{q}) = -d\left[(q_x^2 - q_y^2)\sigma_z + 2q_x q_y \sigma_x\right] + m\sigma_y. \tag{401}$$

The scalar term $(E_\Gamma - cq^2)\mathbb{1}$ has been omitted because it does not affect the eigenstates and therefore does not enter the quantum metric.

The corresponding pseudospin vector is

$$\mathbf{h}(\mathbf{q}) = \left(-2dq_x q_y,\, m,\, -d(q_x^2 - q_y^2)\right), \tag{402}$$

with magnitude

$$R(q) = \sqrt{m^2 + d^2 q^4}. \tag{403}$$

It is particularly transparent to use polar coordinates,

$$q_x = q\cos\theta, \qquad q_y = q\sin\theta. \tag{404}$$

Then

$$q_x^2 - q_y^2 = q^2 \cos 2\theta, \tag{405}$$

$$2q_x q_y = q^2 \sin 2\theta, \tag{406}$$

and hence

$$\widehat{\mathbf{h}} = \frac{1}{R}\left(-dq^2 \sin 2\theta,\; m,\; -dq^2 \cos 2\theta\right). \tag{407}$$

It is convenient to introduce an angle $\beta(q)$ through

$$\cos\beta = \frac{m}{R}, \qquad \sin\beta = \frac{dq^2}{R}. \tag{408}$$

Equation (407) can then be written as

$$\widehat{\mathbf{h}} = \left(-\sin\beta\,\sin 2\theta,\; \cos\beta,\; -\sin\beta\,\cos 2\theta\right). \tag{409}$$

The derivatives with respect to the polar coordinates obey

$$\partial_q \widehat{\mathbf{h}} \cdot \partial_\theta \widehat{\mathbf{h}} = 0, \tag{410}$$

so that the radial and angular directions are principal directions of the metric.
From

$$\tan\beta = \frac{dq^2}{m}, \tag{411}$$

one obtains

$$\frac{d\beta}{dq} = \frac{2dmq}{m^2 + d^2q^4}. \tag{412}$$

Using Eq. (399),

$$\begin{aligned} g_{qq} &= \frac{1}{4}\left(\frac{d\beta}{dq}\right)^2 \\ &= \boxed{\frac{d^2m^2q^2}{(m^2+d^2q^4)^2}}, \end{aligned} \tag{413}$$

while

$$\begin{aligned} g_{\theta\theta} &= \frac{1}{4}\left|\partial_\theta \widehat{\mathbf{h}}\right|^2 \\ &= \sin^2\beta \\ &= \boxed{\frac{d^2q^4}{m^2+d^2q^4}}. \end{aligned} \tag{414}$$

The mixed component vanishes,

$$g_{q\theta} = 0. \tag{415}$$

Notice that $g_{\theta\theta}$ is the metric component associated with the coordinate $\theta$. The corresponding tangential eigenvalue of the metric in an orthonormal Cartesian frame contains an additional factor $q^{-2}$.

### C. Radial and tangential metric eigenvalues

Let $\widehat{\mathbf{e}}_r$ and $\widehat{\mathbf{e}}_\theta$ denote the radial and tangential unit vectors in momentum space. The two eigenvalues of the Cartesian quantum metric are

$$\lambda_r = g_{qq} = \boxed{\frac{d^2m^2q^2}{(m^2+d^2q^4)^2}}, \tag{416}$$

and

$$\lambda_t = \frac{g_{\theta\theta}}{q^2} = \boxed{\frac{d^2q^2}{m^2+d^2q^4}}. \tag{417}$$

Thus

$$g_{ij} = \lambda_r \widehat{e}_{r,i}\widehat{e}_{r,j} + \lambda_t \widehat{e}_{\theta,i}\widehat{e}_{\theta,j}. \tag{418}$$

The Cartesian trace is consequently

$$\begin{aligned}\mathrm{Tr}\, g &= \lambda_r + \lambda_t \\ &= \frac{d^2m^2q^2}{(m^2+d^2q^4)^2} + \frac{d^2q^2}{m^2+d^2q^4} \\ &= \boxed{\frac{d^2q^2\left(2m^2+d^2q^4\right)}{(m^2+d^2q^4)^2}}.\end{aligned} \tag{419}$$

This is the expression quoted in the main text.

The two eigenvalues also clarify the origin of the integrated singularity. For momenta satisfying

$$q \gg q_m, \tag{420}$$

where the mass can be neglected,

$$\lambda_r \simeq \frac{m^2}{d^2q^6} \to 0, \tag{421}$$

$$\lambda_t \simeq \frac{1}{q^2}. \tag{422}$$

The logarithmic divergence of the integrated metric is therefore carried by the tangential quantum metric associated with the winding of the QBT, whereas the radial contribution remains integrable.

### D. Relation to the Berry curvature

For a two-level system the determinant of the quantum metric is related to the Berry curvature by

$$\det g = \frac{\Omega^2}{4}. \tag{423}$$

The QBT provides an explicit check of this identity. From Eqs. (416) and (417),

$$\begin{aligned}\det g &= \lambda_r\lambda_t \\ &= \frac{d^4m^2q^4}{(m^2+d^2q^4)^3}.\end{aligned} \tag{424}$$

The Berry curvature derived in Sec. IV B is

$$\Omega_\pm = \mp\frac{2d^2mq^2}{(m^2+d^2q^4)^{3/2}}, \tag{425}$$

and therefore indeed

$$\sqrt{\det g} = \frac{|\Omega_\pm|}{2}. \tag{426}$$

The Berry curvature changes sign under $m \to -m$, whereas the metric does not:

$$g_{ij}(\mathbf{q}, -m) = g_{ij}(\mathbf{q}, m). \tag{427}$$

The quantum metric therefore probes the magnitude of the QAH scale but not its chirality.

### E. Scaling form and geometric hot spot

The mass defines the characteristic momentum scale

$$\boxed{q_m = \sqrt{\frac{|m|}{d}}.} \tag{428}$$

Introduce the scaling variable

$$x = \frac{q}{q_m}. \tag{429}$$

Since

$$dq_m^2 = |m|, \tag{430}$$

Eq. (419) becomes

$$\boxed{\operatorname{Tr} g(\mathbf{q}, m) = \frac{d}{|m|} F(x),} \tag{431}$$

where

$$\boxed{F(x) = \frac{x^2(2 + x^4)}{(1 + x^4)^2}.} \tag{432}$$

Equation (431) exhibits directly the singular local geometry. The characteristic height scales as

$$\operatorname{Tr} g \sim \frac{d}{|m|}, \tag{433}$$

while the characteristic width scales as

$$q_m \sim |m|^{1/2}. \tag{434}$$

Thus the geometric structure becomes increasingly high and simultaneously increasingly narrow as the QAH gap closes.

The maximum of the scaling function is obtained from

$$\frac{dF}{dx} = 0, \tag{435}$$

which gives

$$x_*^4 = \frac{\sqrt{17} - 3}{2}. \tag{436}$$

Therefore

$$\boxed{x_* = \left( \frac{\sqrt{17} - 3}{2} \right)^{1/4} = 0.8656606\ldots} \tag{437}$$

and

$$F(x_*) = 0.7872003\ldots. \tag{438}$$

The maximum of the geometric hot spot consequently scales as

$$(\mathrm{Tr}\, g)_{\max} = 0.7872003\ldots \frac{d}{|m|}. \tag{439}$$

For $x \ll 1$,

$$F(x) = 2x^2 + O(x^6), \tag{440}$$

so the metric vanishes at the center of the massive QBT. For $x \gg 1$,

$$F(x) = \frac{1}{x^2} + O(x^{-6}), \tag{441}$$

and therefore

$$\mathrm{Tr}\, g \simeq \frac{1}{q^2}. \tag{442}$$

This long $1/q^2$ tail is responsible for the logarithmic integrated metric derived next.

### F. Integrated continuum metric

Define the integrated continuum metric with a momentum cutoff $\Lambda$ inside the regime of validity of the QBT theory,

$$G_{\mathrm{area}}(m;\Lambda) = \int_{q<\Lambda} \frac{d^2q}{(2\pi)^2} \mathrm{Tr}\, g(\mathbf{q}, m). \tag{443}$$

Using the scaling variable $x = q/q_m$ and

$$q\, dq = \frac{|m|}{d}\, x\, dx, \tag{444}$$

Eq. (431) gives

$$G_{\mathrm{area}} = \frac{1}{2\pi} \int_0^{\Lambda/q_m} dx\, xF(x). \tag{445}$$

Since

$$xF(x) = \frac{x^3(2+x^4)}{(1+x^4)^2}, \tag{446}$$

an antiderivative is

$$\int dx\, xF(x) = \frac{1}{4}\ln(1+x^4) - \frac{1}{4(1+x^4)}. \tag{447}$$

Evaluating the lower and upper limits yields

$$\boxed{G_{\mathrm{area}}(m;\Lambda) = \frac{1}{8\pi}\left[\ln\left(1+\frac{d^2\Lambda^4}{m^2}\right) + \frac{d^2\Lambda^4}{m^2+d^2\Lambda^4}\right].} \tag{448}$$

For

$$|m| \ll d\Lambda^2, \tag{449}$$

the weak-mass expansion is

$$\boxed{G_{\rm area}(m;\Lambda) = \frac{1}{4\pi}\ln\frac{d\Lambda^2}{|m|} + \frac{1}{8\pi} + o(1).} \tag{450}$$

The coefficient

$$\boxed{\frac{1}{4\pi}} \tag{451}$$

is independent of $d$. The parameter $d$ enters only through the energy scale $d\Lambda^2$ appearing inside the logarithm.

The origin of the logarithm can be seen directly without performing the full integral. There are three momentum regimes. In the inner region $q \lesssim q_m$, the metric has magnitude $O(d/|m|)$ but occupies an area $O(q_m^2) = O(|m|/d)$, and therefore contributes only $O(1)$. In the intermediate region

$$q_m \ll q \ll \Lambda, \tag{452}$$

Eq. (442) gives

$$\begin{aligned} G_{\rm area}^{\rm intermediate} &\simeq \frac{1}{2\pi}\int_{q_m}^{\Lambda}\frac{dq}{q} \\ &= \frac{1}{2\pi}\ln\frac{\Lambda}{q_m} \\ &= \frac{1}{4\pi}\ln\frac{d\Lambda^2}{|m|}. \end{aligned} \tag{453}$$

The logarithmic enhancement therefore does not arise simply from the height of the geometric hot spot. It is generated by the parametrically broad $1/q^2$ tail extending from the mass scale $q_m$ to the ultraviolet scale of the QBT theory.

### G. Radial and tangential contributions to the integral

The radial–tangential decomposition makes the preceding result even more explicit. Write

$$G_{\rm area} = G_r + G_t, \tag{454}$$

with

$$G_r = \frac{1}{2\pi}\int_0^{\Lambda} q\,dq\,\lambda_r, \tag{455}$$

$$G_t = \frac{1}{2\pi}\int_0^{\Lambda} q\,dq\,\lambda_t. \tag{456}$$

In terms of $x = q/q_m$,

$$G_r = \frac{1}{2\pi}\int_0^{\Lambda/q_m} dx\,\frac{x^3}{(1+x^4)^2}. \tag{457}$$

This integral remains finite for $\Lambda/q_m \to \infty$:

$$\boxed{G_r \longrightarrow \frac{1}{8\pi}.} \tag{458}$$

By contrast,

$$G_t = \frac{1}{2\pi}\int_0^{\Lambda/q_m} dx\,\frac{x^3}{1+x^4}, \tag{459}$$

which gives

$$G_t = \frac{1}{8\pi}\ln\left[1+\left(\frac{\Lambda}{q_m}\right)^4\right]. \tag{460}$$

Hence

$$G_t = \frac{1}{4\pi}\ln\frac{d\Lambda^2}{|m|} + o(1). \tag{461}$$

Thus

$$\boxed{\text{the logarithmic singularity of the integrated QBT metric is entirely tangential.}} \tag{462}$$

It originates from the winding-two angular texture of the QBT eigenstates, whereas radial changes of the pseudospin contribute only a finite amount.

### H. Geometric response to a nonperturbative mass scale

The preceding derivation is independent of how the mass $m$ is generated. If, however, the mass is produced self-consistently by a weak interaction,

$$|m| \sim M\exp\left(-\frac{4\pi d}{\lambda}\right), \tag{463}$$

then Eq. (450) gives

$$\begin{aligned} G_{\text{area}} &= \frac{1}{4\pi}\ln\frac{d\Lambda^2}{|m|} + O(1) \\ &= \frac{d}{\lambda} + O(1). \end{aligned} \tag{464}$$

Thus the exponentially small, beyond-all-orders spectral scale is converted by the singular QBT geometry into an algebraically divergent integrated response,

$$\boxed{|m| \sim e^{-4\pi d/\lambda} \qquad \Longrightarrow \qquad G_{\text{area}} \sim \frac{d}{\lambda}.} \tag{465}$$

This continuum relation is the basic quantum-geometric amplification mechanism discussed in the main text. In the next section we show how the same infrared logarithm is embedded in the full three-band kagome model and how finite contributions from the remainder of the Brillouin zone complete the continuum result.

## VI. LATTICE QUANTUM METRIC AND SINGULARITY SUBTRACTION

We now embed the continuum geometric singularity derived in the previous section into the full three-band kagome model. The lattice quantum metric is a full-Brillouin-zone quantity and therefore contains information about the Bloch eigenstates at all momenta. Nevertheless, its divergent weak-mass dependence originates exclusively from the quadratic band touching at $\Gamma$. This separation can be made explicit by subtracting and adding the analytically known continuum QBT contribution.

Throughout this section we work in the embedded Bloch convention. The relation to the periodic Bloch convention and the corresponding covariant momentum derivatives is discussed in Sec. VII.

### A. Quantum metric of the occupied lattice subspace

At filling two, the occupied subspace of the three-band kagome model is described by the rank-two projector

$$P_{\text{occ}}(\mathbf{q}; m_\Gamma) = \sum_{n\in\text{occ}} |u_{n\mathbf{q}}(m_\Gamma)\rangle\langle u_{n\mathbf{q}}(m_\Gamma)|, \tag{466}$$

where the eigenvectors are those of the full lattice Hamiltonian,

$$H_{\rm lat}(\mathbf{q};m_\Gamma)|u_{n\mathbf{q}}(m_\Gamma)\rangle=\varepsilon_{n\mathbf{q}}(m_\Gamma)|u_{n\mathbf{q}}(m_\Gamma)\rangle. \tag{467}$$

Here $\mathbf{q}=a\mathbf{k}$ is the dimensionless crystal momentum used in the continuum expansion.

The gauge-invariant quantum metric of the occupied subspace is

$$\boxed{g_{ij}^{\rm occ}(\mathbf{q};m_\Gamma)=\frac{1}{2}{\rm Tr}\left[\partial_{q_i}P_{\rm occ}\,\partial_{q_j}P_{\rm occ}\right].} \tag{468}$$

The trace on the right-hand side is over the three orbital components.

Equation (468) is invariant under arbitrary momentum-dependent unitary rotations among the two occupied states. This is important because band crossings within the occupied manifold do not affect the metric of the occupied subspace.

Since at filling two there is a single unoccupied band,

$$P_{\rm occ}=\mathbb{1}-P_{\rm unocc}, \tag{469}$$

and therefore

$$\partial_{q_i}P_{\rm occ}=-\partial_{q_i}P_{\rm unocc}. \tag{470}$$

It follows immediately that

$$\boxed{g_{ij}^{\rm occ}=g_{ij}^{\rm unocc}.} \tag{471}$$

Thus the metric of the rank-two occupied subspace can equivalently be computed from the complementary single unoccupied band.

For momenta where the occupied and unoccupied sectors are separated, Eq. (468) can also be written in the spectral form

$$g_{ij}^{\rm occ}(\mathbf{q})={\rm Re}\sum_{\substack{n\in{\rm occ}\\ \ell\in{\rm unocc}}}\frac{\langle u_{n\mathbf{q}}|\partial_{q_i}H_{\rm lat}|u_{\ell\mathbf{q}}\rangle\langle u_{\ell\mathbf{q}}|\partial_{q_j}H_{\rm lat}|u_{n\mathbf{q}}\rangle}{(\varepsilon_{\ell\mathbf{q}}-\varepsilon_{n\mathbf{q}})^2}. \tag{472}$$

This representation makes explicit that the full lattice metric depends on both the band energies and the Bloch eigenvectors throughout the Brillouin zone. At filling two, the sum contains contributions from both occupied bands to the single unoccupied band. Only the transition between the two bands forming the QBT becomes singular near Γ; all other processes remain finite.

### B. Integrated metric per primitive cell

The integrated lattice metric is defined as

$$\boxed{G_{\rm cell}(m_\Gamma)=\mathcal{A}_c\int_{\rm BZ}\frac{d^2q}{(2\pi)^2}\,{\rm Tr}\,g_{\rm occ}(\mathbf{q};m_\Gamma),} \tag{473}$$

where

$${\rm Tr}\,g_{\rm occ}=g_{xx}^{\rm occ}+g_{yy}^{\rm occ} \tag{474}$$

and

$$\mathcal{A}_c=\frac{A_c^{\rm phys}}{a^2}=\frac{\sqrt{3}}{2}. \tag{475}$$

The normalization is chosen such that

$$\mathcal{A}_c\int_{\rm BZ}\frac{d^2q}{(2\pi)^2}=1. \tag{476}$$

For brevity, define the momentum-resolved metric trace

$$\gamma_{\mathrm{lat}}(\mathbf{q}, m_\Gamma) \equiv \mathrm{Tr}\, g_{\mathrm{occ}}(\mathbf{q}; m_\Gamma). \tag{477}$$

Then

$$G_{\mathrm{cell}}(m_\Gamma) = \mathcal{A}_c \int_{\mathrm{BZ}} \frac{d^2q}{(2\pi)^2} \gamma_{\mathrm{lat}}(\mathbf{q}, m_\Gamma). \tag{478}$$

Equation (478) is the exact full-lattice quantity evaluated numerically. In particular, no continuum approximation is made for momenta away from Γ.

### C. Infrared matching to the continuum QBT

Near Γ, the full lattice Hamiltonian reduces to

$$\begin{aligned} H_{\mathrm{QBT}}(\mathbf{q}) = &\left(E_\Gamma - c_0 q^2\right) \mathbb{1} \\ &- d_0 \left[(q_x^2 - q_y^2)\sigma_z + 2q_x q_y \sigma_x\right] + m_\Gamma \sigma_y + \cdots , \end{aligned} \tag{479}$$

where

$$d_0 = \frac{t}{8}. \tag{480}$$

The ellipsis denotes higher-order terms in momentum and corrections to the lattice mass operator that are regular at the QBT.

The scalar coefficient $c_0$ does not affect the eigenvectors and therefore does not enter the leading quantum metric. The singular continuum metric trace is consequently

$$\boxed{\gamma_{\mathrm{QBT}}(\mathbf{q}, m_\Gamma) = \frac{d_0^2 q^2 \left(2m_\Gamma^2 + d_0^2 q^4\right)}{\left(m_\Gamma^2 + d_0^2 q^4\right)^2}}. \tag{481}$$

The characteristic momentum scale is

$$q_m = \sqrt{\frac{|m_\Gamma|}{d_0}}. \tag{482}$$

For

$$q_m \ll q \ll 1, \tag{483}$$

the lattice metric therefore approaches

$$\gamma_{\mathrm{lat}}(\mathbf{q}, m_\Gamma) \simeq \gamma_{\mathrm{QBT}}(\mathbf{q}, m_\Gamma) \simeq \frac{1}{q^2}. \tag{484}$$

This is the infrared origin of the logarithmic singularity in $G_{\mathrm{cell}}$.

The numerical scaling collapse shown in Fig. 1(c) of the main text directly verifies this matching: as $m_\Gamma$ decreases, the full three-band metric approaches the universal QBT scaling function within a shrinking neighborhood of Γ.

### D. Singularity subtraction

To separate the universal QBT singularity from the regular lattice contribution, choose a fixed disk

$$D_\Gamma(q_c) = \{\mathbf{q} : q < q_c\} \tag{485}$$

contained within the QBT regime. Splitting the Brillouin zone gives

$$\begin{aligned} G_{\rm cell}(m_\Gamma) = \mathcal{A}_c \int_{D_\Gamma} \frac{d^2q}{(2\pi)^2} \gamma_{\rm lat}(\mathbf{q}, m_\Gamma) \\ + \mathcal{A}_c \int_{{\rm BZ}\setminus D_\Gamma} \frac{d^2q}{(2\pi)^2} \gamma_{\rm lat}(\mathbf{q}, m_\Gamma). \end{aligned} \tag{486}$$

Inside $D_\Gamma$ we add and subtract the continuum QBT metric,

$$\begin{aligned} G_{\rm cell}(m_\Gamma) = \mathcal{A}_c \int_{D_\Gamma} \frac{d^2q}{(2\pi)^2} \left[\gamma_{\rm lat}(\mathbf{q}, m_\Gamma) - \gamma_{\rm QBT}(\mathbf{q}, m_\Gamma)\right] \\ + \mathcal{A}_c \int_{{\rm BZ}\setminus D_\Gamma} \frac{d^2q}{(2\pi)^2} \gamma_{\rm lat}(\mathbf{q}, m_\Gamma) \\ + \mathcal{A}_c \int_{D_\Gamma} \frac{d^2q}{(2\pi)^2} \gamma_{\rm QBT}(\mathbf{q}, m_\Gamma). \end{aligned} \tag{487}$$

Only the last term contains the infrared logarithm. The difference in the first line is integrable as $m_\Gamma \to 0$, because its leading QBT singularity has been removed. The second line also has a regular zero-mass limit, since the occupied and unoccupied subspaces are separated everywhere outside the QBT disk.

The last term can be evaluated analytically using the continuum result derived in Sec. V. One obtains

$$\begin{aligned} G^{\rm cell}_{\rm QBT}(m_\Gamma; q_c) \equiv \mathcal{A}_c \int_{D_\Gamma} \frac{d^2q}{(2\pi)^2} \gamma_{\rm QBT}(\mathbf{q}, m_\Gamma) \\ = \frac{\mathcal{A}_c}{8\pi} \left[ \ln\left(1 + \frac{d_0^2 q_c^4}{m_\Gamma^2}\right) + \frac{d_0^2 q_c^4}{m_\Gamma^2 + d_0^2 q_c^4} \right]. \end{aligned} \tag{488}$$

For

$$|m_\Gamma| \ll d_0 q_c^2, \tag{489}$$

this becomes

$$G^{\rm cell}_{\rm QBT} = \frac{\mathcal{A}_c}{4\pi} \ln \frac{d_0 q_c^2}{|m_\Gamma|} + \frac{\mathcal{A}_c}{8\pi} + o(1). \tag{490}$$

It follows that the full lattice metric has the weak-mass asymptotic form

$$\boxed{G_{\rm cell}(m_\Gamma) = \frac{\mathcal{A}_c}{4\pi} \ln \frac{t}{|m_\Gamma|} + B_g + o(1).} \tag{491}$$

The coefficient

$$\boxed{\rho_g \equiv \frac{\mathcal{A}_c}{4\pi}} \tag{492}$$

is fixed entirely by the infrared QBT and is independent of microscopic details away from $\Gamma$.

### E. Finite lattice contribution $B_g$

The finite constant in Eq. (491) is defined by

$$\boxed{B_g = \lim_{m_\Gamma \to 0} \left[ G_{\rm cell}(m_\Gamma) - \frac{\mathcal{A}_c}{4\pi} \ln \frac{t}{|m_\Gamma|} \right].} \tag{493}$$

As for the QAH kernel, it is useful to write $B_g$ as the sum of near-QBT, far-from-QBT, and matching contributions:

$$\boxed{B_g = B_g^{\rm near}(q_c) + B_g^{\rm far}(q_c) + B_g^{\rm match}(q_c).} \tag{494}$$

The regular contribution within the QBT disk is

$$B_g^{\rm near}(q_c) \equiv \mathcal{A}_c \lim_{m_\Gamma\to 0} \int_{D_\Gamma} \frac{d^2q}{(2\pi)^2} \left[\gamma_{\rm lat}(\mathbf{q}, m_\Gamma) - \gamma_{\rm QBT}(\mathbf{q}, m_\Gamma)\right]. \tag{495}$$

The contribution from the remainder of the Brillouin zone is

$$B_g^{\rm far}(q_c) \equiv \mathcal{A}_c \int_{{\rm BZ}\setminus D_\Gamma} \frac{d^2q}{(2\pi)^2} \gamma_{\rm lat}(\mathbf{q}, 0). \tag{496}$$

Finally, comparison of Eq. (490) with the reference logarithm $\ln(t/|m_\Gamma|)$ gives

$$B_g^{\rm match}(q_c) = \frac{\mathcal{A}_c}{4\pi} \ln \frac{d_0 q_c^2}{t} + \frac{\mathcal{A}_c}{8\pi}. \tag{497}$$

The three terms separately depend on the arbitrary matching momentum $q_c$, but their sum does not. Changing $q_c$ merely redistributes finite geometric weight between the continuum and lattice parts.

The role of eigenstates away from $\Gamma$ can be made fully explicit by inserting the spectral representation Eq. (472) into Eq. (496). One finds

$$B_g^{\rm far}(q_c) = \mathcal{A}_c \int_{{\rm BZ}\setminus D_\Gamma} \frac{d^2q}{(2\pi)^2} \sum_{i=x,y} \sum_{\substack{n\in{\rm occ}\\ \ell\in{\rm unocc}}} \frac{\left|\langle u_{\ell\mathbf{q}}^{(0)}|\partial_{q_i} H_{\rm lat}^{(0)}|u_{n\mathbf{q}}^{(0)}\rangle\right|^2}{\left(\varepsilon_{\ell\mathbf{q}}^{(0)} - \varepsilon_{n\mathbf{q}}^{(0)}\right)^2}. \tag{498}$$

Equation (498) shows explicitly that $B_g$ contains information about the Bloch eigenvectors and dispersions throughout the Brillouin zone. In particular, it contains finite transitions from the lower occupied kagome band to the unoccupied band, as well as the regular part of the transition between the two bands forming the QBT.

Thus the weak-mass lattice metric separates naturally into

$$\boxed{\underbrace{\frac{\mathcal{A}_c}{4\pi} \ln \frac{t}{|m_\Gamma|}}_{\text{universal QBT infrared contribution}} + \underbrace{B_g}_{\text{finite full-lattice completion}} + o(1).} \tag{499}$$

For the kagome model and the reference convention used throughout the numerical analysis, we obtain

$$\boxed{B_g = 0.08902304(2).} \tag{500}$$

## F. Numerical implementation of the subtraction

The subtraction in Eq. (487) is also useful numerically. Direct integration of the full lattice metric becomes increasingly difficult as $m_\Gamma \to 0$, because the characteristic momentum scale

$$q_m = \sqrt{\frac{|m_\Gamma|}{d_0}} \tag{501}$$

shrinks to zero while the local metric grows as $1/|m_\Gamma|$.

Instead, the singular QBT contribution is evaluated analytically through Eq. (488), while numerical quadrature is applied only to the regular difference

$$\gamma_{\rm lat} - \gamma_{\rm QBT} \tag{502}$$

inside the QBT disk and to the nonsingular remainder of the Brillouin zone. The resulting integrands remain well behaved in the weak-mass limit and do not require a uniform momentum grid capable of resolving $q_m$ directly.

Equivalently, one may evaluate the finite remainder

$$\Delta G(m_\Gamma) \equiv G_{\rm cell}(m_\Gamma) - \frac{\mathcal{A}_c}{4\pi} \ln \frac{t}{|m_\Gamma|}, \tag{503}$$

which approaches

$$\Delta G(m_\Gamma) \longrightarrow B_g \tag{504}$$

as $m_\Gamma \to 0$.

A complementary diagnostic is the logarithmic derivative

$$\frac{dG_{\rm cell}}{d\ln(t/|m_\Gamma|)} \longrightarrow \boxed{\frac{\mathcal{A}_c}{4\pi}}. \tag{505}$$

This removes the finite constant $B_g$ and provides a direct numerical test of the analytically predicted infrared coefficient. The lattice results and this derivative convergence are shown in Fig.1 (b) of the main text and its inset.

### G. Interaction dependence of the integrated geometry

The geometric singularity becomes particularly important when the mass is not externally imposed but generated self-consistently by the weak interaction. From Sec. III,

$$|m_\Gamma(V_1)| \sim M_{\rm lat} \exp\left(-\frac{A_{\rm lat}}{V_1}\right), \tag{506}$$

with

$$A_{\rm lat} = \frac{4\pi d_0}{\mathcal{A}_c}. \tag{507}$$

Equivalently,

$$\ln\frac{t}{|m_\Gamma|} = \frac{A_{\rm lat}}{V_1} - \ln\frac{M_{\rm lat}}{t} + o(1). \tag{508}$$

Substituting this result into Eq. (491) gives

$$\begin{aligned} G_{\rm cell}(V_1) = {} & \frac{\mathcal{A}_c}{4\pi}\frac{A_{\rm lat}}{V_1} \\ & + B_g - \frac{\mathcal{A}_c}{4\pi}\ln\frac{M_{\rm lat}}{t} + o(1). \end{aligned} \tag{509}$$

Using

$$\frac{\mathcal{A}_c}{4\pi}A_{\rm lat} = d_0, \tag{510}$$

we obtain

$$\boxed{G_{\rm cell}(V_1) = \frac{d_0}{V_1} + C_g + o(1),} \tag{511}$$

where

$$C_g = B_g - \frac{\mathcal{A}_c}{4\pi}\ln\frac{M_{\rm lat}}{t}. \tag{512}$$

For the bare-even kagome benchmark,

$$C_g = 0.08443960\ldots. \tag{513}$$

Equation (511) is the lattice realization of the quantum-geometric amplification mechanism:

$$\boxed{|m_\Gamma| \sim e^{-A_{\rm lat}/V_1} \quad \Longrightarrow \quad G_{\rm cell} \sim \ln\frac{1}{|m_\Gamma|} \sim \frac{1}{V_1}.} \tag{514}$$

For the microscopic kagome model,

$$d_0 = \frac{t}{8}, \tag{515}$$

and therefore

$$\boxed{\frac{V_1}{t}G_{\rm cell}(V_1) \longrightarrow \frac{1}{8}.} \tag{516}$$

### H. Role of the time-reversal-even Fock renormalization

We finally clarify why the coefficient in Eq. (516) is determined by $d_0 = t/8$, even though allowing the real Fock channel to relax renormalizes the quadratic coefficient to

$$d_{\rm eff}(V_1) = \frac{\tau}{8} = d_0 + O(V_1). \tag{517}$$

For the relaxed-even solution discussed above,

$$d_{\rm eff} = d_0 + \frac{V_1}{48} + O(V_1^2). \tag{518}$$

The leading geometric contribution may therefore be written schematically as

$$\frac{d_{\rm eff}(V_1)}{V_1}. \tag{519}$$

However,

$$\frac{d_{\rm eff}(V_1)}{V_1} = \frac{d_0}{V_1} + O(1). \tag{520}$$

In particular, for $d_{\rm eff} = d_0 + V_1/48 + O(V_1^2)$,

$$\frac{d_{\rm eff}}{V_1} = \frac{d_0}{V_1} + \frac{1}{48} + O(V_1). \tag{521}$$

The even-channel Hartree–Fock renormalization therefore modifies the finite part of $G_{\rm cell}$ but not the coefficient of its $1/V_1$ divergence. Consequently,

$$\lim_{V_1\to 0} \frac{V_1}{t} G_{\rm cell} = \frac{d_0}{t} = \frac{1}{8}, \tag{522}$$

irrespective of these $O(V_1)$ corrections to the quadratic QBT coefficient.

This distinction is the reason that the asymptotic coefficient in the main text is expressed in terms of the bare weak-coupling value

$$d_0 = \lim_{V_1\to 0} d_{\rm eff}(V_1), \tag{523}$$

rather than the finite-coupling quantity $d_{\rm eff}(V_1)$.

## VII. ORBITAL EMBEDDING AND COVARIANT DERIVATIVES

In a tight-binding representation with more than one orbital per unit cell, momentum derivatives require some care because the orbital positions inside the unit cell can either be included explicitly in the Bloch phases or absorbed into the Bloch Hamiltonian. The two choices are related by a momentum-dependent unitary transformation and describe the same physical system, but ordinary momentum derivatives do not transform covariantly between them. This distinction is relevant both for the lattice quantum metric discussed in Sec. VI and for the optical matrix elements discussed below.

Let $\mathbf{R}$ denote a Bravais lattice vector and $\mathbf{r}_\alpha$ the position of orbital $\alpha$ within the primitive cell. For the kagome lattice used throughout this work,

$$\mathbf{r}_A = 0, \qquad \mathbf{r}_B = \frac{\mathbf{a}_1}{2}, \qquad \mathbf{r}_C = \frac{\mathbf{a}_2}{2}. \tag{524}$$

We distinguish two standard Bloch conventions. In the periodic convention, the intra-cell positions are not included in the Fourier phase,

$$|\mathbf{k}, \alpha\rangle_{\rm per} = \frac{1}{\sqrt{N_c}} \sum_{\mathbf{R}} e^{i\mathbf{k}\cdot\mathbf{R}} |\mathbf{R}, \alpha\rangle, \tag{525}$$

whereas in the embedded convention

$$|\mathbf{k},\alpha\rangle_{\rm emb} = \frac{1}{\sqrt{N_c}}\sum_{\mathbf{R}} e^{i\mathbf{k}\cdot(\mathbf{R}+\mathbf{r}_\alpha)}|\mathbf{R},\alpha\rangle. \tag{526}$$

The two bases are related by

$$|\mathbf{k},\alpha\rangle_{\rm emb} = \sum_{\beta}|\mathbf{k},\beta\rangle_{\rm per}U_{\beta\alpha}(\mathbf{k}), \qquad U_{\alpha\beta}(\mathbf{k}) = \delta_{\alpha\beta}e^{i\mathbf{k}\cdot\mathbf{r}_\alpha}. \tag{527}$$

Consequently,

$$H_{\rm emb}(\mathbf{k}) = U^\dagger(\mathbf{k})H_{\rm per}(\mathbf{k})U(\mathbf{k}). \tag{528}$$

The kagome Hamiltonian written in Sec. I, with matrix elements proportional to $\cos(\mathbf{k}\cdot\mathbf{a}_i/2)$, is in the embedded convention.

The distinction becomes important upon differentiation with respect to crystal momentum. Introducing the diagonal matrices

$$(r_i)_{\alpha\beta} = (r_\alpha)_i\,\delta_{\alpha\beta}, \tag{529}$$

differentiating Eq. (528) gives

$$\partial_{k_i}H_{\rm emb} = U^\dagger\left[\partial_{k_i}H_{\rm per} - i[r_i, H_{\rm per}]\right]U. \tag{530}$$

It is therefore useful to define the covariant momentum derivative

$$\boxed{\mathcal{D}_iH_{\rm per} \equiv \partial_{k_i}H_{\rm per} - i[r_i, H_{\rm per}]}. \tag{531}$$

The physical velocity operator is then equivalently written as

$$v_i^{\rm emb} = \frac{1}{\hbar}\partial_{k_i}H_{\rm emb}, \qquad v_i^{\rm per} = \frac{1}{\hbar}\mathcal{D}_iH_{\rm per}, \tag{532}$$

with

$$v_i^{\rm emb} = U^\dagger v_i^{\rm per}U. \tag{533}$$

Thus the ordinary derivative of the periodic-gauge Hamiltonian alone is not the physical velocity operator. The commutator term in Eq. (531) restores the contribution from the intra-cell orbital positions.

For the kagome lattice this cancellation can be seen explicitly already at the $\Gamma$ point. For example, the $A$–$B$ matrix element is

$$(H_{\rm emb})_{AB} = -2t\cos\frac{\mathbf{k}\cdot\mathbf{a}_1}{2}, \tag{534}$$

whereas in the periodic convention one may write

$$(H_{\rm per})_{AB} = -t\left(1 + e^{-i\mathbf{k}\cdot\mathbf{a}_1}\right). \tag{535}$$

Although $\partial_{k_i}(H_{\rm emb})_{AB}|_\Gamma = 0$, the ordinary derivative of the periodic-gauge matrix element is nonzero,

$$\partial_{k_i}(H_{\rm per})_{AB}\big|_\Gamma = it(a_1)_i. \tag{536}$$

The additional term in the covariant derivative gives

$$-i[r_i, H_{\rm per}]_{AB}\big|_\Gamma = -it(a_1)_i, \tag{537}$$

and hence

$$(\mathcal{D}_iH_{\rm per})_{AB}\big|_\Gamma = 0, \tag{538}$$

as required. Omitting the commutator term would therefore introduce spurious interband matrix elements that depend only on the chosen Bloch representation.

The same covariance applies to the eigenvectors. If $|u_n^{\rm per}\rangle$ and $|u_n^{\rm emb}\rangle$ denote eigenvectors in the two conventions, then

$$|u_n^{\rm emb}\rangle = U^\dagger |u_n^{\rm per}\rangle, \tag{539}$$

and

$$\partial_{k_i} |u_n^{\rm emb}\rangle = U^\dagger \left(\partial_{k_i} - i r_i\right) |u_n^{\rm per}\rangle. \tag{540}$$

We therefore define the covariant derivative acting on a Bloch vector by

$$\mathcal{D}_i |u_n\rangle \equiv \left(\partial_{k_i} - i r_i\right) |u_n\rangle. \tag{541}$$

For an isolated band, the physical quantum metric can consequently be evaluated either with ordinary derivatives in the embedded representation or with covariant derivatives in the periodic representation,

$$g_{ij}^{(n)} = \mathrm{Re} \left\langle \mathcal{D}_i u_n \left| 1 - |u_n\rangle\langle u_n| \right| \mathcal{D}_j u_n \right\rangle. \tag{542}$$

For the rank-two occupied subspace relevant at filling two, it is more convenient to use the occupied projector

$$P_{\rm occ} = \sum_{n\in \rm occ} |u_n\rangle\langle u_n|. \tag{543}$$

Its covariant derivative is

$$\mathcal{D}_i P_{\rm occ} = \partial_{k_i} P_{\rm occ} - i[r_i, P_{\rm occ}], \tag{544}$$

and the gauge-invariant occupied-subspace metric becomes

$$\boxed{g_{ij}^{\rm occ} = \frac{1}{2}\,\mathrm{Tr}\left[(\mathcal{D}_i P_{\rm occ})(\mathcal{D}_j P_{\rm occ})\right]}. \tag{545}$$

Equation (545) is identical to the ordinary projector expression used in Sec. VI when evaluated in the embedded convention.

The relation to the velocity formulation follows directly by covariantly differentiating the eigenvalue equation. For two distinct bands $n \neq \ell$,

$$\langle u_\ell | \mathcal{D}_i H | u_n\rangle = (\epsilon_n - \epsilon_\ell)\langle u_\ell | \mathcal{D}_i u_n\rangle. \tag{546}$$

The occupied-subspace metric can therefore also be written as

$$g_{ij}^{\rm occ} = \mathrm{Re} \sum_{\substack{n\in \rm occ \\ \ell \in \rm unocc}} \frac{\langle u_n | \mathcal{D}_i H | u_\ell\rangle \langle u_\ell | \mathcal{D}_j H | u_n\rangle}{(\epsilon_\ell - \epsilon_n)^2}. \tag{547}$$

This is the covariant form of the spectral expression used in Sec. VI. In the embedded convention one simply replaces $\mathcal{D}_i H$ by $\partial_{k_i} H_{\rm emb}$.

Several comments are useful. First, Eqs. (532) and (545) concern the fixed physical embedding of the kagome orbitals. The periodic and embedded conventions are merely two representations of this same embedding; changing between them cannot alter any physical metric or optical response. A genuine change of the orbital positions $\mathbf{r}_\alpha$, by contrast, changes the physical tight-binding model and need not leave its quantum geometry unchanged. Second, a common displacement $\mathbf{r}_\alpha \to \mathbf{r}_\alpha + \mathbf{r}_0$ leaves all covariant derivatives unchanged because $[\mathbf{r}_0 \mathbb{1}, H] = 0$, so the construction is independent of the choice of spatial origin. Third, the scalar regularization introduced in Sec. I is proportional to the identity in orbital space and hence commutes with $r_i$. It therefore does not modify interband velocity matrix elements or the quantum metric at fixed QAH mass.

Throughout the lattice calculations in this work we use the embedded Bloch convention of Sec. I. The ordinary derivatives $\partial_{k_i} H_{\rm emb}$ employed in the quantum metric and optical matrix elements are therefore already the physical, gauge-consistent derivatives. Equations (531)– (547) show explicitly how the identical results are obtained in a periodic Bloch convention. In particular, the optical Kubo formula derived in the following section uses the same physical velocity operator as the lattice quantum metric, which ensures the gauge-consistent connection between quantum geometry and optical response.

## VIII. OPTICAL KUBO FORMULA AND SOUZA–WILKENS–MARTIN SUM RULE

We now connect the quantum metric to the longitudinal optical response. Throughout this section we write

$$\Omega = \hbar\omega \tag{548}$$

for the photon energy and define the dimensionless conductivity

$$s_{ij}(\Omega) \equiv \frac{\hbar}{e^2}\,\mathrm{Re}\,\sigma_{ij}(\omega). \tag{549}$$

We focus on the symmetric longitudinal response. The Hall response of the QAH state is controlled by the Berry curvature and is not involved in the Souza–Wilkens–Martin relation considered below.

For a clean band insulator at zero temperature, the absorptive interband conductivity can be written as

$$\mathrm{Re}\,\sigma_{ij}(\omega) = \pi e^2\hbar \int_{\mathrm{BZ}} \frac{d^2k}{(2\pi)^2} \sum_{\substack{n\in\mathrm{occ}\\ \ell\in\mathrm{unocc}}} \frac{\mathrm{Re}\left[v^i_{n\ell}(\mathbf{k})v^j_{\ell n}(\mathbf{k})\right]}{\epsilon_{\ell\mathbf{k}}-\epsilon_{n\mathbf{k}}}\,\delta[\hbar\omega-(\epsilon_{\ell\mathbf{k}}-\epsilon_{n\mathbf{k}})]\,, \tag{550}$$

where

$$v^i_{n\ell} = \langle u_n|v_i|u_\ell\rangle. \tag{551}$$

As discussed in Sec. VII, the physical velocity is

$$v_i = \frac{1}{\hbar}\mathcal{D}_i H, \tag{552}$$

where $\mathcal{D}_i H = \partial_{k_i}H - i[r_i, H]$ in the periodic Bloch convention, while $\mathcal{D}_i H = \partial_{k_i}H_{\mathrm{emb}}$ in the embedded convention used in our numerical calculations.

For $n \neq \ell$, the covariant Feynman–Hellmann relation derived in Sec. VII gives

$$\langle u_\ell|\mathcal{D}_i H|u_n\rangle = (\epsilon_n - \epsilon_\ell)\langle u_\ell|\mathcal{D}_i u_n\rangle. \tag{553}$$

It is therefore natural to introduce the transition-resolved quantum metric

$$g^{n\ell}_{ij}(\mathbf{k}) = \mathrm{Re}\left[\langle \mathcal{D}_i u_n|u_\ell\rangle\langle u_\ell|\mathcal{D}_j u_n\rangle\right] \tag{554}$$

or, equivalently,

$$g^{n\ell}_{ij} = \frac{\mathrm{Re}[\langle u_n|\mathcal{D}_i H|u_\ell\rangle\langle u_\ell|\mathcal{D}_j H|u_n\rangle]}{(\epsilon_\ell-\epsilon_n)^2}. \tag{555}$$

Equation (550) then becomes

$$\boxed{s_{ij}(\Omega) = \pi\Omega \int_{\mathrm{BZ}} \frac{d^2k}{(2\pi)^2} \sum_{\substack{n\in\mathrm{occ}\\ \ell\in\mathrm{unocc}}} g^{n\ell}_{ij}(\mathbf{k})\,\delta[\Omega-(\epsilon_{\ell\mathbf{k}}-\epsilon_{n\mathbf{k}})]}. \tag{556}$$

Thus the longitudinal optical conductivity directly probes the transition-resolved quantum metric on the constant-energy surface selected by the photon frequency. Summing over all unoccupied states gives the occupied-subspace metric,

$$g^{\mathrm{occ}}_{ij} = \sum_{\substack{n\in\mathrm{occ}\\ \ell\in\mathrm{unocc}}} g^{n\ell}_{ij}. \tag{557}$$

### A. Massive-QBT conductivity

We first apply Eq. (556) to the continuum Hamiltonian

$$H_{\rm QBT}(\mathbf{q}) = -cq^2\mathbb{1} - d\left[(q_x^2 - q_y^2)\sigma_z + 2q_xq_y\sigma_x\right] + m\sigma_y. \tag{558}$$

The scalar term proportional to $c$ affects neither the eigenstates nor the direct interband splitting. Defining

$$R(q) = \sqrt{m^2 + d^2q^4}, \tag{559}$$

the two bands are separated by

$$\Delta(q) = \epsilon_+(q) - \epsilon_-(q) = 2R(q). \tag{560}$$

The minimum direct gap is therefore $2|m|$.

As derived in Sec. V, the trace of the quantum metric is

$$\operatorname{Tr} g(\mathbf{q}, m) = \frac{d^2q^2(2m^2 + d^2q^4)}{(m^2 + d^2q^4)^2}. \tag{561}$$

Rotational symmetry implies

$$\int_0^{2\pi} d\theta\, g_{xx} = \int_0^{2\pi} d\theta\, g_{yy} = \pi\ \operatorname{Tr} g. \tag{562}$$

Equation (556) therefore reduces to

$$s_{xx}(\Omega) = \frac{\Omega}{4}\int_0^\infty q\,dq\ \operatorname{Tr} g(q, m)\,\delta[\Omega - 2R(q)]\,. \tag{563}$$

The delta function has support only for

$$\Omega \geq 2|m|. \tag{564}$$

For $\Omega > 2|m|$, the resonant momentum $q_\Omega$ is determined by

$$d^2q_\Omega^4 = \frac{\Omega^2}{4} - m^2. \tag{565}$$

Moreover,

$$\left.\frac{d}{dq}[2R(q)]\right|_{q=q_\Omega} = \frac{8d^2q_\Omega^3}{\Omega}, \tag{566}$$

so that

$$\delta[\Omega - 2R(q)] = \frac{\Omega}{8d^2q_\Omega^3}\delta(q - q_\Omega). \tag{567}$$

On the resonant shell,

$$\operatorname{Tr} g(q_\Omega, m) = \frac{4d^2q_\Omega^2(\Omega^2 + 4m^2)}{\Omega^4}. \tag{568}$$

Substitution into Eq. (563) yields

$$\boxed{s_{xx}(\Omega) = s_{yy}(\Omega) = \frac{1}{8}\left(1 + \frac{4m^2}{\Omega^2}\right)\Theta(\Omega - 2|m|)}. \tag{569}$$

This is the expression quoted in the main text. It is consistent with the optical response of a weakly gapped isotropic QBT discussed in Ref. [1].

Equation (569) makes two characteristic values manifest. Immediately above the absorption threshold,

$$s_{xx}(2|m|^{+}) = \frac{1}{4}, \tag{570}$$

whereas at frequencies large compared with the gap but still within the QBT regime,

$$s_{xx}(\Omega) \longrightarrow \frac{1}{8}, \qquad 2|m| \ll \Omega \ll \Omega_\Lambda. \tag{571}$$

The latter is the universal $d_{\max} = 1$ QBT conductivity. The QAH mass therefore primarily introduces an infrared absorption threshold and a near-threshold correction; it does not suppress the $O(e^2/\hbar)$ optical continuum above the gap.

Introducing the scaled frequency

$$y = \frac{\Omega}{2|m|}, \tag{572}$$

Eq. (569) takes the parameter-free form

$$s_{xx}(y) = \frac{1}{8}\left(1 + \frac{1}{y^2}\right)\Theta(y-1), \tag{573}$$

which is the continuum scaling curve used in Fig. 3 of the main text.

*a. Inverse-frequency optical moment.—* The finite QBT conductivity extending down to the mass scale produces a logarithmically enhanced inverse-frequency optical weight. We define the cumulative dimensionless longitudinal trace

$$\mathcal{W}^{\rm tr}_{-1}(\Omega) = \int_0^{\Omega} \frac{d\Omega'}{\Omega'}\left[s_{xx}(\Omega') + s_{yy}(\Omega')\right]. \tag{574}$$

Using Eq. (569), one obtains for $\Omega \geq 2|m|$

$$\boxed{\mathcal{W}^{\rm tr}_{-1}(\Omega) = \frac{1}{4}\left[\ln\frac{\Omega}{2|m|} + \frac{1}{2} - \frac{2m^2}{\Omega^2}\right]}, \tag{575}$$

while $\mathcal{W}^{\rm tr}_{-1} = 0$ below threshold. Equivalently, in terms of $y = \Omega/(2|m|)$,

$$\mathcal{W}^{\rm tr}_{-1}(y) = \frac{1}{4}\left[\ln y + \frac{1}{2} - \frac{1}{2y^2}\right], \qquad y \geq 1. \tag{576}$$

For $y \gg 1$,

$$\mathcal{W}^{\rm tr}_{-1}(y) = \frac{1}{4}\ln y + O(1). \tag{577}$$

The logarithm therefore arises directly from integrating the finite 1/8 QBT conductivity plateau with the measure $d\Omega/\Omega$. The continuum theory by itself requires an ultraviolet cutoff $\Omega_\Lambda$; in the full lattice model, the finite bandwidth and remote-band transitions provide the ultraviolet completion.

### B. Souza–Wilkens–Martin relation

The same logarithm can be expressed exactly as an integrated quantum metric through the Souza–Wilkens–Martin sum rule [2]. Define the full negative-first longitudinal optical moment

$$W^{\rm tr}_{-1} = \int_0^{\infty} \frac{d\omega}{\omega}\,{\rm Re}\left[\sigma_{xx}(\omega) + \sigma_{yy}(\omega)\right]. \tag{578}$$

Since $d\omega/\omega = d\Omega/\Omega$, integrating Eq. (556) over all positive frequencies removes the energy-conserving delta function:

$$W_{-1}^{\mathrm{tr}} = \frac{\pi e^2}{\hbar} \int_{\mathrm{BZ}} \frac{d^2k}{(2\pi)^2} \sum_{\substack{n\in\mathrm{occ} \\ \ell\in\mathrm{unocc}}} \left(g_{xx}^{n\ell} + g_{yy}^{n\ell}\right)$$
$$= \frac{\pi e^2}{\hbar} \int_{\mathrm{BZ}} \frac{d^2k}{(2\pi)^2} \,\mathrm{Tr}\, g^{\mathrm{occ}}(\mathbf{k}). \tag{579}$$

With the cell-normalized integrated metric introduced in Sec. VI,

$$G_{\mathrm{cell}} = A_c \int_{\mathrm{BZ}} \frac{d^2k}{(2\pi)^2} \,\mathrm{Tr}\, g^{\mathrm{occ}}(\mathbf{k}), \tag{580}$$

we obtain

$$\boxed{W_{-1}^{\mathrm{tr}} = \frac{\pi e^2}{\hbar A_c} G_{\mathrm{cell}}}. \tag{581}$$

Equation (581) is exact for the clean zero-temperature band insulator and includes all interband transitions over the full Brillouin zone. It is therefore not a low-energy approximation.

For the weakly gapped kagome QBT,

$$G_{\mathrm{cell}}(m_\Gamma) = \frac{A_c}{4\pi} \ln \frac{t}{|m_\Gamma|} + B_g + o(1), \tag{582}$$

and hence

$$W_{-1}^{\mathrm{tr}}(m_\Gamma) = \frac{e^2}{4\hbar} \ln \frac{t}{|m_\Gamma|} + O(1). \tag{583}$$

The coefficient of the logarithm is completely fixed by the infrared QBT, whereas the finite term contains the ultraviolet contribution from the rest of the Brillouin zone and from transitions involving the remote kagome band.

Finally, inserting the self-consistent weak-coupling mass

$$|m_\Gamma| \sim M_{\mathrm{lat}} e^{-A_{\mathrm{lat}}/V_1}, \qquad A_{\mathrm{lat}} = \frac{4\pi d_0}{A_c}, \tag{584}$$

or equivalently using $G_{\mathrm{cell}} = d_0/V_1 + O(1)$, gives

$$\boxed{W_{-1}^{\mathrm{tr}}(V_1) = \frac{\pi e^2 d_0}{\hbar A_c} \frac{1}{V_1} + O(1)}. \tag{585}$$

For the kagome model, $d_0 = t/8$ and $A_c = \sqrt{3}/2$, so

$$\boxed{\frac{V_1}{t} \frac{\hbar}{e^2} W_{-1}^{\mathrm{tr}} \longrightarrow \frac{\pi}{4\sqrt{3}}} \qquad (V_1/t \to 0). \tag{586}$$

The optical and geometric descriptions are therefore two equivalent representations of the same infrared amplification. In momentum space, the singular $1/q^2$ quantum-metric tail produces $G_{\mathrm{cell}} \sim \ln(1/|m_\Gamma|)$; in frequency space, the finite QBT optical continuum produces $\int d\Omega\, s(\Omega)/\Omega \sim \ln(1/|m_\Gamma|)$. The Souza–Wilkens–Martin relation guarantees that these two logarithms are identical. The transition-resolved lattice origin of the finite ultraviolet contribution is discussed in Sec. IX.

## IX. TRANSITION-RESOLVED OPTICAL WEIGHT

The exact Souza–Wilkens–Martin relation derived in Sec. VIII contains all interband transitions of the three-band kagome model. It is therefore useful to resolve the optical response into individual band pairs in order to identify which transitions generate the logarithmic infrared enhancement.

At filling two, we label the kagome bands in ascending energy order as $n = 0, 1, 2$, with bands 0 and 1 occupied and band 2 unoccupied. The quadratic touching at $\Gamma$ occurs between bands 1 and 2, whereas band 0 remains separated by an energy of order $t$. Accordingly, the longitudinal optical conductivity decomposes as

$$s_{ij}(\Omega) = s_{ij}^{0\to 2}(\Omega) + s_{ij}^{1\to 2}(\Omega), \tag{587}$$

with

$$s_{ij}^{n\to 2}(\Omega) = \pi\Omega \int_{\mathrm{BZ}} \frac{d^2k}{(2\pi)^2} g_{ij}^{n2}(\mathbf{k})\, \delta[\Omega - \Delta_{n2}(\mathbf{k})]\,, \tag{588}$$

where

$$\Delta_{n2}(\mathbf{k}) = \epsilon_{2\mathbf{k}} - \epsilon_{n\mathbf{k}} \tag{589}$$

and

$$g_{ij}^{n2}(\mathbf{k}) = \frac{\mathrm{Re}\left[\langle u_n|\mathcal{D}_i H|u_2\rangle\langle u_2|\mathcal{D}_j H|u_n\rangle\right]}{\Delta_{n2}^2(\mathbf{k})}. \tag{590}$$

The total occupied-subspace metric is correspondingly

$$g_{ij}^{\mathrm{occ}} = g_{ij}^{02} + g_{ij}^{12}. \tag{591}$$

For each transition channel we define the cell-normalized integrated metric

$$G_{n2} = A_c \int_{\mathrm{BZ}} \frac{d^2k}{(2\pi)^2} \left[g_{xx}^{n2}(\mathbf{k}) + g_{yy}^{n2}(\mathbf{k})\right], \tag{592}$$

so that

$$G_{\mathrm{cell}} = G_{02} + G_{12}. \tag{593}$$

Because the SWM relation is linear in the optical conductivity, it also holds separately for every interband channel:

$$W_{-1}^{n\to 2} = \int_0^\infty \frac{d\omega}{\omega} \mathrm{Re}\left[\sigma_{xx}^{n\to 2}(\omega) + \sigma_{yy}^{n\to 2}(\omega)\right] = \frac{\pi e^2}{\hbar A_c} G_{n2}. \tag{594}$$

Hence

$$W_{-1}^{\mathrm{tr}} = W_{-1}^{0\to 2} + W_{-1}^{1\to 2}. \tag{595}$$

The infrared distinction between the two channels follows directly from their direct excitation gaps. For the transition $1 \to 2$,

$$\Delta_{12}(\mathbf{k}) \longrightarrow 2\sqrt{m_\Gamma^2 + d_0^2 q^4} \tag{596}$$

near $\Gamma$, and therefore its minimum excitation energy collapses as

$$\Delta_{12}^{\min} = 2|m_\Gamma| \longrightarrow 0. \tag{597}$$

The corresponding transition metric inherits the singular QBT structure. In particular,

$$G_{12}(m_\Gamma) = \frac{A_c}{4\pi} \ln \frac{t}{|m_\Gamma|} + B_{12} + o(1), \tag{598}$$

where $B_{12}$ contains the finite difference between the full lattice transition metric and its continuum QBT limit.

By contrast, the lower occupied band remains separated from band 2 by a finite direct energy scale as $m_\Gamma \to 0$,

$$\Delta_{02}(\mathbf{k}) = O(t) \tag{599}$$

in the neighborhood of the QBT. Since neither the energy denominator nor the corresponding matrix elements develop an infrared singularity,

$$G_{02}(m_\Gamma) = B_{02} + o(1). \tag{600}$$

Thus

$$B_g = B_{12} + B_{02}, \tag{601}$$

while the coefficient of the logarithm in $G_{\rm cell}$ originates entirely from the $1 \to 2$ QBT transition.

The same separation is particularly transparent in frequency space. We define the cumulative dimensionless weight of each transition by

$$\mathcal{W}_{-1}^{n\to 2}(\Omega) = \int_0^\Omega \frac{d\Omega'}{\Omega'} \left[ s_{xx}^{n\to 2}(\Omega') + s_{yy}^{n\to 2}(\Omega') \right]. \tag{602}$$

For

$$2|m_\Gamma| \ll \Omega \ll \Omega_{\rm uv}, \tag{603}$$

where $\Omega_{\rm uv} = O(t)$ denotes the scale at which the QBT approximation ceases to apply, the $1 \to 2$ channel approaches the continuum result of Sec. VIII,

$$s_{xx}^{1\to 2} \simeq s_{yy}^{1\to 2} \simeq \frac{1}{8}. \tag{604}$$

Consequently,

$$\mathcal{W}_{-1}^{1\to 2}(\Omega) = \frac{1}{4} \ln \frac{\Omega}{2|m_\Gamma|} + O(1), \tag{605}$$

where the bounded term includes the near-threshold contribution proportional to $m_\Gamma^2/\Omega^2$ and lattice corrections.

The $0 \to 2$ channel has no corresponding low-energy continuum. Its optical absorption starts only at a finite lattice energy scale and therefore contributes a bounded amount as the QAH mass is sent to zero,

$$\mathcal{W}_{-1}^{0\to 2}(\Omega) = O(1) \qquad (m_\Gamma \to 0). \tag{606}$$

It can modify the high-frequency shape of the optical spectrum and the finite part of the total SWM moment, but it cannot change the coefficient of the infrared logarithm.

Integrating over the complete spectrum gives

$$W_{-1}^{1\to 2}(m_\Gamma) = \frac{e^2}{4\hbar} \ln \frac{t}{|m_\Gamma|} + W_{12}^{(0)} + o(1), \tag{607}$$

$$W_{-1}^{0\to 2}(m_\Gamma) = W_{02}^{(0)} + o(1), \tag{608}$$

and hence

$$W_{-1}^{\rm tr}(m_\Gamma) = \frac{e^2}{4\hbar} \ln \frac{t}{|m_\Gamma|} + W_{\rm uv}^{(0)} + o(1), \tag{609}$$

with

$$W_{\rm uv}^{(0)} = W_{12}^{(0)} + W_{02}^{(0)}. \tag{610}$$

The precise value of this finite constant depends on the complete three-band lattice structure, whereas the logarithmic coefficient is fixed solely by the infrared QBT.

This transition decomposition provides the microscopic meaning of the "ultraviolet completion" invoked in the main text. The remote kagome band does contribute to the total optical spectrum and to the finite part of the inverse-frequency moment, but the singular weak-coupling behavior does not originate from remote-band optical transitions. It arises entirely from the pair of bands participating in the quadratic touching, for which the interaction-generated mass moves the lower edge of an $O(e^2/\hbar)$ optical continuum to the exponentially small energy $2|m_\Gamma|$.

Combining Eq. (607) with $|m_\Gamma| \sim M_{\rm lat} e^{-A_{\rm lat}/V_1}$ gives

$$W_{-1}^{1\to 2}(V_1) = \frac{\pi e^2 d_0}{\hbar A_c} \frac{1}{V_1} + O(1), \tag{611}$$

whereas

$$W_{-1}^{0\to 2}(V_1) = O(1). \tag{612}$$

Thus the entire algebraically divergent $1/V_1$ optical weight can be traced microscopically to the transition across the interaction-gapped QBT.

## X. FINITE TEMPERATURE, DOPING, AND BROADENING

The asymptotic results discussed in the main text refer to a clean zero-temperature insulator with the chemical potential inside the interaction-generated QAH gap. Temperature, carrier doping, and finite quasiparticle lifetime introduce additional infrared scales. They do not modify the singular quantum geometry of a given clean Bloch Hamiltonian, but they modify the extent to which this geometry is sampled by the optical response. In particular, once one of these scales exceeds the QAH mass, the logarithmic optical enhancement is cut off before the intrinsic scale $|m_\Gamma|$ is reached.

For a clean system at arbitrary temperature and chemical potential, the interband Kubo formula of Sec. VIII generalizes to

$$s_{ij}(\Omega;T,\mu) = \pi\Omega \int_{\mathrm{BZ}} \frac{d^2k}{(2\pi)^2} \sum_{n<\ell} \mathcal{F}_{n\ell}(\mathbf{k}) g_{ij}^{n\ell}(\mathbf{k}) \delta[\Omega - \Delta_{n\ell}(\mathbf{k})]\,, \tag{613}$$

where

$$\mathcal{F}_{n\ell}(\mathbf{k}) = f[\epsilon_{n\mathbf{k}} - \mu] - f[\epsilon_{\ell\mathbf{k}} - \mu], \tag{614}$$

with

$$f(x) = \frac{1}{e^{x/(k_B T)} + 1}, \tag{615}$$

and $\Delta_{n\ell} = \epsilon_\ell - \epsilon_n > 0$. At $T = 0$ with the chemical potential in a band gap, $\mathcal{F}_{n\ell} = 1$ for occupied-to-unoccupied transitions and Eq. (613) reduces to Eq. (556) of Sec. VIII.

### A. Finite temperature

The effect of temperature can be seen analytically in the particle-hole symmetric massive-QBT limit,

$$\epsilon_\pm(\mathbf{q}) = \pm R(q), \qquad R(q) = \sqrt{m^2 + d^2 q^4}. \tag{616}$$

The scalar term $-cq^2\mathbb{1}$ of the general QBT Hamiltonian does not affect the direct interband energy or the quantum metric, although at finite temperature or finite doping it shifts the individual band energies entering the Fermi factors. We neglect this particle-hole asymmetry in the following analytic expressions; it is retained automatically in the full lattice calculation and does not alter the infrared scaling.

Measuring the chemical potential from the QBT energy, $\delta\mu = \mu - E_\Gamma$, the occupation factor evaluated on the optical shell $2R(q) = \Omega$ is

$$\mathcal{F}(\Omega;T,\delta\mu) = \frac{\sinh[\Omega/(2k_B T)]}{\cosh[\Omega/(2k_B T)] + \cosh[\delta\mu/(k_B T)]}. \tag{617}$$

The massive-QBT conductivity therefore becomes

$$\boxed{s_{xx}(\Omega;T,\delta\mu) = \frac{1}{8}\left(1 + \frac{4m^2}{\Omega^2}\right)\Theta(\Omega - 2|m|)\mathcal{F}(\Omega;T,\delta\mu)} \tag{618}$$

and $s_{yy} = s_{xx}$.

At charge neutrality, $\delta\mu = 0$, Eq. (617) reduces to

$$\mathcal{F}(\Omega;T,0) = \tanh\frac{\Omega}{4k_B T}, \tag{619}$$

so that

$$s_{xx}(\Omega;T) = \frac{1}{8}\left(1 + \frac{4m^2}{\Omega^2}\right)\tanh\frac{\Omega}{4k_B T}\,\Theta(\Omega - 2|m|). \tag{620}$$

For photon energies $\Omega \gg k_BT$, the zero-temperature result is recovered. Conversely, when the relevant transition energy is smaller than the thermal scale,

$$\tanh\frac{\Omega}{4k_BT} \simeq \frac{\Omega}{4k_BT}, \tag{621}$$

and the low-frequency interband response is suppressed by the thermal occupation of both bands.

This suppression cuts off the inverse-frequency logarithm. For example, at charge neutrality the QBT contribution has the asymptotic structure

$$\mathcal{W}_{-1}^{\mathrm{tr}} \sim \frac{1}{4}\ln\frac{\Omega_\Lambda}{\varepsilon_{\mathrm{IR}}} + O(1), \tag{622}$$

where factors of order unity inside the logarithm have been absorbed into the finite term and

$$\varepsilon_{\mathrm{IR}} \sim \max\{|m|, k_BT\}. \tag{623}$$

Thus the zero-temperature logarithm $\ln(\Omega_\Lambda/|m|)$ is observable only when

$$k_BT \ll |m|. \tag{624}$$

There is a second and conceptually distinct thermal effect in the interaction-driven problem: the QAH mass itself is temperature dependent. Within the continuum mean-field theory, the finite-temperature gap equation becomes

$$\frac{1}{\lambda} = \frac{1}{4\pi d}\int_0^{d\Lambda^2} d\varepsilon\, \frac{\tanh\left[\sqrt{\varepsilon^2+m^2(T)}/(2k_BT)\right]}{\sqrt{\varepsilon^2+m^2(T)}}. \tag{625}$$

At the transition temperature, $m(T_c) = 0$, and the logarithmic susceptibility is cut off by $k_BT_c$ rather than by the mass. Hence

$$k_BT_c \propto e^{-4\pi d/\lambda}, \tag{626}$$

with the same essential weak-coupling scale as the zero-temperature gap. For the ideal continuum model with a sharp ultraviolet cutoff,

$$\frac{|m(0)|}{k_BT_c} = \frac{\pi}{e^{\gamma_E}} \simeq 1.764, \tag{627}$$

where $\gamma_E$ is Euler's constant. The numerical prefactor is ultraviolet-model dependent beyond the ideal continuum approximation, whereas the common exponential dependence is fixed by the QBT singularity.

Equation (626) is important for the interpretation of the weak-coupling limit. At any fixed nonzero temperature, decreasing $V_1$ eventually drives the system above its exponentially small transition temperature, so the formal $G_{\mathrm{cell}}, W_{-1}^{\mathrm{tr}} \propto 1/V_1$ asymptotics refers to the ordered low-temperature regime in which $T/T_c$ remains sufficiently small.

### B. Finite chemical potential

At zero temperature, Eq. (617) approaches

$$\mathcal{F}(\Omega; 0, \delta\mu) = \Theta(\Omega - 2|\delta\mu|). \tag{628}$$

Together with the intrinsic direct-gap condition $\Omega \geq 2|m|$, the interband conductivity becomes

$$\boxed{s_{xx}(\Omega; 0, \delta\mu) = \frac{1}{8}\left(1 + \frac{4m^2}{\Omega^2}\right)\Theta[\Omega - 2\max(|m|, |\delta\mu|)]}. \tag{629}$$

When the chemical potential remains inside the gap, $|\delta\mu| < |m|$, the zero-doping result is unchanged. Once the chemical potential enters either QBT band, $|\delta\mu| > |m|$, Pauli blocking raises the lowest allowed interband transition from $2|m|$ to approximately $2|\delta\mu|$.

The corresponding interband inverse-frequency weight therefore behaves as

$$W^{\mathrm{tr}}_{-1,\mathrm{inter}} = \frac{e^2}{4\hbar} \ln \frac{\Omega_\Lambda}{\max(|m|, |\delta\mu|)} + O(1), \tag{630}$$

again up to unimportant factors of order unity in the ultraviolet and infrared reference scales. In particular, if $|\delta\mu| \gg |m|$, the logarithm no longer probes the exponentially small QAH mass.

A doped system is metallic and additionally contains an intraband Drude contribution at $\omega = 0$. The full negative-first moment $\int d\omega\, \mathrm{Re}\,\sigma(\omega)/\omega$ must therefore be treated with care in a metal. Equation (630) refers to the regular interband contribution after separating the Drude weight. Correspondingly, the zero-temperature insulating Souza–Wilkens–Martin relation to the occupied-projector metric does not apply unchanged once the chemical potential crosses a band.

For completeness, at finite temperature the frequency-integrated clean interband response satisfies the occupation-weighted relation

$$W^{\mathrm{tr}}_{-1,\mathrm{inter}}(T, \mu) = \frac{\pi e^2}{\hbar} \int_{\mathrm{BZ}} \frac{d^2k}{(2\pi)^2} \sum_{n<\ell} \mathcal{F}_{n\ell}(\mathbf{k}) \times \left[ g^{n\ell}_{xx}(\mathbf{k}) + g^{n\ell}_{yy}(\mathbf{k}) \right]. \tag{631}$$

At $T = 0$ in an insulator, $\mathcal{F}_{n\ell}$ reduces to the occupied-to-unoccupied projector and Eq. (631) becomes the SWM relation of Sec. VIII. At finite temperature it should instead be regarded as a thermally weighted geometric sum rather than as the zero-temperature localization tensor $G_{\mathrm{cell}}$.

### C. Finite broadening

A finite quasiparticle lifetime introduces a further energy scale $\Gamma$. Phenomenologically, the clean energy-conserving delta function may be replaced by a broadened spectral function,

$$\delta(\Omega - \Delta) \longrightarrow \delta_\Gamma(\Omega - \Delta), \qquad \delta_\Gamma(x) = \frac{1}{\pi} \frac{\Gamma}{x^2 + \Gamma^2}. \tag{632}$$

The sharp absorption edge at $2|m|$ is then rounded over an energy window of order $\Gamma$. For

$$\Omega \gg \Gamma \tag{633}$$

the clean massive-QBT form is recovered, whereas spectral features on energy scales smaller than $\Gamma$ cannot be resolved independently.

At the level of logarithmic accuracy, the experimentally accessible infrared window is therefore controlled by

$$\boxed{\varepsilon_{\mathrm{IR}} \sim \max\{|m|, k_B T, |\delta\mu|, \Gamma\}.} \tag{634}$$

Whenever

$$\varepsilon_{\mathrm{IR}} \ll \Omega_\Lambda, \tag{635}$$

there remains a parametrically broad regime in which the QBT conductivity is approximately constant,

$$\varepsilon_{\mathrm{IR}} \ll \Omega \ll \Omega_\Lambda, \tag{636}$$

and the cumulative interband optical weight grows as

$$\mathcal{W}^{\mathrm{tr}}_{-1} = \frac{1}{4} \ln \frac{\Omega}{\varepsilon_{\mathrm{IR}}} + O(1). \tag{637}$$

A constant Lorentzian broadening should not, however, be extrapolated literally to $\Omega \to 0$ inside the negative-first moment. Such a phenomenological replacement produces spectral tails below the clean absorption edge and does not by itself constitute a complete gauge-consistent treatment of dc transport. In the presence of scattering or metallic carriers, the experimentally relevant quantity is therefore the regular interband conductivity, with any Drude contribution separated, or a cumulative moment evaluated above a lower frequency larger than the transport

broadening scale. The exact projector form of the SWM relation used in the main text applies to the clean zero-temperature insulating limit.

The different infrared perturbations can thus be summarized by the hierarchy

$$|m_\Gamma| \quad \longrightarrow \quad \varepsilon_{\mathrm{IR}} \sim \max\{|m_\Gamma|, k_B T, |\delta\mu|, \Gamma\}\,. \tag{638}$$

The interaction-generated mass controls the optical amplification only when it is the largest relevant infrared cutoff of the coherent low-energy QBT problem. Otherwise the logarithmic response remains, but its lower limit is determined by temperature, Pauli blocking, or lifetime broadening rather than by the exponentially small spectral gap.

## XI. NUMERICAL CONVERGENCE AND ASYMPTOTIC SOLVERS

The weak-coupling regime considered in this work presents a numerical difficulty that is absent in conventional mean-field calculations. The self-consistent QAH mass becomes exponentially small,

$$|m_\Gamma| \sim M_{\mathrm{lat}} e^{-A_{\mathrm{lat}}/V_1}, \tag{639}$$

so that both the characteristic momentum scale $q_m \sim \sqrt{|m_\Gamma|/d_0}$ and the direct optical gap $2|m_\Gamma|$ rapidly become far smaller than any uniform numerical resolution. Direct iteration of the order parameter or direct Brillouin-zone sampling of the shrinking QBT region therefore becomes inefficient already at moderately weak coupling. We instead combine full-lattice calculations with explicit infrared subtraction and solve the self-consistency equation in a logarithmic mass variable.

### A. Logarithmic mass solver

For a prescribed QAH mass $m_\Gamma$, the full three-band lattice Hamiltonian is diagonalized over the Brillouin zone and the induced QAH response is evaluated from the finite-source occupied eigenstates. The cell-normalized lattice kernel introduced in Sec. III is

$$K_{\mathrm{lat}}^{\mathrm{cell}}(m_\Gamma) = -\frac{1}{m_\Gamma}\int_{\mathrm{BZ}} d\mu_{\mathbf{k}} \sum_{n\in\mathrm{occ}} \langle u_{n\mathbf{k}}(m_\Gamma)|\widehat{O}_m(\mathbf{k})|u_{n\mathbf{k}}(m_\Gamma)\rangle, \tag{640}$$

where

$$d\mu_{\mathbf{k}} = A_c \frac{d^2k}{(2\pi)^2}. \tag{641}$$

The Hartree–Fock self-consistency condition is the scalar equation

$$\frac{1}{V_1} = K_{\mathrm{lat}}^{\mathrm{cell}}(m_\Gamma). \tag{642}$$

Importantly, Eq. (640) is evaluated from the eigenvectors of the *finite-mass* lattice Hamiltonian. The zero-source spectral susceptibility given in Sec. III is used only to analyze the regular $m_\Gamma \to 0$ contribution away from the QBT and is not substituted for the finite-source kernel in the self-consistency calculation.

Rather than solving Eq. (642) directly for $m_\Gamma$, we introduce

$$u = \ln\frac{t}{|m_\Gamma|}, \qquad |m_\Gamma| = te^{-u}. \tag{643}$$

The root equation becomes

$$F(u;V_1) \equiv \frac{1}{V_1} - K_{\mathrm{lat}}^{\mathrm{cell}}(te^{-u}) = 0. \tag{644}$$

This change of variable removes the exponentially small unknown from the root-finding problem. In particular, the weak-coupling solution grows only algebraically,

$$u = \frac{A_{\mathrm{lat}}}{V_1} - \ln\frac{M_{\mathrm{lat}}}{t} + o(1), \tag{645}$$

and can therefore be followed over many decades in $|m_\Gamma|$ without requiring a momentum grid fine enough to resolve $q_m$ directly.

### B. Singularity subtraction for the lattice kernel

The numerical stability of the weak-mass calculation is further improved by separating the analytically known QBT logarithm. We write

$$K_{\mathrm{lat}}^{\mathrm{cell}}(m) = \rho_{\mathrm{lat}}^{\mathrm{cell}} \ln \frac{t}{|m|} + B_{\mathrm{lat}}^{\mathrm{cell}} + \Delta K(m), \tag{646}$$

with

$$\rho_{\mathrm{lat}}^{\mathrm{cell}} = \frac{A_c}{4\pi d_0}, \qquad \Delta K(m) \longrightarrow 0 \quad (m \to 0). \tag{647}$$

The remainder $\Delta K$ is regular in the weak-mass limit and can be interpolated or evaluated numerically without resolving the logarithmic singularity itself. Equation (644) can equivalently be written as

$$F(u; V_1) = \frac{1}{V_1} - \rho_{\mathrm{lat}}^{\mathrm{cell}} u - B_{\mathrm{lat}}^{\mathrm{cell}} - \Delta K(te^{-u}). \tag{648}$$

In the asymptotic regime the last term vanishes, giving directly

$$A_{\mathrm{lat}} = \frac{1}{\rho_{\mathrm{lat}}^{\mathrm{cell}}} = \frac{4\pi d_0}{A_c}, \tag{649}$$

and

$$M_{\mathrm{lat}} = t \exp\left(\frac{B_{\mathrm{lat}}^{\mathrm{cell}}}{\rho_{\mathrm{lat}}^{\mathrm{cell}}}\right). \tag{650}$$

Thus neither the coefficient of the essential singularity nor its prefactor needs to be obtained by fitting exponentially small numerical masses.

For the regularized kagome model used in the figures, $t = 1$ and $r/t = 0.05$, we obtain

$$B_{\mathrm{lat}}^{\mathrm{cell}} = 0.036667506(6)\, t^{-1}, \tag{651}$$

which gives

$$M_{\mathrm{lat}} = 1.0687690(1)\, t. \tag{652}$$

The leading coefficient is fixed analytically by $d_0 = t/8$ and $A_c = \sqrt{3}/2$,

$$\frac{A_{\mathrm{lat}}}{t} = \frac{\pi}{\sqrt{3}} = 1.813799364\ldots. \tag{653}$$

### C. Integrated quantum metric

An analogous subtraction is used for the Brillouin-zone integral of the occupied-projector quantum metric. We evaluate

$$G_{\mathrm{cell}}(m) = A_c \int_{\mathrm{BZ}} \frac{d^2k}{(2\pi)^2} \operatorname{Tr} g_{\mathrm{occ}}(\mathbf{k}, m) \tag{654}$$

using the gauge-consistent spectral representation of Secs. VI and VII. Inside a disk $D_\Gamma$ surrounding the quadratic touching, the continuum singular contribution

$$\gamma_{\mathrm{QBT}}(\mathbf{q}, m) = \frac{d_0^2 q^2 (2m^2 + d_0^2 q^4)}{(m^2 + d_0^2 q^4)^2} \tag{655}$$

is added and subtracted:

$$\begin{aligned} G_{\rm cell}(m) &= A_c \int_{D_\Gamma} \frac{d^2q}{(2\pi)^2} \left[\gamma_{\rm lat}(\mathbf{q},m) - \gamma_{\rm QBT}(\mathbf{q},m)\right] \\ &+ A_c \int_{{\rm BZ}\setminus D_\Gamma} \frac{d^2q}{(2\pi)^2} \gamma_{\rm lat}(\mathbf{q},m) \\ &+ A_c \int_{D_\Gamma} \frac{d^2q}{(2\pi)^2} \gamma_{\rm QBT}(\mathbf{q},m). \end{aligned} \tag{656}$$

The first two terms are regular as $m \to 0$, while the last term is evaluated analytically:

$$G^{\rm cell}_{\rm QBT}(m;q_c) = \frac{A_c}{8\pi}\left[\ln\left(1+\frac{d_0^2 q_c^4}{m^2}\right) + \frac{d_0^2 q_c^4}{m^2 + d_0^2 q_c^4}\right]. \tag{657}$$

Consequently,

$$G_{\rm cell}(m) = \frac{A_c}{4\pi}\ln\frac{t}{|m|} + B_g + o(1). \tag{658}$$

For the same regularized kagome model we find

$$B_g = 0.08902304(2). \tag{659}$$

The value of the auxiliary matching radius $q_c$ affects separately the near-QBT, far-zone, and analytic matching contributions, but their sum is independent of $q_c$ within numerical accuracy. This provides a direct convergence check on the singularity-subtraction procedure.

### D. Asymptotic coefficients without fitting

The logarithmic slope of the metric is fixed by the continuum theory,

$$\frac{dG_{\rm cell}}{d\ln(t/|m|)} \longrightarrow \frac{A_c}{4\pi}, \tag{660}$$

and is therefore tested directly rather than extracted from a free linear fit. The inset of Fig. 2(a) of the main text shows this logarithmic derivative approaching the parameter-free value $A_c/(4\pi)$.

Combining Eq. (658) with the independently determined lattice mass law gives

$$G_{\rm cell}(V_1) = \frac{d_0}{V_1} + C_g + o(1), \tag{661}$$

where the finite constant contains the ultraviolet metric contribution, the mass prefactor, and regular interaction-induced renormalizations. For the bare-even reference lattice calculation,

$$C_g = 0.08443960\ldots. \tag{662}$$

The quantity plotted in Fig. 2(b) is therefore expected to satisfy

$$\frac{V_1}{t}G_{\rm cell} \longrightarrow \frac{d_0}{t} = \frac{1}{8}, \tag{663}$$

again with no fitted leading coefficient.

The same logic applies to the optical moment. Using the exact Souza–Wilkens–Martin relation,

$$W^{\rm tr}_{-1} = \frac{\pi e^2}{\hbar A_c} G_{\rm cell}, \tag{664}$$

the weak-coupling limit is predicted before any optical fit is performed:

$$\frac{V_1}{t}\frac{\hbar}{e^2}W^{\rm tr}_{-1} \longrightarrow \frac{\pi}{4\sqrt{3}}. \tag{665}$$

### E. Convergence with Brillouin-zone discretization

The subtracted quantities entering Eqs. (646) and (656) are smooth functions of momentum. Their numerical integrals therefore converge with ordinary Brillouin-zone refinement even when $q_m$ is far smaller than the mesh spacing. We verify convergence by repeating the calculation for successively denser momentum meshes and by varying the radius $q_c$ of the analytic QBT patch. The quantities used as convergence diagnostics are

$$B_{\rm lat}^{\rm cell}, \qquad B_g, \qquad \frac{dG_{\rm cell}}{d\ln(t/|m|)}, \qquad \frac{V_1}{t}G_{\rm cell}, \tag{666}$$

rather than the exponentially small mass itself. The latter is obtained from the converged regular kernel through the logarithmic solver.

This procedure is essential at very weak coupling. A uniform grid would need a spacing

$$\Delta q \ll q_m = \sqrt{\frac{|m_\Gamma|}{d_0}} \tag{667}$$

to resolve the QBT core directly. Since $q_m \propto \exp[-A_{\rm lat}/(2V_1)]$, the required number of momentum points would itself grow exponentially as $V_1 \to 0$. Explicit singularity subtraction removes this numerical infrared catastrophe: the universal singular part is treated analytically, while only a smooth full-lattice remainder is integrated numerically.

### F. Extremely small masses and numerical precision

For sufficiently weak coupling, direct floating-point representation of $|m_\Gamma| = te^{-u}$ eventually underflows even though the logarithmic variable $u$ remains perfectly well behaved. In that regime all asymptotic quantities are therefore evaluated directly in terms of $u$. For example,

$$\ln\frac{t}{|m_\Gamma|} = u, \tag{668}$$

so neither the kernel asymptotics nor the integrated metric requires the explicit construction of the exponentially small number $m_\Gamma$. When an explicit value of the mass is required, arbitrary-precision arithmetic can be used after the root in $u$ has been determined.

This distinction is important for interpreting the extremely weak coupling data. The calculation does not infer the blue points in Fig. 1(b) from the asymptotic law $m_\Gamma \propto e^{-A_{\rm lat}/V_1}$. Rather, the full-lattice finite-source kernel is first evaluated and its scalar self-consistency equation is solved in the variable $u$; the analytic asymptotic law is then compared with these independently obtained lattice solutions.

The numerical strategy can therefore be summarized as a separation of infrared and ultraviolet information. The universal QBT singularities are integrated analytically, fixing the coefficients of $\ln(1/|m_\Gamma|)$ and $1/V_1$, while the full three-band lattice calculation supplies the finite ultraviolet constants and the nonuniversal mass prefactor. This separation permits controlled access to the asymptotic weak-coupling regime without requiring exponentially fine momentum grids.

## XII. EXACTLY-FLAT ENDPOINT

The nearest-neighbor kagome model possesses an exactly flat band that touches a dispersive band quadratically at $\Gamma$. In the main text we introduce a weak symmetry-preserving scalar regularization $r > 0$ in order to separate the physics of a generic isolated quadratic band touching from additional degeneracies associated with the exactly flat band. We clarify here the $r \to 0$ endpoint and show that the geometric and optical singularities discussed in this work have a smooth and particularly simple limit.

As derived in Sec. I, the regulator is

$$H_{\rm reg}(\mathbf{k}; r) = 2r\left[3 - \cos(\mathbf{k}\cdot\mathbf{a}_1) - \cos(\mathbf{k}\cdot\mathbf{a}_2) - \cos[\mathbf{k}\cdot(\mathbf{a}_2 - \mathbf{a}_1)]\right]\mathbb{1}_3. \tag{669}$$

It is proportional to the identity in sublattice space and vanishes at $\Gamma$. Expanding around the touching point gives

$$H_{\rm reg}(\mathbf{q}; r) = \frac{3r}{2}q^2\mathbb{1}_3 + O(q^4). \tag{670}$$

The low-energy QBT parameters are consequently

$$c_0 = \frac{t}{8} - \frac{3r}{2}, \qquad d_0 = \frac{t}{8}. \tag{671}$$

At zero QAH mass the two touching branches therefore satisfy

$$E_+(\mathbf{q}) - E_\Gamma = (d_0 - c_0)q^2 + O(q^4) = \frac{3r}{2}q^2 + O(q^4), \tag{672}$$

$$E_-(\mathbf{q}) - E_\Gamma = -(d_0 + c_0)q^2 + O(q^4). \tag{673}$$

For any $r > 0$, the upper branch bends upward and the lower branch bends downward, so that the $\Gamma$ point is an ordinary isolated QBT.

The exactly-flat endpoint is reached at

$$r = 0, \qquad c_0 = d_0 = \frac{t}{8}. \tag{674}$$

Then

$$E_+(\mathbf{q}) = E_\Gamma, \qquad E_-(\mathbf{q}) = E_\Gamma - 2d_0 q^2 + O(q^4), \tag{675}$$

and, for the nearest-neighbor kagome Hamiltonian, the upper branch is in fact exactly flat throughout the Brillouin zone. Thus the regularized model interpolates continuously between a conventional isolated QBT and the flat–dispersive quadratic touching of the ideal kagome lattice.

### A. Why the geometric sector is insensitive to the regulator

The crucial observation is that $r$ enters only through a scalar term. For the continuum Hamiltonian

$$H_{\mathrm{QBT}} = -c_0 q^2 \mathbb{1} - d_0\left[(q_x^2 - q_y^2)\sigma_z + 2q_xq_y\sigma_x\right] + m\sigma_y, \tag{676}$$

the eigenvectors depend only on the pseudospin vector

$$\mathbf{h}(\mathbf{q}) = \left(-2d_0 q_x q_y,\, m,\, -d_0(q_x^2 - q_y^2)\right), \tag{677}$$

and not on $c_0$. The direct interband separation is likewise

$$\Delta(\mathbf{q}) = 2\sqrt{m^2 + d_0^2 q^4}, \tag{678}$$

which is independent of $c_0$ and hence of $r$.

It follows immediately that the quantum metric, Berry curvature, and interband optical matrix elements are unchanged by the scalar regularization at fixed $m$. In particular,

$$\operatorname{Tr} g(\mathbf{q}, m) = \frac{d_0^2 q^2 (2m^2 + d_0^2 q^4)}{(m^2 + d_0^2 q^4)^2} \tag{679}$$

holds equally at $r > 0$ and at the exactly-flat endpoint $r = 0$. Consequently,

$$G_{\mathrm{QBT}}(m) = \frac{1}{4\pi} \ln \frac{E_{\mathrm{uv}}}{|m|} + O(1) \tag{680}$$

and the coefficient of the logarithmic divergence is completely insensitive to the scalar curvature $c_0$.

The same statement holds for the optical response. Since both the transition matrix elements and the direct excitation energy (678) are independent of $c_0$,

$$s_{xx}(\Omega) = s_{yy}(\Omega) = \frac{1}{8}\left(1 + \frac{4m^2}{\Omega^2}\right)\Theta(\Omega - 2|m|) \tag{681}$$

is unchanged upon taking $r \to 0$. The universal $1/8$ QBT plateau and its inverse-frequency logarithm therefore survive smoothly at the flat-band endpoint.

This invariance is even stronger in the full lattice model. Since Eq. (669) is proportional to $\mathbb{1}_3$, adding $r$ leaves the Bloch eigenvectors and all same-$\mathbf{k}$ interband energy differences exactly unchanged:

$$|u_{n\mathbf{k}}(r)\rangle = |u_{n\mathbf{k}}(0)\rangle, \qquad \epsilon_{\ell\mathbf{k}}(r) - \epsilon_{n\mathbf{k}}(r) = \epsilon_{\ell\mathbf{k}}(0) - \epsilon_{n\mathbf{k}}(0). \tag{682}$$

Therefore, as long as the occupied space is represented by the same rank-two band projector, the lattice quantum metric, transition-resolved interband optical conductivity, and corresponding SWM weight at fixed QAH source are unchanged by $r$. The regulator changes the absolute dispersion and hence the indirect energetic placement of states, but not their local Bloch geometry or direct interband excitation energies.

### B. Interaction-generated mass

The same distinction appears in the QAH susceptibility. The singular part of the self-consistency kernel is controlled by the direct splitting of the two QBT bands and therefore does not involve the scalar coefficient $c_0$:

$$K^{\mathrm{cell}}_{\mathrm{QBT}}(m) = \frac{A_c}{4\pi d_0}\,\mathrm{asinh}\,\frac{d_0 q_c^2}{|m|}. \tag{683}$$

Thus

$$K^{\mathrm{cell}}_{\mathrm{QBT}}(m) = \frac{A_c}{4\pi d_0}\ln\frac{1}{|m|} + O(1), \tag{684}$$

with the same logarithmic coefficient at $r = 0$ and $r > 0$. The corresponding essential scale in the QAH channel is therefore

$$|m_\Gamma| \sim M_{\mathrm{lat}} \exp\left[-\frac{4\pi d_0}{A_c V_1}\right], \tag{685}$$

and, since $d_0 = t/8$ is itself independent of $r$,

$$A_{\mathrm{lat}} = \frac{4\pi d_0}{A_c} = \frac{\pi t}{\sqrt{3}} \tag{686}$$

remains unchanged at the exactly-flat endpoint.

Combining Eqs. (680) and (685) therefore gives exactly the same leading geometric amplification,

$$G_{\mathrm{cell}}(V_1) = \frac{d_0}{V_1} + O(1), \tag{687}$$

and hence the same optical law

$$W^{\mathrm{tr}}_{-1}(V_1) = \frac{\pi e^2 d_0}{\hbar A_c}\frac{1}{V_1} + O(1). \tag{688}$$

The leading asymptotic coefficients are therefore properties of the quadratic pseudospin texture rather than of the scalar curvature of the two touching bands.

### C. Why we nevertheless keep $r > 0$

Although the geometric and interband-response sector has a smooth $r \to 0$ limit, setting $r = 0$ from the outset changes the many-body starting point in an important way. At the exactly-flat nearest-neighbor kagome point, an entire one-particle band lies at the QBT energy. The noninteracting system therefore contains an extensive flat-band degeneracy in addition to the local quadratic touching at $\Gamma$. Interaction channels associated with this macroscopic degeneracy can compete with, or preempt, the marginal QBT instability.

This issue is distinct from the infrared mechanism studied in the main text. Our aim is to isolate the sequence

$$V_1 \longrightarrow |m_\Gamma| \sim e^{-A_{\mathrm{lat}}/V_1} \longrightarrow G_{\mathrm{cell}} \sim \ln\frac{1}{|m_\Gamma|} \longrightarrow \frac{1}{V_1}, \tag{689}$$

which is controlled by the logarithmically singular susceptibility of an isolated QBT. Taking a weak $r > 0$ removes the extensive flat-band degeneracy while leaving the eigenvectors, the direct QBT splitting, $d_0$, and all leading geometric coefficients unchanged. It therefore provides a clean diagnostic that the asymptotic amplification reported here is a QBT effect rather than a consequence of an exactly flat band.

There is accordingly an order-of-limits distinction. For

$$r \to 0^+ \tag{690}$$

taken after identifying the QBT instability, or at any fixed nonzero QAH mass, the geometric and optical results approach the exactly-flat endpoint smoothly. By contrast, setting $r = 0$ first at the noninteracting point and then asking for the unrestricted interacting ground state exposes the full macroscopic flat-band degeneracy and may allow additional weak-coupling instabilities not contained in the isolated-QBT description. We do not make a claim here about the complete competition among such orders at the strictly flat microscopic endpoint.

The regulator should therefore not be viewed as an ingredient responsible for the amplification mechanism. Its purpose is precisely the opposite: it removes an additional flat-band singularity while preserving the quadratic pseudospin texture. The persistence of the mass law, quantum metric singularity, universal optical continuum, and SWM enhancement as $r \to 0^+$ demonstrates that the results of this work originate from the singular quantum geometry of the QBT itself.



\end{document}